\documentclass[12pt,letterpaper, titlepage]{article}

\usepackage[margin=1in]{geometry}
\usepackage{pdflscape}

\usepackage{amsmath, amsthm, bm}

\graphicspath{{exhibits/}}
\makeatletter
\def\input@path{{exhibits/}}
\makeatother

\usepackage[justification=justified, labelfont={color=black, bf}, labelsep=period, singlelinecheck=false, font=footnotesize]{caption}
\usepackage[justification=justified, labelfont={color=black, bf}, font=footnotesize]{subcaption}
\usepackage{graphicx, booktabs, multicol, multirow, tikz, adjustbox, tabularx, rotating, dcolumn, array, threeparttable}
\usepackage{makecell}
\usepackage{longtable}
\usepackage{afterpage}
\usepackage{siunitx}
\usetikzlibrary{positioning}
\usepackage{verbatim, setspace}
\usepackage{enumitem}

\usepackage{eurosym}

\usepackage[colorlinks=true, linkcolor=prussianblue, citecolor=prussianblue, urlcolor=prussianblue]{hyperref}
\usepackage[stable]{footmisc}

\usepackage[T1]{fontenc}
\usepackage{txfonts}

\usepackage{xcolor,colortbl}
\definecolor{LightGray}{gray}{0.95}
\definecolor{moonstoneblue}{rgb}{0.45, 0.66, 0.76}
\definecolor{prussianblue}{RGB}{0, 49, 82}
\definecolor{yaleblue}{RGB}{14, 77, 146}

\usepackage{titlesec}

\usepackage[backend=biber, maxnames=5, style=authoryear, doi=false, isbn=false, url=false, eprint=false, uniquelist=false, uniquename=false]{biblatex}

\AtEveryBibitem{\clearfield{month} \clearfield{day} \clearfield{urlyear} \clearfield{urlmonth}}
\renewbibmacro{in:}{}

\usepackage{framed, titling}

\newtheorem{lemma}{Lemma}[section]
\newtheorem{proposition}{Proposition}[section]
\newtheorem{assumption}{Assumption}[section]

\newcolumntype{d}[1]{D{.}{.}{#1}}

\DeclareCiteCommand{\cite}
{\usebibmacro{prenote}}
{\usebibmacro{citeindex}%
  \printtext[bibhyperref]{\usebibmacro{cite}}}
{\multicitedelim}
{\usebibmacro{postnote}}

\DeclareCiteCommand*{\cite}
{\usebibmacro{prenote}}
{\usebibmacro{citeindex}%
  \printtext[bibhyperref]{\usebibmacro{citeyear}}}
{\multicitedelim}
{\usebibmacro{postnote}}

\DeclareCiteCommand{\parencite}[\mkbibparens]
{\usebibmacro{prenote}}
{\usebibmacro{citeindex}%
  \printtext[bibhyperref]{\usebibmacro{cite}}}
{\multicitedelim}
{\usebibmacro{postnote}}

\DeclareCiteCommand*{\parencite}[\mkbibparens]
{\usebibmacro{prenote}}
{\usebibmacro{citeindex}%
  \printtext[bibhyperref]{\usebibmacro{citeyear}}}
{\multicitedelim}
{\usebibmacro{postnote}}

\DeclareCiteCommand{\footcite}[\mkbibfootnote]
{\usebibmacro{prenote}}
{\usebibmacro{citeindex}%
  \printtext[bibhyperref]{ \usebibmacro{cite}}}
{\multicitedelim}
{\usebibmacro{postnote}}

\DeclareCiteCommand{\footcitetext}[\mkbibfootnotetext]
{\usebibmacro{prenote}}
{\usebibmacro{citeindex}%
  \printtext[bibhyperref]{\usebibmacro{cite}}}
{\multicitedelim}
{\usebibmacro{postnote}}

\DeclareCiteCommand{\textcite}
{\boolfalse{cbx:parens}}
{\usebibmacro{citeindex}%
  \printtext[bibhyperref]{\printnames{labelname}\printtext{ (\usebibmacro{citeyear}\printtext{)}}}}
{\ifbool{cbx:parens}
  {\bibcloseparen\global\boolfalse{cbx:parens}}
  {}%
  \multicitedelim}
{\usebibmacro{textcite:postnote}}

\hypersetup{hypertexnames=false}

\newtheorem{definition}{Definition}[section]

\newcommand{\E}{\mathbb{E}}
\renewcommand{\Pr}{\operatorname{Pr}}

\newcommand{\rbar}{\bar{r}}

\newcommand{\ind}{\mathbf{1}}

\newcommand{\syutitle}{Optimal Trading Speed on Blockchain}

\usepackage{pdflscape}
\begin{document}

\title{\syutitle\thanks{We thank Ying Chen, Data Always, Fay\c{c}al Drissi (discussant), Andrey Malenko, Katya Malinova (discussant), Christian Opp, Burak \"{O}z, Christoph Schlegel, Elvira Sojli (discussant), Mihir Wadekar, and participants at the SMU Finance Summer Camp 2026, the MEV Workshop at the 2026 Science of Blockchain Conference, the 2026 Nanyang Blockchain Conference, the 10th SAFE Market Microstructure Conference, and the 2026 WBS Gillmore Centre Conference on DeFi and Digital Currencies for helpful comments. We gratefully acknowledge financial support from a Flashbots MEV Fellowship Grant. This study was funded by the Singapore Management University through a research grant (grant approval number \mbox{23-LKCSB-SMU-047}) from the Ministry of Education Academic Research Fund Tier 1.}}
\author{Ruizhe Jia\thanks{Department of Management Science and Engineering, Stanford University; ruizhe@stanford.edu.} \and Shihao Yu\thanks{Lee Kong Chian School of Business, Singapore Management University; shihaoyu@smu.edu.sg.}}
\date{Current version: November 2, 2026}

\begingroup
\thanksmarkseries{fnsymbol}
\maketitle
\endgroup
\setcounter{footnote}{0}

\begin{titlepage}
    \null\vfil
    \begin{center}
        {\LARGE\syutitle\par}
        \vspace{2.5em}
        {\bfseries\abstractname\par}
    \end{center}
    \vspace{1em}
    \begingroup\singlespacing\noindent On blockchain-based exchanges, faster blocks weaken the priority-fee competition among arbitrageurs and shift rents from the block producer to them.
    Equilibrium liquidity peaks at an interior block speed; beyond a threshold, an active liquidity provider cream-skims and passive capital shrinks.
    Evidence from the Base Flashblocks upgrade, a tenfold block-speed increase, confirms the predictions.
    Priority-fee bids decrease sharply and arbitrageurs' net profit rises.
    The active liquidity provider rebalances faster, avoids toxic flow, and earns a higher profit.
    Pool depth falls and effective spreads widen despite closer tracking of external prices.
    \par\endgroup
    \vfil\null
\end{titlepage}

\clearpage

\doublespacing

\section{Introduction}\label{sec:introduction}

Blockchain-based financial infrastructure is moving from a crypto-native experiment toward a broader layer for asset representation, trading, and settlement.
This shift reached U.S. stock trading on September 17, 2026, when the Securities and Exchange Commission (SEC) issued temporary, conditional exemptions that let U.S. stocks trade as blockchain tokens in automated market maker (AMM) liquidity pools \parencite{SEC2026TSV}.
Such a pool is a program that holds a shared reserve of tokens and prices each trade by a formula.
The exemptions require these pools to be deployed on a public, permissionless blockchain, which anyone can read and write to.
The design of that blockchain therefore becomes part of the design of the market for tokenized U.S. stocks.

Trading on blockchains is already large, and it is no longer confined to cryptocurrencies.
In August 2026, decentralized exchanges recorded approximately \$233.5 billion in spot trading volume and \$553 billion in perpetual futures trading volume.\footnote{Spot volume is the sum of the 31 daily observations in DeFiLlama's all-chain decentralized-exchange volume series for August 2026, totaling \$233,480,726,515.97: \url{https://defillama.com/dexs}. Perpetual volume is reported in CryptoRank's August 2026 market recap, which cites CryptoRank and DeFiLlama: \url{https://cryptorank.io/insights/reports/crypto-market-recap-august-2026}. Both figures describe tracked decentralized-exchange activity rather than an exhaustive census, and use the same complete calendar month.}
Tokenized U.S. equities now trade on Solana, on Robinhood Chain, and on Hyperliquid's on-chain order books.\footnote{Robinhood Chain's stock tokens are documented at \url{https://docs.robinhood.com/chain/stock-tokens/}. RWA.xyz's tokenized-stocks dashboard reports \$13.0 billion in on-chain transfer volume over the 30 days to September 16, 2026, and \$2.9 billion in tokenized-equity value outstanding (\url{https://app.rwa.xyz/stocks}, accessed September 16, 2026). Transfer volume counts all on-chain token transfers, not only trades.}
The exemptions are designed to facilitate the trading of tokenized U.S. stocks on public blockchains, which can expand this market.

Trading on a blockchain differs from trading on an exchange in a basic way: transactions do not execute as they arrive, but wait to execute together in the next block, and blocks come at intervals set by the chain's protocol.
Within each block, the sequencer, the party that builds the block, decides which transaction goes first and usually sells that priority through fees.
The time between blocks is therefore a design choice of the chain, and every pool on the chain inherits it as its execution clock.

\begin{figure}[t]
    \centering
    \begin{subfigure}[b]{0.48\textwidth}
        \centering
        \includegraphics[width=\textwidth]{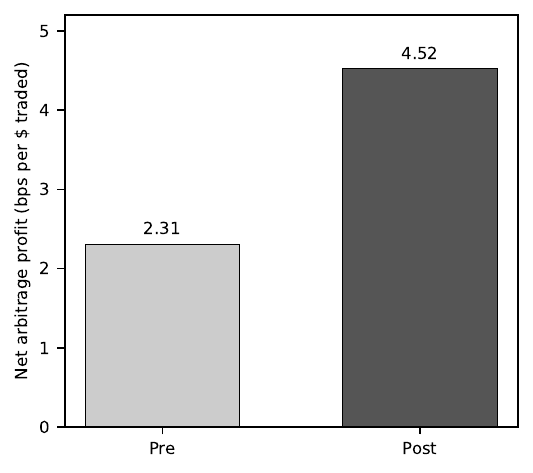}
        \caption{Net arbitrage profit (bps per dollar traded)}
        \label{fig:flashblocks_puzzle_profit}
    \end{subfigure}
    \hfill
    \begin{subfigure}[b]{0.48\textwidth}
        \centering
        \includegraphics[width=\textwidth]{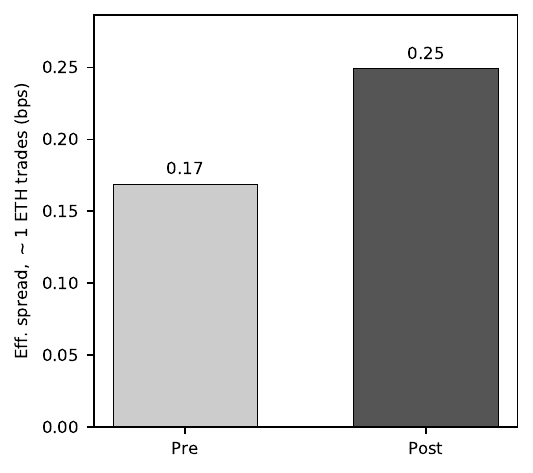}
        \caption{Effective spread, $\sim$1 ETH trades (bps)}
        \label{fig:flashblocks_puzzle_spread}
    \end{subfigure}
    \caption{\textbf{Arbitrage profits and effective spreads on Base around Flashblocks.} This figure plots, for Base, arbitrageurs' net profit per dollar traded and the effective spread, before and after the public launch on July~16, 2025; the post-launch sample begins on July~17. Panel~(a) plots the net markout on arbitrage swaps in basis points (bps): the 10-second price move against the Binance mid-quote (the midpoint of Binance's best bid and ask), net of the blockchain fee. Panel~(b) plots the effective spread (the execution price relative to the pool's price just before the trade, excluding the fee) in bps for trades of $\sim$1 ETH. Results for the comparison chain, Arbitrum, are reported in the empirical section. Data: Uniswap~v3 WETH--USDC 5~bp pool on Base, May~16 to September~15, 2025; the July~7--16 transition window is excluded.}
    \label{fig:flashblocks_puzzle}
\end{figure}

Blockchains are now competing to shorten the time between blocks, the block interval.
In July 2025, Base, a blockchain operated by Coinbase, cut its effective block interval from two seconds to 200 milliseconds by introducing Flashblocks, which release a sub-block of transactions every 200 milliseconds within each two-second block.\footnote{Base describes the shorter transaction-ordering window in \url{https://blog.base.dev/accelerating-base-with-flashblocks}.}
One would expect faster blocks to help liquidity providers: they can adjust their positions every 200 milliseconds instead of every two seconds, pool prices stay fresher, and fewer liquidity traders give up waiting.
The chain should gain as well, with more liquidity, more users, and more fee revenue.
Base's evidence points the other way.
As Figure~\ref{fig:flashblocks_puzzle} shows, after Flashblocks launched, arbitrageurs' profit per dollar traded, net of the blockchain fee, nearly doubled, and the effective spread, the cost of trading against the pool, widened.
Faster execution went together with larger arbitrage rents and worse liquidity, the opposite of what faster price updating would predict.

This paper shows that faster blocks fundamentally change how a blockchain allocates priority.
That change reshapes how arbitrageurs and liquidity providers compete.
Among arbitrageurs, competition moves from a priority auction toward first come, first served, which shifts arbitrage rents from the sequencer to the arbitrageurs.
Among liquidity providers, an active provider can withdraw ahead of arbitrage trades while sharing the fees from non-arbitrage ones, so she cream-skims the passive providers.
Beyond a certain block speed, the pool's total liquidity falls as well.
Faster blocks can thus undermine what makes an AMM pool attractive for equities: letting long-term asset holders supply liquidity passively \parencite{kp2023dex}.
We test these predictions using Base's switch to Flashblocks as a quasi-experiment, and the evidence is consistent with each of them.

In our model, arbitrageurs, liquidity traders, and an active liquidity provider send transactions to a sequencer, who groups them into blocks.
Liquidity traders abandon transactions that wait too long for a block, so faster blocks let more of their trades execute.
Within each block, transactions execute in priority order.
Across blocks, the earlier block executes first.
This ordering shapes two competitions.
First, when news makes the pool price stale, arbitrageurs bid for priority to trade against it.
Each arbitrageur chooses the percentage of the total arbitrage prize to pay the sequencer.
Second, the active liquidity provider tries to withdraw her capital before the arbitrage trade, while passive liquidity providers keep their capital in the pool.
Block speed is the only thing we vary: it shifts how much the ordering turns on priority rather than on arrival, and with it every player's incentives, strategy, and payoff.

\begin{figure}[t!]
    \centering
    \includegraphics[width=\textwidth]{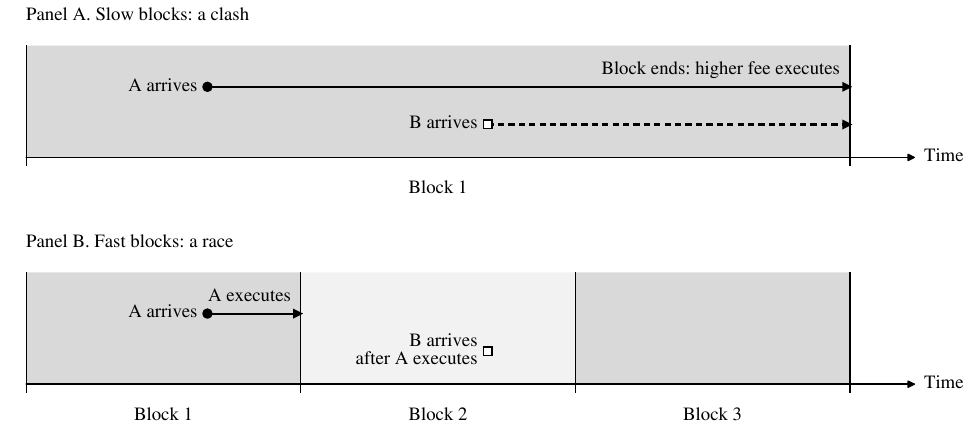}
    \caption{\textbf{Block timing and the form of competition.}
        Arbitrage transactions A and B arrive at the same times in both panels; each arbitrageur chooses the fee it pays the sequencer, as a percentage of the arbitrage prize, when it sends its transaction, without seeing whether its rival has arrived.
        In Panel A, both arrive within one long block and the higher fee executes.
        In Panel B, the same arrival times fall across shorter blocks: A executes at the end of the first block, and B arrives after the opportunity is taken.
        Both outcomes can occur at any block speed; faster blocks make clashes less likely.}
    \label{fig:clash-race-rents-review}
\end{figure}

The first result is that faster blocks lower the sequencer's share of the arbitrage prize and raise the share that arbitrageurs keep.
The central mechanism is that block speed changes how often arbitrageurs \emph{clash} (arrive in the same block and compete for priority) or \emph{race} (compete across blocks for who lands first).
When blocks are slow, the time between blocks is long, so two arbitrageurs' transactions are more likely to arrive before the same block.
They clash and enter a priority auction.
The auction pushes bids toward the arbitrage prize, so the sequencer captures a larger share of the prize.
When blocks are fast, the time between blocks is short: one arbitrageur is more likely to execute before its rival arrives, and the rival cannot outbid an already completed trade.
Competition therefore turns increasingly on which transaction reaches the sequencer first, moving outcomes closer to first come, first served (Figure~\ref{fig:clash-race-rents-review}).
The shift comes from the chain's clock alone: no arbitrageur becomes faster, yet faster blocks turn their competition from an auction into a race.

The shift from clashes to races also explains why active liquidity management emerges on fast chains.
On a slow chain, a liquidity provider's withdrawal and an arbitrage trade are likely to wait for the same block.
When they clash, her withdrawal loses to arbitrage and fails to protect her capital.
Faster blocks give her withdrawal a greater chance to execute in an earlier block, protecting her capital from the arbitrage loss.
Once blocks are sufficiently fast, this protection makes withdrawing and redepositing worthwhile, even though her transactions travel no faster than those of the arbitrageurs.

The second result is that sufficiently fast blocks let an active liquidity provider (active LP) cream-skim the passive providers.\footnote{Our active liquidity provider resembles a high-frequency market maker who monitors public prices and cancels stale quotes before they are picked off \parencite{budish2015high}.
    In an AMM, she manages this exposure by withdrawing capital and redepositing after an arbitrage trade brings the pool price back to the reference price.
    Passive providers instead leave their capital deposited, earning trading fees while bearing arbitrage losses \parencite{LeharParlour2023,zhang2023amm,capponi2025liquidity}.
    This also differs from just-in-time (JIT) provision: a JIT provider observes a particular pending swap and places liquidity immediately before it, then removes that liquidity immediately afterward.
    Those transactions can be arranged within one block, so JIT provision can operate on slow-block chains through control of transaction ordering \parencite{WanAdams2022,CapponiJiaZhu2024JIT}.
    Our active provider instead responds to price news and, in our model, does not bid for priority over arbitrageurs.
    With slow blocks, her withdrawal and the arbitrage transaction are likely to await the same block, where arbitrage executes first.
    Faster blocks give her an opportunity to withdraw in an earlier block, making active management worthwhile.
    Separately, in practice, frequent withdrawals and redeposits must justify their gas costs; cheaper transactions on Layer~2 (L2) chains make such repeated adjustments more economical \parencite{fan2023strategic,AdamsL2scaling2024}.}
With slow blocks she stays deposited; with fast blocks she withdraws as soon as a stale price becomes profitable to arbitrage.
Fees and arbitrage losses are shared pro rata among the liquidity deposited at the time of each trade, so by being absent for some arbitrage losses while still sharing the fees, she avoids some losses that passive providers still bear.
The faster the blocks, the more often her withdrawal executes before the arbitrage trade.
Faster blocks also reduce passive providers' share of the pool while she is deposited.

The third result is that pool depth, the liquidity the pool offers traders, is greatest at a unique intermediate block speed.
Faster blocks serve more liquidity traders, generating fee income that attracts passive capital.
At higher speeds, this fee benefit levels off.
Meanwhile, passive providers face more frequent arbitrage losses.
At sufficiently high block speeds, faster blocks raise the active LP's income even as passive capital retreats and pool liquidity falls.

We test the three main model predictions using a difference-in-differences design around the public launch of Base Flashblocks on July 16, 2025, which shortened the effective block interval from two seconds to 200 milliseconds while preserving the priority gas auction ordering rule within each Flashblock. The post-upgrade sample begins on July 17, 2025. Arbitrum, another Ethereum Layer~2 blockchain whose block interval did not change during the sample, serves as the comparison chain. The analysis focuses on the Uniswap~v3 WETH--USDC 5~bp pool (wrapped ether traded against the USDC dollar stablecoin, with a 5~basis-point swap fee), using on-chain transaction data matched to Binance order-book snapshots from May to September 2025.

The evidence on arbitrage supports the first prediction: faster blocks redistribute rents from the sequencer to arbitrageurs.
Gas bids fall most on large arbitrage trades, where executing ahead of rivals is worth the most and arbitrageurs had paid the highest gas prices.
The decline grows with trade size across the top three size deciles, reaching $73\%$ in the top decile.
Smaller trades, which pay only the baseline gas price set by network congestion, show no change.

Arbitrageurs earn more per dollar traded because they pay less in blockchain fees.
The blockchain fee falls by $3.26$~bps, while the raw markout before the blockchain fee falls by only $0.53$~bps, a statistically insignificant change.
The net markout therefore rises by $2.73$~bps.
The same holds in dollars: the gross profit extracted from the pool falls by about $29\%$ on Base, yet net profit rises by 40\% because the gas saving more than offsets the smaller gross opportunity.
Faster blocks shrink the prize and, by more, the cost of capturing it, moving rents from the sequencer to arbitrageurs.

The evidence on liquidity provision supports the second prediction: faster blocks let the active LP avoid some arbitrage losses while continuing to share trading fees.
She withdraws and redeposits her capital faster without bidding more for priority.
Her rebalancing time falls by about 66\% after Flashblocks, while her gas price does not change relative to Arbitrum.
Faster rebalancing pays off: her per-dollar profit rises by $0.73$~bps, from approximately break-even to positive, and she scales up her liquidity capital substantially.

Her gain comes from cream-skimming: she increasingly bears order flow that does not adversely select her and avoids the flow that does.
We measure her exposure to such toxic flow: trades for which the price gap against her exceeds the fee she earns.
Her share of exposure to these trades falls by $8.93$ percentage points more than the pool's toxic-flow share.
Comparing her share with the pool's nets out the change in the mix of flow reaching the pool, so what remains is her own withdrawal from toxic flow.

The evidence on pool depth is consistent with the third prediction for sufficiently fast blocks: even as the active LP gains, pool depth falls.
Depth at the current pool price falls by about 19\%.
Liquidity also becomes more concentrated: the largest single liquidity provider's share of pool liquidity rises by $4.20$ percentage points.
Shallower depth goes with higher transaction costs.
The effective spread on trades of about 1~ETH widens by $0.07$~bps, a $47\%$ increase, with larger increases for larger trades.
Trading volume also falls by 18\%.

This paper builds on the literature on trading speed in traditional markets.
This strand has established that faster markets are not unambiguously better: continuous execution creates arbitrage rents and a socially wasteful speed race among high-frequency traders \parencite{budish2015high}; faster exchange speed can either improve or worsen liquidity depending on the news-to-liquidity-trader ratio \parencite{Menkveld2017NeedLiquidity}; and private investment in speed can be excessive \parencite{biais2011}.
In periodic-auction models, an interior optimal trading frequency balances quicker responses to information against less aggressive demand schedules in each auction \parencite{DuZhu2017}.
Here block timing changes competition for priority and the capital that active and passive liquidity providers keep available for trading, even though participants' message speeds remain fixed.
Blockchains also differ in block speed and change it in production, which provides a rare laboratory for testing theories of trading speed.

A second strand of literature studies adverse selection and liquidity provision on AMMs.
\textcite{LeharParlour2023}, \textcite{zhang2023amm}, and \textcite{capponi2025liquidity} characterize the stale-price losses borne by passive LPs; \textcite{HasbrouckRiveraSaleh2023} study concentrated-liquidity range choice; and \textcite{park2023} examines how invariant-based pricing invites toxic flow.
A related set of papers documents active strategies: \textcite{fan2023strategic} analyze strategic concentrated provision under gas costs; \textcite{WanAdams2022} document just-in-time liquidity and \textcite{CapponiJiaZhu2024JIT} show that JIT LPs can crowd out passive LPs; \textcite{kp2023dex} analyze the risk-sharing benefits of AMM pooling.
We contribute to this literature with an equilibrium model in which active liquidity provision emerges endogenously as block speed rises, and in which the active provider can earn more while passive capital and total depth fall.

Finally, our paper connects to the literature on Maximal Extractable Value (MEV), ordering rents, and L2 market structure.
\textcite{Daian2020} document the prevalence of priority-gas-auction arbitrage on decentralized exchanges, and the proposer-builder separation literature \parencite{Schwarz-SchillingTime23, BahraniGarimidiRoughgarden2024} formalizes the market for block-construction rents.
\textcite{AdamsL2scaling2024}, \textcite{caparros2023blockchain}, and \textcite{gogol2024crossrollup} examine how cheaper gas and faster blocks on L2s affect Uniswap~v3 execution, LP repositioning, and cross-rollup arbitrage.
\textcite{WuOz2026} study the same Flashblocks upgrade through the lens of atomic cyclic arbitrage, showing that faster sub-blocks reshape the mix of on-chain search strategies.
A parallel mechanism-design literature proposes alternative AMM and blockchain execution designs to reduce stale-price rents \parencite{CanidioFritsch2023, AdamsMoallemiReynoldsRobinson2024}.
We share with these strands the view that infrastructure is not neutral to AMMs and that transaction ordering is economically valuable.
Our contribution is to provide a tractable equilibrium model in which block speed jointly determines the allocation of rents across arbitrageurs, liquidity providers, and the sequencer.

The remainder of the paper is organized as follows.
Section~\ref{sec:institutional} describes the institutional setting: how blockchain settlement, AMM pricing, and L2 execution clocks shape the economic environment.
Section~\ref{sec:model} presents the model.
Section~\ref{sec:equilibrium} examines rent allocation, liquidity management, passive capital supply, and the conditions for an interior liquidity maximum.
Section~\ref{sec:empirical} takes the model's predictions to data using the Base Flashblocks quasi-experiment.
Section~\ref{sec:conclusion} concludes.
\section{Institutional Details}\label{sec:institutional}

This section describes the institutional features of blockchain-based trading that motivate and inform the model.
Section~\ref{sec:inst-settlement} explains how a blockchain executes transactions in blocks and who decides their order, and Section~\ref{sec:inst-l2} explains how Layer~2 rollups work and compares their activity and block intervals with Ethereum.
Section~\ref{sec:inst-amm} describes how automated market makers price trades from pooled liquidity, and Section~\ref{sec:inst-lp} explains why their quotes become stale, why a liquidity provider cannot simply withdraw ahead of an arbitrage when both wait for the same block, and why faster blocks reopen that question.

\subsection{Blockchain execution and transaction ordering}\label{sec:inst-settlement}
\begin{figure}[t!]
    \centering
    \includegraphics[width=\textwidth]{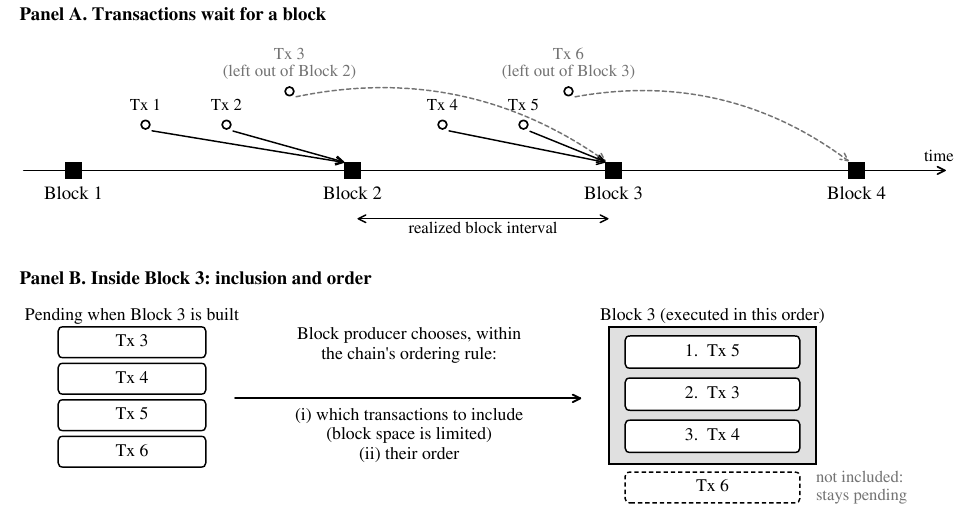}
    \caption{\textbf{Transaction execution on a blockchain.}
        This figure illustrates how transactions wait for blocks and how they are ordered within a block.
        Panel~A places six transactions on a timeline: each can be included no earlier than the first block after it arrives, and Transactions~3 and~6 are left out of that block and included in a later one.
        The protocol sets a target block interval, and the realized interval can vary.
        Panel~B shows the two choices of the block producer for Block~3: which pending transactions to include, given limited block space, and in what order to execute them.
        The order shown is one order that the chain's ordering rule allows.
        The figure is a schematic.}\label{fig:inst-execution}
\end{figure}

On a traditional exchange, a matching engine processes incoming orders continuously: each order is matched against the standing book as soon as it arrives.
On a blockchain, execution is batched.
Submitted transactions wait for inclusion in a \emph{block}, which applies them to the shared ledger in a specified sequence \parencite{bitcoin}.
The interval between blocks therefore determines the finest granularity at which trades can execute.
A transaction submitted between two blocks can be included no earlier than the next block, and it waits longer if that block leaves it out (Figure~\ref{fig:inst-execution}, Panel~A).

Within each block, a designated entity, the \emph{block producer}, exercises temporary monopoly control over two decisions: which pending transactions to include, and in what order to arrange them.%
\footnote{On Ethereum, a validator selected as the block proposer proposes the block to the network.
    The proposer may delegate block construction, including the selection and ordering of transactions, to a specialized builder \parencite{BahraniGarimidiRoughgarden2024}.}
Users pay a transaction fee, commonly referred to as a \emph{gas fee}, for executing transactions on the blockchain.\footnote{On Ethereum, the gas fee is calculated by multiplying the gas price by the amount of gas needed to execute the transaction.
    The gas price represents the amount of ETH paid per unit of gas, and the gas amount indicates the computational resources required.
    The unit for gas price is Gwei, where 1 Gwei equals $10^{-9}$ ETH.}
Blocks are constrained by size, and where priority is allocated through gas bids, transactions with higher fees are typically prioritized to maximize the block producer's revenue; transactions with lower fees may remain pending for a longer time before being included.
This fee-based priority rule within a block is known as a \emph{priority gas auction} (PGA).
Under \emph{first-come, first-served} (FCFS) ordering, the block producer instead commits to processing transactions in the order they arrive.
On a decentralized exchange, every action, whether a trade, a deposit of liquidity, or a withdrawal of liquidity, is a transaction that waits for a block and takes a position within it (Figure~\ref{fig:inst-execution}, Panel~B).
The block interval and the ordering rule are both design choices of a blockchain: the protocol sets a target interval, and the realized spacing between blocks can vary.\footnote{Shorter block intervals can create trade-offs with security and decentralization because participants need time to receive and verify blocks \parencite{croman2016scaling, gervais2016security}.
    We study how block speed affects trading within a given blockchain architecture.}
This ordering discretion (the block producer's ability to choose which transactions execute and in what sequence) is the institutional feature that determines the division of the economic rents central to our analysis.\footnote{A block is the blockchain analog of a batch, and the ordering rule within it plays the role of the matching rule on a traditional exchange: continuous price-time priority rewards speed, whereas a batch auction removes the reward \parencite{budish2015high}.
    On a blockchain the rule is a choice of the block producer, so it can differ across chains and change over time.}

\subsection{Layer~2 blockchains}\label{sec:inst-l2}
The demand for faster and less costly transactions has motivated the development of Layer~2 (L2) blockchains such as Ethereum rollups.
A rollup is a separate execution environment that runs on top of Ethereum, a Layer~1 (L1) blockchain: it processes transactions on its own chain, with its own block interval, throughput capacity, and ordering rule, and periodically posts compressed transaction data and commitments to its resulting state to Ethereum for settlement and security \parencite{ethereum2026rollups}.
Block intervals shorter than Ethereum's reduce the wait for execution, and processing transactions outside Ethereum and batching their data lower transaction fees.
Rollups now carry most of Ethereum's transaction activity: in August 2025, nine Ethereum rollups together processed about 15.8~million transactions per day, compared with 1.7~million on Ethereum itself.\footnote{The nine rollups are Arbitrum, Base, Ink, Linea, OP Mainnet, Scroll, Starknet, Unichain, and ZKsync Era; the set does not include all rollups.
    Counts are daily means over August 1--31, 2025, of user and smart-contract transactions, excluding system transactions, as reported by growthepie \parencite{growthepie2026txcount}.
    They count transactions, not dollar trading volume.}

Execution on an L2 is typically managed by a single entity, the \emph{sequencer}, which acts as its block producer.
Because one operator produces every block, an L2 can set its block interval and ordering rule directly, within the limits that its protocol sets.
Ethereum produces at most one block every 12 seconds \parencite{ethereum2025specs}.
Rollup blocks arrive at intervals of two seconds or less, and some rollups now commit transactions in smaller batches within each block, so the effective block interval is shorter: 200 milliseconds on Base since July 2025 and 250 milliseconds on OP Mainnet since September 2025 \parencite{base2025deepdive, optimism2025flashblocks}.\footnote{OP Mainnet's two-second block interval is set in its protocol specification, and ZKsync Era's one-second blocks are documented by its developers \parencite{optimism2026derivation, zksync2026blocks}.}
Ordering rules differ as well.
Base uses a PGA \parencite{base2025deepdive}, while Arbitrum used FCFS ordering, with an auction for an express lane that skipped a short delay, from April 2025 until September 2026, when it too adopted a PGA \parencite{arbitrum2026timeboost, arbitrum2026pga}.
Section~\ref{sec:emp-institutional} describes both chains' rules during our sample period.

\subsection{Decentralized exchanges and automated market makers}\label{sec:inst-amm}
The majority of trading on blockchain occurs through decentralized exchanges (DEXs), smart contracts deployed on the blockchain that execute trades without the need for a centralized operator.\footnote{In 2025, spot trading volume on the ten largest DEXs was approximately one-fifth of that on the ten largest centralized exchanges (CEXs) \parencite[p.~50]{coingecko2025annual}.
    Uniswap reported more than \$1 trillion in trading volume that year \parencite{uniswap2025recap}.}
The dominant form of DEX is the automated market maker (AMM), a market structure characterized by liquidity pools and algorithmic pricing: an AMM holds reserves of tokens and prices trades algorithmically rather than matching orders in a traditional order book \parencite{capponi2025liquidity}.
The SEC's 2026 conditional exemptions for trading tokenized U.S. stocks require the AMM pools' smart contracts to be deployed on a public, permissionless blockchain \parencite{SEC2026TSV}.\footnote{Permissionless means that anyone can read or write to the blockchain without authorization.
    Access to a pool itself can still be restricted to approved participants \parencite{SEC2026TSV}.}

An AMM holds reserves of two tokens in a \emph{liquidity pool}.
In a standard open pool, any participant can act as a \emph{liquidity provider} (LP) by depositing an equivalent value of both tokens into the pool.
In return, the LP receives a proportional share of the pool's liquidity, a claim on the pool's reserves at the time of withdrawal.
For example, if the pool initially contains 100 units of each token and an LP deposits 10 units of each, the LP holds $10/110$, or about $9.1\%$, of the pool.
The value of this share depends on the pool's composition at withdrawal, which may differ from the composition at deposit if trades have occurred in the interim.
In this simple pool, LPs earn revenue from trading fees: each trade against the pool incurs a fee proportional to the trade size, and the fee accrues to all LPs in proportion to their pool shares.%
\footnote{Common fee tiers on Uniswap~v3 are 1, 5, 30, and 100 basis points.
    On Uniswap~v3, where liquidity providers choose price ranges, fees are shared pro rata among the liquidity active at the current price, and a position whose range excludes that price stops earning fees \parencite{uniswap2026fees, uniswap2026liquidity}.}
Liquidity provision on an AMM is therefore analogous to market making on a traditional exchange, with an important structural difference: on a limit order book, each market maker submits and manages individual quotes, while on an AMM, all liquidity is \emph{pooled}.
Gains and losses are shared pro rata across the liquidity deposited at the time of each trade, regardless of whether its provider actively monitors the market.

The AMM determines exchange rates algorithmically through a \emph{pricing function} that relates the pool's token reserves.
A common specification is the constant product rule, $x \cdot y = K_{\mathrm{pool}}$, where $x$ and $y$ denote the two reserves and $K_{\mathrm{pool}}$ is set by the total liquidity deposited \parencite{capponi2025liquidity}.
Ignoring trading fees, a trader who takes one token out must put in enough of the other to keep this product unchanged.

For example, consider a pool holding 100 units of each token, so the product of its reserves is 10,000.
A trader makes a large purchase of 20 units of the first token, leaving 80 in the pool.
To preserve the product, the pool must then hold $10{,}000/80 = 125$ units of the second token.
The trader therefore pays 25 units of the second token, an average price of 1.25 per unit of the first.

The purchase also raises the pool's quote for the first token.
Its marginal price, measured in units of the second token, equals $y/x$: it rises from 1 to $125/80 = 1.5625$.
This is the price of buying a further small amount; the completed trade cost 1.25 on average.
Each trade moves the pool along the pricing curve, with greater price impact when the trade is large relative to pool depth.

Newer AMM designs build on this basic mechanism.
Uniswap~v3 extends inventory-based pricing by allowing liquidity providers to choose price ranges \parencite{HasbrouckRiveraSaleh2023}, while Uniswap~v4 permits additional programmed rules that can change fees and pricing in response to other market conditions \parencite{BachuHasbrouckSalehWan2025}.

\subsection{Automated quoting and liquidity provision}\label{sec:inst-lp}

In the basic reserve-based pool just explained, this automated pricing mechanism has two consequences for liquidity providers.
First, the pool offers a quote while capital remains deposited, so liquidity is available without any provider standing ready to trade; how much is available depends on the capital that providers keep in the pool.
Second, unlike a limit order book, in which market makers actively update their quotes in response to news, a pool using the reserve-based rule described above adjusts its price when a trade changes its reserves.
Between trades, the pool's marginal quote can remain unchanged even as prices move on external venues such as Binance; the quote is then \emph{stale}.

The pricing example above omits trading fees.
The reserve-implied price plays the role of a dealer's mid-quote, and the pool's trading fee plays a role similar to the half-spread: for a small trade, buyers pay above that price and sellers receive below it.
The fee compensates liquidity provision and creates a price gap that an arbitrageur must overcome.
It therefore defines a no-arbitrage band around the pool's price, whose edges we call the pool's \emph{price boundaries}.
The pool's trading fee is separate from the transaction fee paid for execution on the blockchain.

\begin{figure}[t!]
    \centering
    \includegraphics[width=\textwidth]{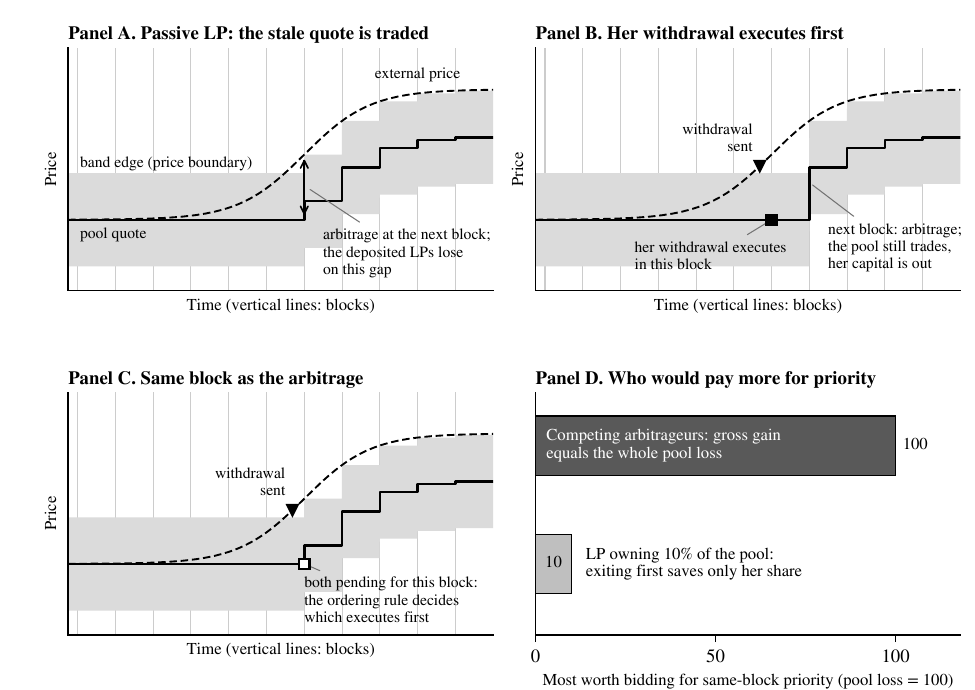}
    \caption{\textbf{Stale quotes and the timing of a withdrawal.}
        This figure illustrates, with a stylized price path, how a pool's quote becomes stale and why a withdrawal must execute in time.
        In each price panel, the dashed line is the external price, the solid step line is the pool's quote, the shaded band around it is the no-arbitrage band set by the pool's trading fee, and the vertical lines mark blocks.
        In Panel~A, once the external price leaves the band, an arbitrage trade at the next block moves the quote toward the external price until the remaining gap no longer covers the fee, and the liquidity providers deposited at that moment bear the loss.
        In Panel~B, an active provider sends a withdrawal after the gap opens, and it executes one block before the arbitrage, so her capital has left the pool when the stale quote is traded; the remaining liquidity still bears the loss.
        In Panel~C, her withdrawal and the arbitrage are pending for the same block, and the chain's ordering rule decides which executes first.
        Panel~D illustrates the commons argument of \textcite{capponi2025liquidity} for a pool loss of 100 before priority fees: competing arbitrageurs are willing to bid up to their gross gain of 100 for priority, whereas a provider who owns 10\% of the pool avoids at most 10 by exiting first.
        The price path, the band width, and the block intervals are illustrative and not calibrated.}\label{fig:inst-stale}
\end{figure}

The staleness of the pool's quote creates a systematic adverse-selection cost for liquidity providers.
Arbitrageurs monitor the deviation between the AMM price and external reference prices, and when this deviation exceeds the pool's trading fee, they can submit a transaction to trade against the stale quote.
An arbitrage trade moves the pool's quote toward the external price, stopping when the remaining gap no longer covers the pool's trading fee, and transaction fees further reduce the arbitrageur's net profit.
The arbitrageur profits from the price discrepancy, while the pool, and therefore its deposited liquidity providers, bears the corresponding loss.
Valued at the external reference price, the arbitrageur's gain after the pool's trading fee but before payment to the block producer equals the pool's loss.
Figure~\ref{fig:inst-stale}, Panel~A, illustrates this sequence for a pool whose liquidity providers stay deposited.
This adverse-selection cost is the primary friction in liquidity provision on AMMs \parencite{capponi2025liquidity, LeharParlour2023, zhang2023amm}.

A passive liquidity provider deposits capital into the pool and leaves it for days, weeks, or longer, earning fees and bearing adverse-selection losses on the trades its liquidity serves.
An active liquidity provider, by contrast, monitors the reference price and adjusts her position frequently.
She aims to earn fees while the external price stays inside the pool's price boundaries, to withdraw before an arbitrage executes when the external price approaches or crosses a boundary, and to re-enter once the external price is back inside the band.
This operational distinction (the frequency and speed of deposit-and-withdrawal cycling) is orthogonal to the price range over which liquidity is supplied.\footnote{Recent AMM designs, notably Uniswap~v3, allow LPs to specify a concentrated price range \parencite{HasbrouckRiveraSaleh2023}.
    Concentrated liquidity increases capital efficiency but does not by itself determine whether an LP is active or passive in our operational sense.
    An LP can supply concentrated liquidity passively (set a range and leave) or actively (reposition the range every few minutes).
    The endpoints of such a range differ from the pool's price boundaries defined above: when the pool price leaves an LP's range, her position stops earning fees and holds only one token, but no transaction removes it from the pool.}
An active LP who repositions every few minutes resembles a traditional high-frequency market maker more than she resembles a passive liquidity provider.

An active liquidity provider can protect her deposited capital by withdrawing it before an arbitrage trade executes against the pool, but her withdrawal is itself a transaction.
Nothing in the pool cancels her quote when the external price moves: she must submit a withdrawal, and the withdrawal protects her only once a block includes it.
If her withdrawal reaches the block producer before the arbitrage does and a block is produced between their arrivals, that block can execute her withdrawal first.
Her capital has then left the pool when the stale quote is traded, while the liquidity that remains deposited bears the loss (Figure~\ref{fig:inst-stale}, Panel~B).
If both transactions are pending for the same block, the chain's ordering rule decides which executes first, and her capital bears its share of the loss whenever the arbitrage executes first (Panel~C).
Each deposit and withdrawal also pays a transaction fee, which adds to the cost of active management.

Long block intervals make it difficult for liquidity providers to protect their capital through active management.
A withdrawal can arrive before an arbitrage trade yet wait with it for the same block.
If transactions in that block are ordered by fees, arriving first does not ensure that her withdrawal executes first.
She must instead win the fee competition against the arbitrageurs.

That competition favors arbitrageurs because they compete for the gain from trading against the entire pool, whereas a provider who withdraws avoids only her own share of the corresponding loss.
A provider holding 10\% of the pool would therefore pay at most one-tenth of what competing arbitrageurs are willing to pay for priority, as Panel~D illustrates.
Although the providers collectively bear the whole loss, no individual provider can profitably outbid the competing arbitrageurs.
\textcite{capponi2025liquidity} call this a tragedy of the commons.
They argue that it provides a rationale for models that hold liquidity positions fixed during trading, such as \textcite{HasbrouckRiveraSaleh2023} and \textcite{zhang2023amm}.

Shorter block intervals open a route to protecting capital without winning this fee competition.
A withdrawal that arrives before the arbitrage trade is more likely to execute in a separate, earlier block, removing the provider's capital before the stale quote is traded.
This opportunity makes active liquidity management a question worth revisiting on faster chains.
Faster blocks also give arbitrageurs earlier opportunities to trade.
Our model asks when active management becomes worthwhile and how the providers' choices affect the liquidity available to traders.

\section{Model}\label{sec:model}

A single risky asset trades on a blockchain AMM, where time is continuous but transactions execute only when a block is produced.
Blocks arrive as a Poisson process with rate $B>0$, so the expected block interval is $1/B$.
The block rate, set by the protocol and taken as given by market participants, is the only rate our comparisons vary.
A higher block rate means faster blocks, so we also call it block speed.
All participants are risk neutral and evaluate expected long-run average income.

\subsection{Trading environment and players}

The asset has an external reference price, the price observed in a deep external market.
The pool price, the AMM's marginal exchange rate, can remain stale between transactions even though traders observe the reference price continuously.
We call the difference between the reference price and the pool price the price gap.

\paragraph{Sequencer.}

A sequencer orders transactions and collects priority fees.
It has no strategic choice and executes transactions in a fixed order within each block.

\paragraph{Arbitrageurs.}

Two arbitrageurs monitor the price gap and compete to profit from trading against the pool's stale price.
We call a gap profitable when it exceeds the pool's trading fee.
Once the gap becomes profitable, each arbitrageur may send one transaction for that opportunity.
The transaction reaches the sequencer after a random latency, exponentially distributed with rate $\lambda>0$.%
\footnote{In practice, an arbitrageur must receive the external price update, prepare and submit a transaction, and have it reach the sequencer. Each communication step takes time: cross-region transmission can take on the order of 100 milliseconds, while colocated servers can communicate much faster (\url{https://docs.sequencemkts.com/architecture/}). These delays precede the wait for block inclusion. The modeled latency covers transaction submission to sequencer receipt.}
It executes in the first block after its arrival unless the rival's transaction executes first.

Each arbitrageur chooses the percentage of the total arbitrage prize to pay the sequencer.
The prize is the amount the trade extracts from the pool at execution, before that payment.
The arbitrageur chooses the percentage $b\in[0,1]$ when sending the transaction and cannot change it before the gap closes.

\paragraph{Liquidity providers.}

Liquidity is supplied by many small passive LPs and one active LP.
Before trading begins, passive LPs together deposit capital $P\geq0$ and leave it in the pool throughout trading.
They bear a holding cost $k>0$ per unit of capital per unit of time, representing the return that outside capital requires to enter the pool.
Passive capital can enter freely.

We think of the active LP as a professional market-making firm that moves capital into and out of the pool as the reference price changes.
She has a fixed capital endowment $q>0$, which she deposits or withdraws in full.
While deposited, her capital earns trading fees in proportion to its share of the pool's deposited capital.
Outside the pool, it earns nothing.

Her withdrawal and redeposit transactions reach the sequencer after a latency with the same rate $\lambda$ as each arbitrageur's transaction, so she has no speed advantage over the arbitrageurs.
She chooses when to send a withdrawal while a price gap is open and when to send a redeposit after it closes, trading off fees earned in the pool against arbitrage losses avoided by withdrawing.

\paragraph{Liquidity traders.}

Liquidity traders arrive as a Poisson process, and each pays a fixed fee to the liquidity providers only if its transaction executes.
We write $\rbar$ for the fee flow they offer per unit of time, which equals their arrival rate times the fee.
Each liquidity-trader fee is shared among deposited providers in proportion to their capital in the pool before that block's withdrawals and redeposits execute.

Each liquidity-trader transaction has an independent exponential deadline with rate $\delta>0$.
It executes at the first block after its submission if its deadline has not passed and the pool holds capital, and otherwise it leaves without retrying.

Block speed therefore affects their executed trading volume and the fee income available to liquidity suppliers.
When the pool holds capital, the expected fee flow served per unit of time is
\begin{equation}\label{eq:customer-fee-flow}
    r(B)=\rbar\,\frac{B}{B+\delta},
\end{equation}
because $B/(B+\delta)$ is the probability that a transaction reaches a block before its deadline.

\subsection{Price gap and information arrival}

Public information can leave the pool price stale relative to the reference price, but trading fees make small price discrepancies unprofitable to arbitrage.
Let $p^*$ and $p^{\mathrm{AMM}}$ denote the reference price and the pool's marginal price.
We allow the absolute gap to take three values,
\[
    |p^*-p^{\mathrm{AMM}}|\in\{0,M,H\},\qquad 0<M<f<H,
\]
where $f$ is the fee threshold for profitable arbitrage.
The small gap $M$ lies within the no-arbitrage band; the large gap $H$ offers a profitable arbitrage trade.

Let $J>0$ be the rate of public information arrival.
When prices are aligned, a news arrival creates the small gap $M$.
At $M$, the next news arrival either eliminates the discrepancy or widens it to $H$, with equal probability:
\[
    0\xrightarrow{J}M,\qquad M\xrightarrow{J/2}0,\qquad M\xrightarrow{J/2}H.
\]

At $H$, further news leaves the gap unchanged.
An arbitrage transaction closes the gap when it executes in a block and brings the pool price back to the reference price \parencite{HasbrouckRiveraSaleh2023}.
Liquidity trades do not change the gap.
News arrivals, blocks, liquidity-trader arrivals, and message latencies are independent.

Everyone observes the gap, the blocks, the transactions that execute, and the capital in the pool.
The rates of news arrivals, blocks, liquidity-trader arrivals and message latencies are common knowledge.
Before execution, each agent privately observes when its own transactions reach the sequencer, but not whether a rival's transaction has arrived or what percentage the rival has bid.
There is no private information about fundamentals; adverse selection arises from ``public information asymmetry'' \parencite{budish2015high}.

\subsection{Timing and stages of play}\label{sec:timing}

\paragraph{Sending a withdrawal.}

While deposited, the active LP may send a withdrawal transaction whenever a small or a large gap is open.
If the gap widens from $M$ to $H$, the withdrawal remains valid and any delivery already in progress continues.
The withdrawal is canceled when the gap returns to zero, either because news removes the discrepancy or because an arbitrage trade brings the pool price back to the reference price.

A valid withdrawal executes at the first block after it reaches the sequencer.%
\footnote{While the price gap is unchanged, the LP cannot recall a submitted withdrawal or postpone its execution. She cannot speed it up with a duplicate. She may send a redeposit only after her withdrawal has executed.}
Withdrawing capital, unlike arbitrage, leaves the price gap unchanged.

\paragraph{Block execution.}

Among arbitrage transactions that have reached the sequencer, the one offering the highest priority fee executes.
When an arbitrage trade and a withdrawal are included in the same block, the arbitrage trade executes first.%
\footnote{Every block has room for all transactions. Under this ordering, paying a priority fee would not speed up the active LP's withdrawal. If both arbitrage transactions offer the same priority fee, each executes first with probability one half.}
The active LP's capital therefore bears the loss, and the executed arbitrage trade cancels her withdrawal.
To avoid that loss, her withdrawal must execute in an earlier block.

\paragraph{Redeposit.}

The active LP may send a redeposit once the pool price equals the reference price.%
\footnote{An empty pool earns no liquidity-trader fees. If her withdrawal leaves the pool empty, either arbitrageur can reset its stale price to the reference price with a zero-size transaction that extracts nothing. We assume arbitrageurs send these transactions as promptly as ordinary arbitrage transactions.}
It travels at the same speed $\lambda$ as her withdrawal and executes in the first block after reaching the sequencer, provided prices remain aligned.%
\footnote{If a small gap opens before execution, the redeposit is canceled; she may send a fresh transaction once prices align again. A delayed deposit faces a related risk on Uniswap~v3, where providers specify token amounts and a slippage tolerance. While the transaction waits, the pool price can change, altering the required mix of tokens. If the token amounts the deposit would use at the new price fall outside her limits, the deposit fails automatically \parencite{uniswap2026minting}.}

\paragraph{Strategies.}

\begin{figure}[!tbp]
    \centering
    \includegraphics[width=\textwidth]{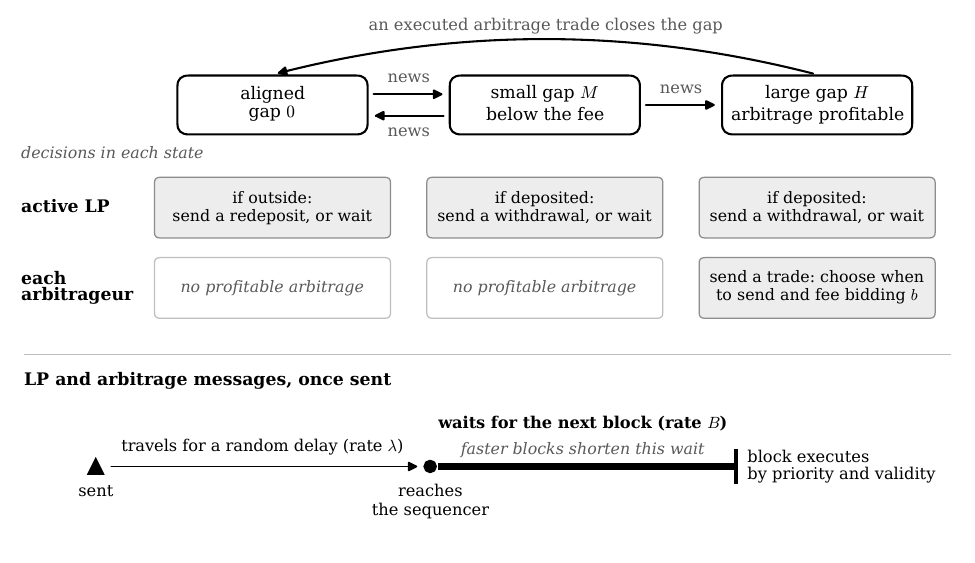}
    \caption{\textbf{The game.}
        Top: public news moves the price gap between three states, and an executed arbitrage trade closes it.
        Middle: the decisions available to the active LP and to each arbitrageur in each state.
        Bottom: execution requires message delivery and a subsequent block; priority and validity determine which transactions execute.
        Play runs in continuous time.}\label{fig:s3-game}
\end{figure}

Her strategy specifies, given everything she has observed, when she sends each withdrawal and each redeposit.
Figure~\ref{fig:s3-game} summarizes the sequence of decisions and events.

\subsection{Payoffs}\label{sec:model-payoffs}

Let $I=\ind\{\text{the active LP's capital is deposited}\}$, so that the capital in the pool is $Q=P+qI$.
An executed arbitrage trade extracts the arbitrage prize $\ell Q$, so every deposited unit of capital loses $\ell>0$ at each such trade.%
\footnote{Trading gains and losses are settled separately from the pool's capital and do not change the capital stocks $P$ and $q$.}

\paragraph{Arbitrageur payoff.}

The arbitrageur whose transaction executes keeps $(1-b)\ell Q$, and the sequencer receives $b\ell Q$.
A transaction that does not execute pays nothing, and sending one carries no separate cost.

\paragraph{Active LP payoff.}

For a strategy with well-defined long-run averages, two long-run quantities determine the active LP's income: $\pi$, the fraction of calendar time her capital is deposited, and her exposed-trade rate, the rate of executed arbitrage trades while it is deposited.
Each such trade costs her $\ell q$.

While she is deposited, her share of the pool's capital is $q/(P+q)$.
Capital changes only at a block, after that block's liquidity traders have executed.
Blocks arrive as a Poisson process that her strategy does not affect.
Hence her fee income per unit of time is $\pi$ times the fee flow she earns while deposited.
Her trading income $U_A$, fee income less arbitrage losses per unit of time, is therefore
\begin{equation}\label{eq:lp-income}
    U_A=\pi\,r(B)\,\frac{q}{P+q}-\ell q\times\bigl(\text{her exposed-trade rate}\bigr).
\end{equation}

The active LP does not pay the holding cost $k$.
Withdrawing protects only her own capital.
When she is out, the arbitrage prize shrinks by exactly the loss she avoids.

\paragraph{Passive LP payoff.}

Passive LPs receive the fraction $P/(P+q)$ of liquidity-trader fees while the active LP is deposited and all fees earned by a nonempty pool while she is out.
They also bear the loss $\ell P$ at every executed arbitrage trade and pay the holding cost $kP$ per unit of time.

\paragraph{Sequencer payoff.}

The sequencer's share of the arbitrage prize, $s$, is its expected revenue divided by the expected total prize, both per unit of time, so that the arbitrageurs keep the share $1-s$.
These priority fees are transfers from arbitrageurs, distinct from the liquidity-trader fees that LPs earn.

\subsection{Equilibrium concept}

Each arbitrageur and the active LP evaluate a strategy by its expected long-run average income.%
\footnote{We use expected cumulative income divided by elapsed time, taking the limit superior as the horizon grows. This measure equals the ordinary limit whenever that limit exists.}
Passive LPs deposit their capital before trading begins, anticipating the active LP's strategy.

\begin{definition}[Equilibrium]\label{def:equilibrium}
    An operating equilibrium consists of passive capital $P\geq0$, stationary symmetric arbitrageur strategies and a pure active-LP strategy satisfying:%
    \footnote{We describe each equilibrium by its passive capital, the long-run fraction of time the active LP's capital is deposited and her exposed-trade rate. We count equilibria by these descriptions, so two equilibria that differ only during a finite initial period count as the same equilibrium. We do not classify equilibria that rely on punishing past behavior, on asymmetric bidding by the arbitrageurs, or on histories whose long-run averages do not converge.}
    \begin{enumerate}[nosep]
        \item Each arbitrageur's bidding and submission strategy maximizes its expected long-run average income, given the other agents' strategies.
        \item Given $P$, the active LP maximizes her expected long-run average trading income over all admissible strategies, including randomized deviations.
        \item Passive LPs earn zero expected net profit per unit of time if $P>0$. At $P=0$, a marginal entrant cannot earn a positive expected net profit.
    \end{enumerate}
\end{definition}

When several equilibria exist, we select the one that gives the active liquidity provider the highest income.

\subsection{Assumptions}

We impose four restrictions on the model's parameters.

\begin{assumption}\label{ass:main}
    \begin{enumerate}[label=(\roman*),ref=\theassumption(\roman*),leftmargin=*,nosep]
        \item\label{ass:clock} Arbitrageurs are not too slow relative to the arrival of public information: $2\lambda>J$.
        \item\label{ass:carry} Passive LPs' holding costs are not so high as to severely discourage their participation: $k<\ell\,\frac{4J\lambda^2}{(J+8\lambda)(J+2\lambda)}$.
        \item\label{ass:patience} Liquidity traders are not so impatient that only extremely fast chains can execute their trades: $0<\delta<k/(20\ell)$.%
              \footnote{The patience assumption's constant twenty is sufficient for our proofs, not an estimated market threshold.}
        \item\label{ass:fee} Fee income is sufficiently high relative to the active LP's capital to make participation worthwhile: $\rbar\left(k-\frac{2\ell J\lambda^2}{(J+8\lambda)(J+\lambda)}\right)>q\left(k+\frac{2\ell J\lambda}{J+8\lambda}\right)^2$.
    \end{enumerate}
\end{assumption}

\makeatletter
\@ifundefined{theorem}{\newtheorem{theorem}[proposition]{Theorem}}{}
\@ifundefined{example}{\newtheorem{example}[proposition]{Example}}{}
\makeatother

\section{Equilibrium Analysis}\label{sec:equilibrium}

Block speed changes the division of arbitrage rents and the capital available to liquidity traders.
We vary the block rate $B$ while holding the other primitives fixed.
Section~\ref{sec:within-block} analyzes competition between arbitrageurs and the division of the arbitrage prize.
Section~\ref{sec:lp-behavior} determines the active LP's withdrawal policy together with passive liquidity provision.
Section~\ref{sec:liquidity-maximum} examines how faster blocks affect active LP income and passive participation.
It then establishes how pool liquidity varies with block speed and where it is greatest.
All proofs are in the Internet Appendix.

\subsection{Block speed and arbitrage competition}\label{sec:within-block}

How does block speed determine who captures the arbitrage rent, the prize that each executed arbitrage trade extracts from the deposited capital?
Arbitrageurs can compete within a block or across blocks.
In a clash, both transactions reach the sequencer before the next block, and the higher bid wins.
In a race across blocks, the first transaction executes before its rival arrives, so earlier arrival determines who captures the arbitrage prize.

Slow blocks leave more time for both transactions to arrive, making clashes more likely.
Faster blocks shorten that wait, so the first transaction is more likely to execute before its rival arrives.
Faster blocks therefore shift competition away from a priority auction and toward first come, first served.

When a large price gap opens, the arbitrageurs compete to capture the arbitrage prize.
Each chooses when to send its transaction and how much of the prize to bid for priority.
At smaller gaps, the price discrepancy does not cover the pool's trading fee, so no profitable arbitrage trade is available.

Formally, we characterize symmetric bidding and submission strategies satisfying the arbitrageurs' best-response condition in Definition~\ref{def:equilibrium}.

\begin{proposition}[Priority Competition and Rent Division]\label{prop:pfp}\label{prop:rent-share}
    There are symmetric equilibrium strategies under which each arbitrageur submits immediately when the large gap opens and draws its percentage bid $b$ from the distribution
    \begin{equation}\label{eq:bid-distribution}
        \Pr(\text{bid}\le b)=\frac{B}{2\lambda}\,\frac{b}{1-b},\qquad 0\leq b\leq\frac{2\lambda}{B+2\lambda}.
    \end{equation}
    The sequencer's share of the arbitrage prize and the arbitrageurs' combined share are
    \begin{equation}\label{eq:revenue}
        s(B)=\frac{\lambda}{B+\lambda},\qquad 1-s(B)=\frac{B}{B+\lambda}.
    \end{equation}
    The sequencer's share decreases with block speed, approaching one as blocks become arbitrarily slow and zero as blocks become arbitrarily fast.%
    \footnote{The equilibrium results below concern equilibria with the arbitrageur strategies in this proposition.}
\end{proposition}

Neither arbitrageur sees whether its rival's transaction has arrived, so each bids against a rival that may or may not be present.
A higher percentage wins more same-block auctions but keeps less of the prize when the transaction executes alone.

Once the first arbitrage transaction reaches the sequencer, the next block arrives at rate $B$ and the rival's transaction arrives at rate $\lambda$.
If the block arrives first, the earlier transaction executes alone but still pays its bid.
If the rival arrives first, both transactions compete in the same block and the higher bid wins.
Because the waiting times are independent and exponential, each event occurs first with probability equal to its rate divided by the combined rate $B+\lambda$:
\[
    \begin{aligned}
        \Pr(\text{clash})
         & =\Pr(\text{rival arrives before the block})
        =\frac{\lambda}{B+\lambda},                    \\[4pt]
        \Pr(\text{race})
         & =\Pr(\text{block arrives before the rival})
        =\frac{B}{B+\lambda}.
    \end{aligned}
\]
Faster blocks therefore make a higher bid less likely to change who wins, while it still raises the payment whenever the transaction wins.
Equilibrium percentage bids fall, leaving arbitrageurs a larger share of the prize.

\begin{figure}[ht]
    \centering
    \includegraphics[width=\textwidth]{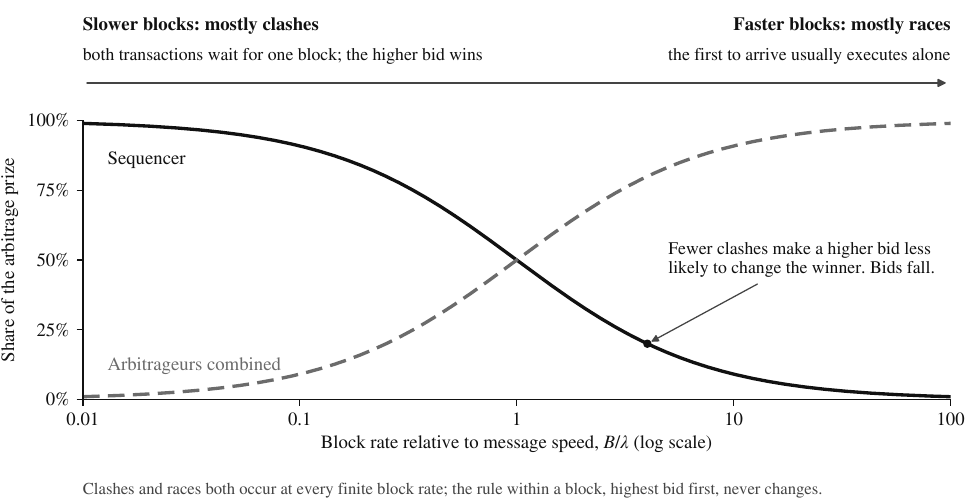}
    \caption{Block speed and the division of the arbitrage prize.
        The solid curve is the sequencer's share of the arbitrage prize, $s(B)=\lambda/(B+\lambda)$, and the dashed curve is the arbitrageurs' combined share, $1-s(B)$, against the block rate relative to the message speed, $B/\lambda$, on a log scale.
        Faster blocks make it less likely that the rival's transaction arrives before the next block, so a higher bid less often changes which transaction executes; equilibrium bids fall and the arbitrageurs keep a larger share of the prize.
        Curves show expected shares of the prize, not revenues per unit of time.}\label{fig:block-speed-spectrum}
\end{figure}

Figure~\ref{fig:block-speed-spectrum} shows the resulting shift in the division of the arbitrage prize.

The sequencer's revenue per unit of time depends on its share of the arbitrage prize, how often arbitrage trades execute, and the capital they trade against.
How often arbitrage trades execute also determines how often passive LPs bear arbitrage losses.
The following lemma describes how block speed affects both arbitrage frequency and the average gap between the pool price and the reference price.

\begin{lemma}[Arbitrage Frequency and Price Adjustment]\label{lem:frequency}
    Under the arbitrageur strategies of Proposition~\ref{prop:pfp}, let $T=4/J+1/(2\lambda)$ denote the combined mean waiting time for news and transaction delivery.
    \begin{enumerate}[label=(\roman*),leftmargin=*]
        \item Arbitrage trades execute at rate
              \begin{equation}\label{eq:correction-rate}
                  e=\frac{B}{1+TB}.
              \end{equation}
              This rate increases with $B$ and converges to $e_\infty=1/T=2J\lambda/(J+8\lambda)$ as blocks become arbitrarily fast.
        \item The time-average absolute gap between the pool price and the reference price is
              \begin{equation}\label{eq:calendar-gap}
                  H-\frac{e(4H-2M)}{J},
              \end{equation}
              which decreases with $B$.
    \end{enumerate}
\end{lemma}

Faster blocks shorten the wait from a transaction reaching the sequencer to its execution.
Arbitrage trades therefore execute sooner, so the pool price spends less time away from the reference price.
But passive capital bears a loss at every arbitrage trade, whether or not the active LP is deposited.
Each passive unit loses the fixed amount $\ell$ per trade, so its expected losses per unit of time, $\ell e$, increase with block speed.

\subsection{Liquidity management}\label{sec:lp-behavior}

When does active liquidity management become worthwhile?
The active LP can protect her capital by withdrawing before an arbitrage trade executes, but she gives up fee income until she redeposits.
On a slow chain, a withdrawal and an arbitrage trade are likely to execute in the same block, when arbitrage takes priority.
Even if she withdraws in time, she must wait for her redeposit to execute before earning fees again.

\begin{figure}[!tbp]
    \centering
    \includegraphics[width=\textwidth]{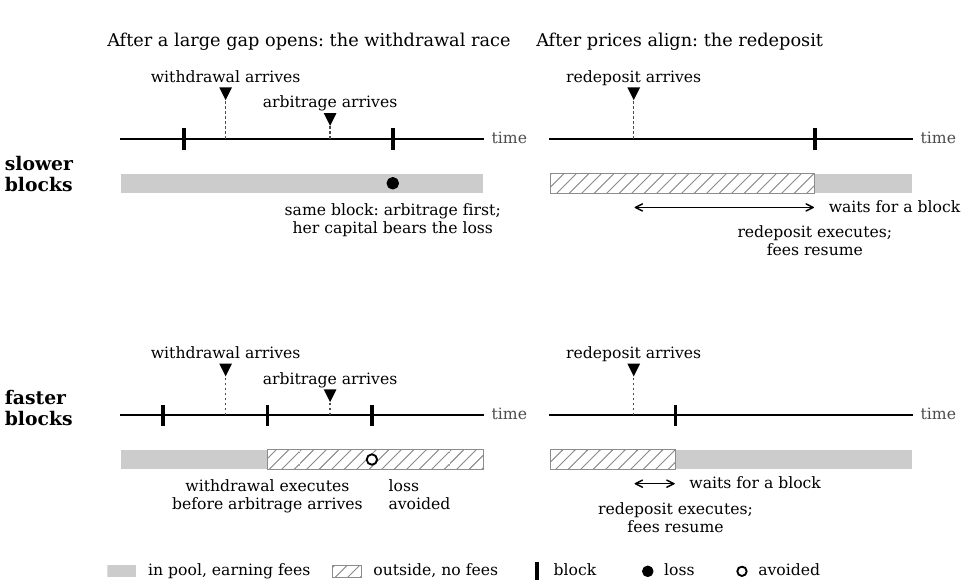}
    \caption{Two benefits of faster blocks.
        With message arrival times held fixed, faster blocks can execute a withdrawal before the arbitrage trade (left) and a redeposit sooner after prices align (right).
        Shaded intervals earn trading fees; hatched intervals are outside the pool and earn none.
        The times are illustrative.}\label{fig:two-benefits}
\end{figure}

Faster blocks offer two benefits for active liquidity management.
First, her withdrawal is more likely to execute in a block before the arbitrage trade, protecting her capital from the loss.
Second, once prices align, her redeposit can execute sooner, allowing her to resume earning trading fees earlier.
Figure~\ref{fig:two-benefits} illustrates both benefits with message arrival times held fixed.

Her decision also depends on how much passive capital shares the pool's fees.
More passive capital lowers the fee income she gives up by withdrawing, while her absences leave a larger share of fees to passive LPs.
Her strategy and passive participation must therefore be determined together.

\begin{theorem}[Equilibrium]\label{thm:map}
    At every block rate, there is a unique selected equilibrium with the arbitrageur strategies of Proposition~\ref{prop:pfp}.
    There is a threshold $B_2>0$ such that:
    \begin{enumerate}[label=(\roman*),ref=\thetheorem(\roman*),leftmargin=*]
        \item Each arbitrageur sends its transaction immediately when a large gap opens and independently draws its percentage bid from the distribution in Proposition~\ref{prop:pfp}.
        \item For $0<B\le B_2$, the active LP is always deposited.
              For $B>B_2$, she withdraws late: whenever deposited, she sends a withdrawal as soon as a large gap opens and, once outside, a redeposit as soon as the gap returns to zero.
              If the redeposit is canceled, she sends a fresh one when the gap next returns to zero.
        \item For $0<B\le B_2$, passive capital is
              \[
                  P(B)=\max\left\{\frac{r(B)}{k+\ell e(B)}-q,\;0\right\}.
              \]
              For $B>B_2$, passive capital is the unique positive solution to
              \[
                  r(B)\left[\frac{\pi_L(B)}{P+q}+\frac{1-\pi_L(B)}{P}\right]=k+\ell e(B),
              \]
              where $\pi_L(B)$ is the fraction of time the active LP is deposited under the withdrawal and redeposit policy in part~(ii).
    \end{enumerate}
\end{theorem}

The threshold $B_2$ marks the point at which the gains from withdrawing offset the fees forgone.
Withdrawing only after a large gap opens lets her earn fees while the gap is small and seek protection when an arbitrage trade becomes profitable.

When she remains deposited and passive capital is positive, entry sets fee income per unit of capital at $k+\ell e(B)$, covering passive LPs' holding costs and arbitrage losses.
As blocks become faster, the losses she can avoid per unit of time outside rise by more than this fee income.
Above $B_2$, she therefore prefers to withdraw at the passive capital that continuous deposit would attract.

\subsection{Cream-skimming and pool liquidity}\label{sec:liquidity-maximum}\label{sec:capital-equilibrium}\label{sec:liquidity-results}

Section~\ref{sec:lp-behavior} characterizes liquidity management and passive entry.
We now examine the implications for pool liquidity and for active and passive LPs: how the active LP cream-skims passive providers, how her income changes, and how passive capital responds.

\begin{proposition}[Cream-Skimming]\label{prop:r7}\label{prop:cream}
    There exists a block rate $\bar B\ge B_2$ such that, for every $B>\bar B$, late withdrawal is the unique equilibrium and:
    \begin{enumerate}[label=(\roman*),ref=\theproposition(\roman*),leftmargin=*,nosep]
        \item The active LP's arbitrage losses per unit of time strictly decrease with $B$, and her trading income $U_A$ strictly increases.
        \item Passive capital $P$ and passive LPs' share $P/(P+q)$ when the active LP is deposited strictly decrease with $B$.
    \end{enumerate}
\end{proposition}

Active and passive LPs have different exposure to arbitrage.
The active LP sends a withdrawal when a large gap opens, seeking to remove her capital before the arbitrage trade executes.
Passive capital remains deposited and bears every arbitrage loss.
Sharing trading fees while selectively avoiding some arbitrage losses is cream-skimming.
When the active LP is deposited, passive LPs receive a smaller share of trading fees while remaining exposed to every arbitrage trade.

Faster blocks give the active LP's withdrawal a greater chance to execute before the arbitrage trade.
Once blocks are sufficiently fast, the active LP's arbitrage losses fall even as arbitrage becomes more frequent.
These loss savings more than offset any additional fees forgone while outside, raising her trading income.

At these block speeds, most liquidity trades already execute before expiry.
Faster blocks therefore add little fee income while exposing passive capital to more frequent arbitrage losses.
Equilibrium passive capital falls, and so does passive LPs' share of the pool when the active LP is deposited.

How does block speed affect the total liquidity available to traders?
We measure liquidity by time-average depth, $D=P+q\pi$, where $\pi$ is the fraction of time the active LP's capital is deposited.

\begin{proposition}[Pool Liquidity]\label{prop:liquidity-maximum}
    Depth has a unique global maximum at an intermediate block rate $B_d<B_2$.
\end{proposition}

When blocks are slow, faster execution allows more liquidity trades to complete before expiry, generating fees that support liquidity provision.
These gains diminish once most trades execute before expiry, while arbitrage becomes more frequent.
Faster blocks also help the active LP withdraw before arbitrage.
At high block speeds, passive capital declines, as shown above; any additional liquidity from the active LP is too small to offset that decline, so total liquidity falls.

Active management also has an observable implication for how long capital remains in the pool.
We measure the duration of liquidity provision from an executed redeposit to the next successful withdrawal.

\begin{proposition}[Duration of Liquidity Provision]\label{prop:spells}
    When the active LP follows late withdrawal, faster blocks strictly shorten the duration of liquidity provision in the sense of first-order stochastic dominance.
    The mean and median duration therefore fall.
\end{proposition}

The active LP may lose several withdrawal races before one succeeds.
Faster blocks shorten the wait for each attempt to resolve and increase her chance of withdrawing first.

\section{Empirics}\label{sec:empirical}

We test the model's predictions on the Uniswap~v3 WETH--USDC 5~bp pool, using Base's Flashblocks rollout as a quasi-natural experiment and Arbitrum as the control chain.
The analysis is organized around three sets of predictions.
Section~\ref{sec:emp-arb} tests whether faster blocks reduce gas-fee competition and redistribute rents from the sequencer to arbitrageurs.
Section~\ref{sec:emp-lp} tests whether active LPs exploit faster blocks to cream-skim.
Section~\ref{sec:emp-mq} turns to market quality and the model's liquidity predictions.

\subsection{Empirical predictions}\label{sec:emp-predictions}

The model yields three sets of predictions for the effects of faster blocks.
\begin{enumerate}
    \item \emph{Arbitrage rents.} Faster blocks reduce priority-fee bids and increase the share of arbitrage rents retained by arbitrageurs (Proposition~\ref{prop:pfp}).
    \item \emph{Active liquidity management.} Under late withdrawal, faster blocks shorten liquidity-provision durations and reduce the active LP's average loss per arbitrage trade in the pool. At sufficiently high block speeds, her trading income also rises (Section~\ref{sec:liquidity-maximum} and Proposition~\ref{prop:spells}).
    \item \emph{Price alignment and liquidity.} Faster blocks reduce the price gap averaged over calendar time, while pool depth falls at sufficiently high block speeds (Lemma~\ref{lem:frequency} and Section~\ref{sec:liquidity-maximum}).
\end{enumerate}

\subsection{Setting and identification}\label{sec:emp-institutional}\label{sec:emp-methodology}

\paragraph{Base.}
Base is an Ethereum L2 rollup operated by Coinbase with a centralized sequencer.
Transactions within each block are ordered by a priority gas auction (PGA): users bid gas prices, and the sequencer places higher-bidding transactions earlier in the block.
Base rolled out Flashblocks internally on July 7, 2025 and launched it publicly on July 16, reducing the effective block interval from two seconds to 200 milliseconds by streaming sub-blocks within each two-second slot.
The ordering rule within each block remains PGA; Flashblocks changes only the frequency at which new transaction batches are committed.

\paragraph{Arbitrum.}
Arbitrum is another Ethereum L2 rollup, also with a centralized sequencer.
Transactions are ordered first-come, first-served (FCFS) by arrival time at the sequencer, with one modification: since April 17, 2025, a sealed-bid second-price auction (Timeboost) allocates an exclusive ``express lane'' within each block.
Express-lane winners are sequenced without delay, while all other transactions face a 200-millisecond delay; gas price plays no role in ordering, in contrast to Base's PGA.
Arbitrum serves as the control chain in our analysis: Timeboost was already in effect before our sample began, and Arbitrum did not implement Flashblocks or any other change to its sequencing rule during the sample period.
Arbitrum retired the express lane and adopted a priority gas auction in September 2026, after our sample \parencite{arbitrum2026pga}.

\paragraph{Sample periods.}
Flashblocks was enabled on Base's internal block builder on July 7, 2025, made available to key partners on July 14, and launched publicly with a public RPC endpoint on July 16.\footnote{The timeline was confirmed through our conversations with engineers at Coinbase.}
We define the \emph{pre-upgrade} period as May 16--July 6, 2025, and the \emph{post-upgrade} period as July 17--September 15, 2025.
The window starts 15 days after our data begin on May~1, which initialize the position reconstruction, and ends 15 days before the data end on September~30.
The transition period July 7--16, during which Flashblocks was live internally but not publicly accessible, is excluded.

\paragraph{Identification.}
We test the model's predictions using a difference-in-differences (DiD) design that compares Base (treated) with Arbitrum (control) around the Flashblocks launch:
\begin{equation}\label{eq:did-panel}
    Y_{it} = \alpha + \beta_1 \text{Post}_t + \beta_2 \text{Base}_i + \beta_3 (\text{Post}_t \times \text{Base}_i) + \varepsilon_{it},
\end{equation}
where $\text{Post}_t$ indicates the post-upgrade period ($t \geq$ July~17, 2025), $\text{Base}_i$ indicates the Base chain, and $\beta_3$ is the coefficient of interest.

Panel structure, fixed effects, and clustering vary by outcome and are specified in each table caption; we flag two common choices here. For LP outcomes we cluster only by date, because the number of entity clusters (one Base LP, four Arbitrum LPs) is too small for two-way inference. Because both gas bids and trading costs vary steeply with trade size, and the upgrade itself shifts the trade-size distribution, we estimate swap-level outcomes (gas prices, effective spreads) within fixed trade-size bins rather than pooling across sizes.
Appendix~\ref{app:parallel-trends} plots daily differences for selected outcomes to help assess the parallel-trends assumption.
The toxic-exposure comparison is between the active Base LP and the pool; the other plotted series compare Base with Arbitrum.
Appendix~\ref{app:window-robustness} complements this evidence with a four-week window on either side of the upgrade and a placebo launch date within the pre-upgrade period.

\subsection{Data}\label{sec:emp-data}

\paragraph{On-chain DEX data.}
We obtain Uniswap~v3 event-level data from Allium for the WETH--USDC pool at the 5 basis point (0.05\%) fee tier on Arbitrum and Base. The data come in two complementary layers.

\textit{Pool events.} The pool contract emits Swap, Mint, and Burn events recording all trades and liquidity adjustments. We obtain the full set of such events for the sample pools and period. For every event we use the initiator address (\texttt{transaction\_from\_address}) and the block number and log index that pin down its ordering within a block. Swap events additionally carry the signed token amounts (\texttt{amount0}, \texttt{amount1}) and the post-swap square-root pool price (\texttt{sqrt\_price\_x96}), from which we recover trade direction, size, and execution price. Mint and Burn events additionally carry the position's price-range bounds (\texttt{tick\_lower}, \texttt{tick\_upper}) and the liquidity amount added (Mint) or removed (Burn) by that event, not the position's remaining liquidity.\footnote{We drop Burn events with \texttt{liquidity} $= 0$, which are fee-collection checkpoints emitted by Uniswap~v3 and do not represent actual liquidity withdrawals.}

\textit{NFT position manager events.} Uniswap~v3 represents liquidity positions as ERC-721 non-fungible tokens (NFTs) managed by the Non-Fungible Position Manager (NFPM) contract. Each NFT carries a unique \texttt{tokenId} that serves as a position identifier, analogous to an order ID in a limit order book, and allows us to track the full lifecycle of each liquidity position (from opening through incremental adjustments to final closure) at the position level rather than merely at the address level. The NFPM does not emit separate Mint and Burn events. Instead, it emits an \emph{IncreaseLiquidity} event whenever liquidity is added to a position (covering both the opening of a new position and subsequent additions) and a \emph{DecreaseLiquidity} event whenever liquidity is reduced. We retrieve these events from Allium for the same period. The key fields are the \texttt{tokenId} and the liquidity amount, which together with the transaction hash allow us to link each NFPM event back to its pool-layer Mint or Burn counterpart.

\textit{Linking pool and NFPM events.} We match pool-layer Mint and Burn events to the corresponding NFPM IncreaseLiquidity and DecreaseLiquidity events using the transaction hash, event type, and liquidity amount as a composite key, thereby recovering the \texttt{tokenId} for each pool-layer position adjustment. A non-trivial subset of pool-layer Mint and Burn events has no matching NFPM record. These are direct pool interactions by sophisticated LPs (typically high-frequency market makers, several of whom are among the largest LPs on Base) who interact with the pool contract directly rather than through the NFPM. For these LPs we cannot recover a \texttt{tokenId}, so we identify positions by the composite key of LP address and tick range (\texttt{transaction\_from\_address}, \texttt{tick\_lower}, \texttt{tick\_upper}) and reconstruct each position's cumulative liquidity from the pool-layer Mint and Burn events directly.

\paragraph{Centralized-exchange (CEX) reference prices.}
We obtain Binance ETHUSDC spot market data from Tardis. Specifically, we use Level-2 order book snapshots at 5-level depth, sampled approximately every 100 milliseconds and stamped with the Binance matching-engine timestamp at microsecond resolution throughout the sample period. The Binance mid-quote, computed as the average of the best bid and best ask prices, serves as our reference for the efficient price.

\paragraph{Summary statistics.}
Table~\ref{tab:market-activity} reports daily market activity statistics for the WETH--USDC 5~bp pool on Base and Arbitrum, separately for the pre- and post-upgrade periods. Four observations stand out. First, Arbitrum is the larger venue throughout, with daily dollar volume roughly two to three times Base's in both periods. Second, mint and burn activity is more intensive on Base, with one and a half to three times as many daily events as Arbitrum. Third, Base attracts a substantially larger and faster-growing trader base: about 4{,}300 unique addresses per day pre-upgrade, rising to nearly 14{,}000 post-upgrade, compared with about 1{,}300 to 1{,}800 on Arbitrum. Fourth, the two volume rows move in opposite directions across the upgrade: dollar volume rises on both chains while ETH volume falls, reflecting the $66\%$ rise in the mean ETH price between the periods (from \$2{,}511 to \$4{,}178).

\begin{table}[t!]
    \centering\small
    \caption{\textbf{Daily market activity for the WETH--USDC 5~bp pool on Uniswap~v3.} This table reports daily trading activity for the pool; each cell is the mean across days, with the median in parentheses. Pre-upgrade: May 16--July 6, 2025; post-upgrade: July 17--September 15, 2025 (July 7--16 transition excluded). Dollar volume is valued at contemporaneous market prices; ETH volume is the sum of absolute WETH amounts. Mint/burn counts exclude fee-checkpoint burns (zero liquidity). Unique traders counted by initiating address.}
    \label{tab:market-activity}
    \begin{tabular}{lcccc}
\toprule
 & \multicolumn{2}{c}{Base} & \multicolumn{2}{c}{Arbitrum} \\
\cmidrule(lr){2-3} \cmidrule(lr){4-5}
 & Pre & Post & Pre & Post \\
\midrule
Daily volume (\$M) & 69.3 & 85.6 & 168.9 & 257.9 \\
 & (63.4) & (86.0) & (158.9) & (249.3) \\[2pt]
Daily volume (k ETH) & 27.2 & 20.8 & 66.6 & 62.5 \\
 & (25.5) & (21.0) & (63.1) & (59.1) \\[2pt]
Daily swaps & 30,679 & 92,149 & 33,387 & 43,310 \\
 & (31,173) & (86,571) & (31,255) & (41,365) \\[2pt]
Unique traders & 4,344 & 13,913 & 1,318 & 1,755 \\
 & (4,329) & (14,351) & (1,304) & (1,651) \\[2pt]
Mint events & 3,873 & 4,492 & 1,825 & 2,842 \\
 & (4,054) & (4,694) & (1,774) & (2,692) \\[2pt]
Burn events & 3,424 & 4,022 & 1,090 & 2,016 \\
 & (3,680) & (4,340) & (989) & (1,863) \\[2pt]
\bottomrule
\end{tabular}

\end{table}

\subsection{Arbitrage competition and profit}\label{sec:emp-arb}

The model predicts that faster blocks reduce the probability that multiple arbitrageurs arrive in the same block, making same-block auctions for priority rarer and lowering equilibrium gas-fee bids (Proposition~\ref{prop:pfp}). The sequencer's rent share falls, and the savings accrue to the arbitrageurs themselves. We test both sides of this prediction: first, whether gas-fee bidding declines; second, whether the net markout rises through the fee savings.

\subsubsection{Identifying arbitrage swaps}

We classify a swap as an \emph{arbitrage trade} if its signed CEX--DEX log price gap at execution exceeds a threshold $\tau$:
\begin{equation}\label{eq:arblike}
    (p^{\text{CEX}}_t - p^{\text{DEX}}) \times d \times 10^4 > \tau,
\end{equation}
where $p^{\text{CEX}}_t$ is the contemporaneous Binance mid-quote, $p^{\text{DEX}}$ is the reference DEX log price (the post-swap price of the last swap in the previous block), and $d \in \{+1,-1\}$ is the trade direction; a positive value means the trade moves the pool toward the CEX price.
This is a property of the swap, applies symmetrically before and after the upgrade, and does not depend on the identity of the trader.

We use $\tau = 10$ basis points as the baseline. Appendix~\ref{app:arb-classification} reports the gap distribution motivating this choice (a $10$~bp cutoff isolates comparably aggressive directional flow on both chains, whereas lower cutoffs classify different populations because Arbitrum's near-zero gas puts its arbitrage break-even at the pool fee) and shows that the arbitrage-side results are robust to alternative thresholds of $6$ and $8$ basis points.

\subsubsection{Gas-fee bidding}

\begin{figure}[t!]
    \centering
    \includegraphics[width=0.48\textwidth]{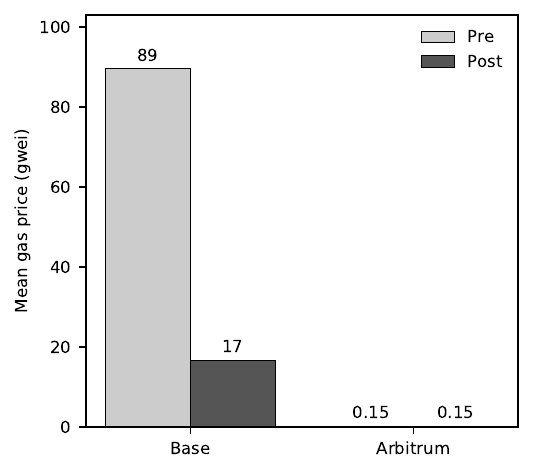}
    \caption{\textbf{Gas price on the largest arbitrage trades.} This figure plots the mean gas price (gwei) on arbitrage swaps above the 90th percentile of the pre-upgrade trade-size distribution, before and after Flashblocks, for Base and Arbitrum on a common scale; Arbitrum's no-priority gas floor of roughly $0.15$~gwei is visually indistinguishable from zero. Sample: WETH--USDC 5~bp pool, arbitrage swaps (signed CEX--DEX gap $>10$~bp), May~16--September~15, 2025 (July~7--16 excluded).}
    \label{fig:arb-gas-collapse}
\end{figure}

Under PGA ordering, arbitrageurs bid gas prices for early block positions. The model predicts that faster blocks reduce the collision probability (two arbitrageurs in the same block), making same-block auctions rarer and lowering equilibrium bids; and because priority is only worth buying when the opportunity is large, the decline should be concentrated among large trades. Figure~\ref{fig:arb-gas-collapse} shows the collapse directly in raw terms. Among the largest arbitrage trades (above the 90th percentile of the pre-upgrade size distribution), where priority bidding is most aggressive, the mean gas bid on Base falls from $89$ to $17$~gwei after Flashblocks, while on Arbitrum it sits near $0.15$~gwei throughout.\footnote{Arbitrum's near-zero gas bids reflect that gas plays no role in ordering under FCFS, not that priority is free. Arbitrageurs who value priority on Arbitrum purchase it through Timeboost express-lane auctions (Section~\ref{sec:emp-institutional}), whose sealed bids accrue to the sequencer outside per-transaction gas fees and are therefore invisible in swap-level gas data. Because Timeboost operates on unchanged terms throughout our sample, this channel is absorbed by the chain fixed effect and does not affect the difference-in-differences estimates.}

Table~\ref{tab:did-consolidated-arb-behavior} estimates the difference-in-differences separately within the top five deciles of the pre-upgrade trade-size distribution (pooled across chains), with the corresponding ETH ranges shown in the column headers. The columns trace the mechanism through the size distribution. Trades in the 50th to 70th percentile range ($2.9$--$5.7$~ETH), large enough to arbitrage but rarely worth prioritizing, show no change in either bin. From the 70th percentile up, bids collapse, monotonically in trade size: $-0.75$ for $5.7$--$9.8$~ETH trades, $-0.86$ for $9.8$--$17$~ETH, and $-1.31$ (a $73\%$ decline) for the top decile. Figure~\ref{fig:app-pt-arb-gas} plots the daily Base $-$ Arbitrum gap within each bin.

Appendix~\ref{app:arb-classification} reports the full size profile with a split of the top decile into P90--P95 and P95--P100, which confirms that the wide top bin's estimate is not driven by within-bin composition: the decline deepens further into the tail ($-1.03$ for P90--P95 and $-1.47$ for P95--P100).

\begin{table}[t!]
    \centering\small
    \caption{\textbf{Arbitrageur gas price by trade size: difference-in-differences estimates.} This table reports difference-in-differences estimates of the log gas price on arbitrage swaps (signed CEX--DEX gap $> 10$~bp), estimated separately within the top five deciles of the pre-upgrade trade-size distribution (both chains), with the corresponding ETH ranges shown in the column headers. Each column is a swap-level regression with chain and date fixed effects; standard errors clustered by date. Appendix~\ref{app:arb-classification} reports the full size profile, including the bottom half of the distribution and a split of the top decile. Sample: WETH--USDC 5~bp pool on Base and Arbitrum, May~16--September~15, 2025 (July~7--16 excluded).}
    \label{tab:did-consolidated-arb-behavior}
    \begin{tabular}{lD{.}{.}{2.2}D{.}{.}{2.2}D{.}{.}{2.2}D{.}{.}{2.2}D{.}{.}{2.2}}
\toprule
 & \multicolumn{1}{c}{(1)} & \multicolumn{1}{c}{(2)} & \multicolumn{1}{c}{(3)} & \multicolumn{1}{c}{(4)} & \multicolumn{1}{c}{(5)} \\
 & \multicolumn{1}{c}{P50--P60} & \multicolumn{1}{c}{P60--P70} & \multicolumn{1}{c}{P70--P80} & \multicolumn{1}{c}{P80--P90} & \multicolumn{1}{c}{P90--P100} \\
 & \multicolumn{1}{c}{2.9--4.0 ETH} & \multicolumn{1}{c}{4.0--5.7 ETH} & \multicolumn{1}{c}{5.7--9.8 ETH} & \multicolumn{1}{c}{9.8--17 ETH} & \multicolumn{1}{c}{$\geq$17 ETH} \\
\midrule
Post $\times$ Base & 0.37 & 0.04 & -0.75^{**} & -0.86^{***} & -1.31^{***} \\
 & (0.26) & (0.37) & (0.35) & (0.32) & (0.22) \\[4pt]
\midrule
Chain FE & \multicolumn{1}{c}{Yes} & \multicolumn{1}{c}{Yes} & \multicolumn{1}{c}{Yes} & \multicolumn{1}{c}{Yes} & \multicolumn{1}{c}{Yes} \\
Date FE   & \multicolumn{1}{c}{Yes} & \multicolumn{1}{c}{Yes} & \multicolumn{1}{c}{Yes} & \multicolumn{1}{c}{Yes} & \multicolumn{1}{c}{Yes} \\
$N$   & \multicolumn{1}{c}{16,631} & \multicolumn{1}{c}{15,073} & \multicolumn{1}{c}{13,685} & \multicolumn{1}{c}{10,834} & \multicolumn{1}{c}{9,250} \\
\bottomrule
\end{tabular}

\end{table}

\subsubsection{Arbitrage profit}\label{sec:arb-profit-metrics}

Arbitrageurs now bid less for priority, but faster blocks also shrink the price gaps they trade on. Does the fee saving outweigh the smaller prize? We measure arbitrage profitability through two markout-based measures.\footnote{Throughout, all prices are in logs and all profit metrics are in basis points ($\times 10^4$).} The \emph{raw markout} is the signed change from the swap's execution price to the CEX mid-quote at a 10-second horizon, minus the 5~bp swap fee:
\begin{equation}
    \text{Raw markout} = (p^{\text{CEX}}_{t+10\text{s}} - p^{\text{exec}}) \times d \times 10^4 - \text{Swap fee},
\end{equation}
where $p^{\text{exec}}$ is the fee-adjusted execution price on the DEX and $d \in \{+1,-1\}$ is the trade direction. The \emph{net markout} subtracts the blockchain fee (the total transaction cost in bps of trade size):
\begin{equation}
    \text{Net markout} = \text{Raw markout} - \text{Blockchain fee}.
\end{equation}

Table~\ref{subtab:arb-markout} reports the estimates, and Figure~\ref{fig:app-pt-arb-profit} plots the daily series behind them. The raw markout falls by $0.53$~bps but is not statistically significant; the blockchain fee falls by far more, $3.26$~bps; and the net markout therefore rises by $2.73$~bps.
The lower blockchain-fee burden is consistent with the model's prediction that faster blocks let arbitrageurs retain a larger share of the prize.

Table~\ref{subtab:arb-dollar} values these flows in dollars at contemporaneous market prices, the same convention as Table~\ref{tab:market-activity}.\footnote{The mean ETH price rises from \$2{,}511 to \$4{,}178 across the periods, so within-chain dollar changes mix quantities with the ETH price. The Base-versus-Arbitrum contrast is essentially unchanged by the denomination: both chains face the same price path on every day. The log-volume difference-in-differences of Section~\ref{sec:emp-mq}, which carries the identified comparison, is exactly invariant, since the common daily price is absorbed by the date fixed effects.} Two patterns stand out. First, Flashblocks fragments arbitrage into more, smaller trades: on Base the number of arbitrage trades more than doubles ($683$ to $1{,}728$ per day) while the mean trade shrinks from \$22{,}200 to \$7{,}400. Arbitrage's share of total pool volume, a measure invariant to the ETH-price appreciation, falls from $22\%$ to $15\%$ on Base and from $5\%$ to $3\%$ on Arbitrum. Second, the gross profit extracted from the pool (the raw markout) falls from \$10{,}200 to \$7{,}200 per day on Base while rising from \$8{,}800 to \$10{,}500 on Arbitrum with the broader market, so the adverse-selection loss borne by liquidity providers shrinks sharply on the faster chain. Net dollar profit on Base rises, from \$4{,}500 to \$6{,}300 per day: the gas saving more than offsets the smaller gross opportunity.

The ratio of blockchain fees to gross arbitrage profit in Table~\ref{subtab:arb-dollar} is the sequencer's share of the rent, the object of the first prediction: the faster the chain, the smaller that share (Proposition~\ref{prop:rent-share}). On 2-second Base, blockchain fees took \$5{,}700 of the \$10{,}200 daily gross arbitrage profit, a $56\%$ share; under 200-millisecond Flashblocks, they took \$900 of \$7{,}200, or $12\%$.\footnote{Because the blockchain fee includes the cost of posting transaction data to Ethereum, which the sequencer passes through, these shares are an upper bound on the sequencer's own rent share. We do not compute a comparable share for Arbitrum: under Timeboost, priority is sold through sealed-bid express-lane auctions whose proceeds are not part of per-transaction gas fees and are not observed in our data. Public aggregate figures nonetheless suggest the auction channel is material: Timeboost raised roughly \$2 million chain-wide in its first three months \parencite{chawla2025timeboost}, on the order of \$20{,}000 per day. The two numbers are not directly comparable (Timeboost revenue covers priority for all activity on Arbitrum and is bid per 60-second round in advance, whereas our figures cover realized arbitrage in a single pool), but they indicate that Arbitrum's sequencer also captures priority rents, through auction proceeds rather than gas.} A third, slower execution speed extends the pattern: on 12-second Ethereum, arbitrageurs hand roughly $96\%$ of their profits to validators \parencite{capponi2025liquidity}.

\begin{table}[t!]
    \centering\small
    \caption{\textbf{Arbitrageur profit, by chain and period.} This table reports arbitrageur profitability by chain and period. Subtable~(a): per-dollar markout at the chain-day level, volume-weighted basis points at the 10-second horizon; the DiD column is the Post~$\times$~Base coefficient from a two-way fixed-effects regression with date-clustered standard errors. Raw markout $= (p^{\text{CEX}}_{t+10\text{s}} - p^{\text{exec}}) \times d \times 10^4 -$ swap fee; net markout $=$ raw markout $-$ blockchain fee. Subtable~(b): aggregate activity and profit (descriptive, no DiD); dollar amounts valued at contemporaneous market prices, matching Table~\ref{tab:market-activity}. Sample: WETH--USDC 5~bp pool on Base and Arbitrum, May~16--September~15, 2025 (July~7--16 excluded).}
    \label{tab:did-consolidated-arb-profit}
    \begin{subtable}{\linewidth}
        \centering
        \caption{Per-dollar markout}\label{subtab:arb-markout}
        \begin{tabular}{lD{.}{.}{2.2}D{.}{.}{2.2}D{.}{.}{2.2}D{.}{.}{2.2}D{.}{.}{2.3}@{\hskip 1.5em}l}
\toprule
 & \multicolumn{2}{c}{Base} & \multicolumn{2}{c}{Arbitrum} & \multicolumn{2}{c}{} \\
\cmidrule(lr){2-3}\cmidrule(lr){4-5}
 & \multicolumn{1}{c}{Pre} & \multicolumn{1}{c}{Post} & \multicolumn{1}{c}{Pre} & \multicolumn{1}{c}{Post} & \multicolumn{2}{c}{DiD} \\
\midrule
Raw markout & 6.23 & 5.18 & 8.90 & 8.38 & -0.53 & (1.30) \\
Blockchain fee & 3.92 & 0.66 & 0.04 & 0.04 & -3.26^{***} & (0.21) \\
Net markout & 2.31 & 4.52 & 8.86 & 8.34 & 2.73^{**} & (1.35) \\
\bottomrule
\end{tabular}

    \end{subtable}

    \vspace{1em}

    \begin{subtable}{\linewidth}
        \centering
        \caption{Aggregate activity and profit}\label{subtab:arb-dollar}
        \begin{tabular}{lrrrr}
\toprule
 & \multicolumn{2}{c}{Base} & \multicolumn{2}{c}{Arbitrum} \\
\cmidrule(lr){2-3}\cmidrule(lr){4-5}
 & Pre & Post & Pre & Post \\
\midrule
Arb trades/day & 683 & 1,728 & 746 & 560 \\
Mean trade size (\$000) & 22.2 & 7.4 & 11.1 & 16.0 \\
Volume/day (\$M) & 15.2 & 12.8 & 8.3 & 9.0 \\
Gross profit/day (\$000) & 10.2 & 7.2 & 8.8 & 10.5 \\
Net profit/day (\$000) & 4.5 & 6.3 & 8.7 & 10.4 \\
\bottomrule
\end{tabular}

    \end{subtable}
\end{table}

\subsection{Liquidity provision and cream-skimming}\label{sec:emp-lp}

On the liquidity provision side, the model predicts that faster blocks enable active LPs to time liquidity provision, selectively avoiding toxic order flow while collecting fee income on non-toxic flow (Proposition~\ref{prop:r7}). We test this prediction by tracking the rebalancing behavior, flow composition, and payoff of active LPs.

\subsubsection{Identifying active LPs}

We define an \emph{active LP} as a wallet that supplies liquidity through frequent, tightly ranged position cycling rather than passive buy-and-hold provision, and we identify the cohort from pre-upgrade behavior using four criteria applied symmetrically on both chains.

Specifically, starting from the top~10 addresses by total mint/burn activity on each chain, we retain a wallet only if it meets all four. First, it rebalances quickly: its 25th-percentile elapsed time between opening and closing a position is under 60 seconds. Second, it quotes tight ranges: its 25th-percentile price-range width is under 100 basis points, consistent with concentrated provision around the current price. Third, it maintains a balanced position lifecycle: its burn-to-mint ratio lies within 0.5 of unity, ruling out one-sided liquidity additions. Fourth, it makes markets rather than trading directionally: its order-to-trade ratio exceeds~2. We further require each wallet to be active on at least 10 distinct days in the pre-upgrade window, so the cohort is selected on pre-treatment information alone and reflects persistent market-making rather than burst-and-exit activity. The two wallets this last requirement excludes stop transacting in the pool by late May, roughly two months before the upgrade. These filters leave one Base address and four Arbitrum addresses.

\subsubsection{Active LP behavior and payoff}

\begin{figure}[t!]
    \centering
    \includegraphics[width=0.6\textwidth]{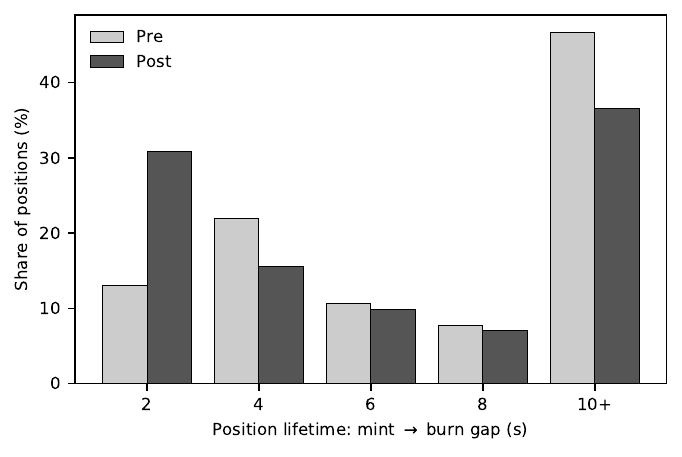}
    \caption{\textbf{Active Base LP position lifetime.} This figure plots the share of the active Base LP's positions by lifetime (the mint-to-burn elapsed time), in two-second buckets, before and after Flashblocks. Sample: WETH--USDC 5~bp pool on Base, May~16--September~15, 2025 (July~7--16 excluded).}
    \label{fig:lp-rebalancing-dist}
\end{figure}

\paragraph{Rebalancing speed.}
We measure rebalancing behavior through the log of the LP's median daily rebalancing time (the mint-to-burn elapsed time) and the median daily gas price. Table~\ref{tab:did-consolidated-lp} reports the difference-in-differences estimates (daily series in Figure~\ref{fig:app-pt-lp}). The log rebalancing time falls by $1.07$ on Base relative to Arbitrum, roughly $66\%$ faster after Flashblocks,\footnote{Rebalancing time is measured using block timestamps, which have 2-second resolution even after Flashblocks; mint-burn pairs within the same two-second block are rare on Base (0.03\% both pre and post), so the resolution does not materially affect the median.} while the gas-price difference-in-differences is insignificant. This contrast is consistent with the model's prediction: on a faster chain, the active LP's edge comes from speed, not aggressive fee bidding. The rebalancing time is the empirical counterpart of the duration of liquidity provision in Proposition~\ref{prop:spells}, the time from a redeposit to the next successful withdrawal, which the model predicts faster blocks shorten. Figure~\ref{fig:lp-rebalancing-dist} shows the distribution behind the rebalancing result. After Flashblocks, position lifetimes of the active Base LP shift sharply toward two seconds: she mints a position and burns it in the very next block much more often.

\begin{table}[t!]
    \centering\small
    \caption{\textbf{LP rebalancing behavior: difference-in-differences estimates.} This table reports LP-day difference-in-differences estimates of active-LP rebalancing behavior. Gas price = median daily gas price (gwei); Log reb.\ time = log median daily rebalancing time (minutes). Both at the LP-day level. Standard errors clustered by date. Sample: WETH--USDC 5~bp pool on Base and Arbitrum, May~16--September~15, 2025 (July~7--16 excluded).}
    \label{tab:did-consolidated-lp}
    \begin{tabular}{lD{.}{.}{2.2}D{.}{.}{2.2}}
\toprule
 & \multicolumn{1}{c}{(1)} & \multicolumn{1}{c}{(2)} \\
 & \multicolumn{1}{c}{Gas price} & \multicolumn{1}{c}{Log reb.\ time} \\
\midrule
Post $\times$ Base & 0.03 & -1.07^{***} \\
 & (0.02) & (0.33) \\[4pt]
\midrule
LP FE & \multicolumn{1}{c}{Yes} & \multicolumn{1}{c}{Yes} \\
Date FE   & \multicolumn{1}{c}{Yes} & \multicolumn{1}{c}{Yes} \\
$N$   & \multicolumn{1}{c}{314} & \multicolumn{1}{c}{314} \\
$R^2$ & 0.008 & 0.060 \\
\bottomrule
\end{tabular}

\end{table}

\paragraph{Cream-skimming.}
Faster rebalancing enables the active LP to selectively provide liquidity, bearing non-toxic order flow while avoiding the flow that adversely selects her. We measure this through her \emph{toxic-exposure share}: the fraction of her total dollar exposure that falls on swaps whose signed execution-time CEX--DEX gap exceeds the $5$~bp pool fee (the point beyond which the price move outruns the fee she earns). We benchmark it against the same share computed on the pool's total flow. The cream-skimming estimate is then a difference-in-differences measure, i.e., her share minus the pool's, which nets out the shift in the flow mix itself.

\begin{figure}[t!]
    \centering
    \includegraphics[width=0.62\textwidth]{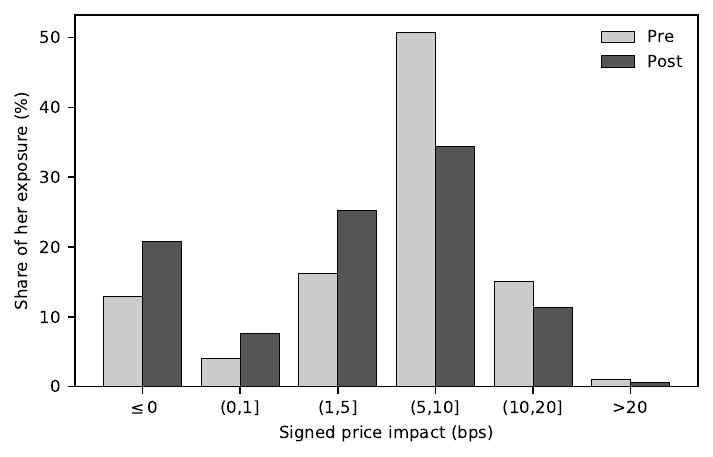}

    \vspace{1.2em}
    {\small\begin{tabular}{lD{.}{.}{2.1}D{.}{.}{2.1}D{.}{.}{2.2}}
\toprule
 & \multicolumn{1}{c}{Pre} & \multicolumn{1}{c}{Post} & \multicolumn{1}{c}{$\Delta$ (Post $-$ Pre)} \\
\midrule
Active LP toxic-exposure share & 62.9 & 47.5 & -15.40^{***} \\
Pool toxic-flow share & 83.0 & 76.5 & -6.48^{***} \\
Difference (LP $-$ pool) & -20.1 & -29.0 & -8.93^{***} \\
\bottomrule
\end{tabular}
}

    \caption{\textbf{Active Base LP cream-skimming.} This figure plots the active Base LP's exposure across signed price-impact buckets (top) and reports her toxic-exposure share (bottom). In the top panel, each bar is the share of her total exposure in a disjoint signed execution-time CEX--DEX price-impact bucket (bps); positive impact is toxic to her. The bottom panel reports the toxic-exposure share, the fraction of her total exposure falling on toxic flow ($>5$~bp gap), as daily means in percentage points; $\Delta$ is the date-clustered change, and for the Difference row it is the cream-skimming difference-in-differences (her share minus the pool's). Sample: WETH--USDC 5~bp pool on Base, May~16--September~15, 2025 (July~7--16 excluded).}
    \label{fig:lp-cream-skimming}
\end{figure}

Figure~\ref{fig:lp-cream-skimming} shows the shift directly. The plot bins her dollar exposure by the signed CEX--DEX price impact at execution; positive impact is toxic to her, negative is non-toxic, and the $5$~bp pool fee marks the break-even point. Before Flashblocks the bulk of her exposure sits in the toxic buckets beyond the fee; after, it migrates toward the non-toxic buckets near and below zero, thinning the mass she carries in toxic flow.

The bottom panel of Figure~\ref{fig:lp-cream-skimming} summarizes this as her toxic-exposure share, the fraction of her exposure beyond the $5$~bp fee. It falls from $62.9\%$ to $47.5\%$, while the pool's toxic-flow share falls only from $83.0\%$ to $76.5\%$; the difference-in-differences is a statistically significant $-8.93$ percentage points. Because the measure nets out both her scaling-up and the shifting flow mix, the remaining gap isolates her withdrawal from toxic flow, the model's cream-skimming mechanism. Her pre-period series is not entirely flat: she already avoids toxic flow sharply on a handful of days late in the pre-period, so we probe the estimate in Appendix~\ref{app:window-robustness}. The change remains $-6.81$ percentage points (significant at the $1\%$ level) in a $\pm$4-week window around the launch, and the post-launch series is flat after an immediate level drop. Before Flashblocks such strong avoidance was the exception; after, it is the everyday level.

\paragraph{Payoff.}
The cream-skimming documented above translates into a marked improvement in the active LP's payoff. We define the LP's net payoff per swap, the mirror image of the arbitrageur's markout, as the swap fee minus the signed change from the execution price to the CEX mid-quote at the 10-second horizon, volume-weighted by her dollar exposure $|V_j| \cdot s_j$, where $V_j$ is the dollar volume of swap $j$ and $s_j$ is her share of pool liquidity at the active tick.\footnote{We reconstruct her position from the cumulative sequence of mint and burn events for each (address, tick-lower, tick-upper) tuple, initializing from May~1 (15~days before the analysis window).}

Table~\ref{subtab:lp-payoff} reports the result, with the daily series in Figure~\ref{fig:app-pt-lp}. Her per-dollar net payoff rises by $0.73$~bps on Base relative to Arbitrum, shifting from approximately break-even ($0.26$~bps) to clearly positive ($1.28$~bps).\footnote{Throughout Table~\ref{tab:did-lp-payoff}, LP-days on which an LP's dollar exposure falls below $5\%$ of her own pre-upgrade median are excluded: a volume-weighted payoff on negligible exposure is noise, not economics. The excluded days form a well-separated near-zero cluster in the exposure distribution, so the estimate is similar for any cutoff between $2\%$ and $25\%$ of the pre-upgrade median; without the filter the estimate is $0.70$~bps, significant at the 5\% level.}
This improvement is consistent with Proposition~\ref{prop:r7}: at sufficiently high block speeds, the active LP's arbitrage losses fall and her trading income rises.

Table~\ref{subtab:lp-dollar} puts this in dollars. The active Base LP's mean daily profit rises from about \$9 to about \$652, and her share of profitable days climbs from $51\%$ to $89\%$. The four active Arbitrum LPs improve on neither margin: their combined daily profit stays negative and they are profitable on $12\%$ of days before and $3\%$ after. The Base LP's increased profitability is therefore specific to the faster chain, not a market-wide trend.

\begin{table}[t!]
    \centering\small
    \caption{\textbf{Active LP payoff, by chain and period.} This table reports the active LP's per-dollar payoff and dollar profit by chain and period. Subtable~(a): per-dollar net payoff (the swap fee minus the signed 10-second price move against the Binance mid-quote, as defined in the text) at the LP-day level, volume-weighted basis points at the 10-second horizon; the DiD column is the Post~$\times$~Base coefficient from a regression with date-clustered standard errors. The cohort is the single active Base LP versus the four active Arbitrum LPs. In both subtables, LP-days on which the LP's dollar exposure falls below 5\% of her own pre-upgrade median are excluded (negligible-exposure outage days; see footnote in the text). Subtable~(b): dollar profit (descriptive, no scale-matched control), valued at contemporaneous market prices, matching Tables~\ref{tab:market-activity} and~\ref{subtab:arb-dollar}; the Base column is the single active LP, the Arbitrum column the daily sum across its four active LPs, so the dollar levels are not per-LP comparable across chains (the profitable-days share is scale-free and is comparable). Sample: WETH--USDC 5~bp pool on Base and Arbitrum, May~16--September~15, 2025 (July~7--16 excluded).}
    \label{tab:did-lp-payoff}
    \begin{subtable}{\linewidth}
        \centering
        \caption{Per-dollar net payoff}\label{subtab:lp-payoff}
        \begin{tabular}{lD{.}{.}{2.2}D{.}{.}{2.2}D{.}{.}{2.2}D{.}{.}{2.2}D{.}{.}{2.2}@{\hskip 1.5em}l}
\toprule
 & \multicolumn{2}{c}{Base (active LP)} & \multicolumn{2}{c}{Arbitrum} & \multicolumn{2}{c}{} \\
\cmidrule(lr){2-3}\cmidrule(lr){4-5}
 & \multicolumn{1}{c}{Pre} & \multicolumn{1}{c}{Post} & \multicolumn{1}{c}{Pre} & \multicolumn{1}{c}{Post} & \multicolumn{2}{c}{DiD} \\
\midrule
Net payoff (markout, bps) & 0.26 & 1.28 & -0.83 & -0.67 & 0.73^{**} & (0.36) \\
\bottomrule
\end{tabular}

    \end{subtable}

    \vspace{1em}

    \begin{subtable}{\linewidth}
        \centering
        \caption{Dollar profit, active-LP cohort}\label{subtab:lp-dollar}
        \begin{tabular}{lrrrr}
\toprule
 & \multicolumn{2}{c}{Base} & \multicolumn{2}{c}{Arbitrum} \\
\cmidrule(lr){2-3}\cmidrule(lr){4-5}
 & Pre & Post & Pre & Post \\
\midrule
Mean (\$/day) & 9 & 652 & -26 & -58 \\
Median (\$/day) & 18 & 440 & -21 & -39 \\
Profitable days (\%) & 51 & 89 & 12 & 3 \\
\bottomrule
\end{tabular}

    \end{subtable}
\end{table}

\subsection{Market quality}\label{sec:emp-mq}

Faster blocks move the two sides of market quality through different channels. Faster arbitrage execution keeps the pool price closer to the CEX, improving price efficiency. For liquidity, the model's prediction depends on where the chain sits relative to the depth peak: faster blocks raise fee income by executing more liquidity-trader transactions before they expire, but they also realize stale-price losses more often, and once blocks are fast enough the losses dominate, passive capital retreats, and depth falls (Propositions~\ref{prop:r7} and~\ref{prop:liquidity-maximum}). Table~\ref{tab:did-consolidated-mq} reports both sides, and Figure~\ref{fig:app-pt-mq} plots the daily series for every outcome.

Price efficiency improves unambiguously. For each block, we compute the block-end price deviation, the absolute log gap between the post-swap DEX price and the contemporaneous Binance mid-quote, then summarize it as the daily time-weighted average (weighting each block by the time until the next block). The time-weighted average deviation falls by $0.11$~bps on Base relative to Arbitrum.

Liquidity moves the other way, as the model predicts once blocks are fast enough: log pool depth (the daily time-weighted average pool liquidity at the active tick) falls by $0.21$ on Base, a roughly $19\%$ decline. Liquidity also concentrates: the largest single LP's time-weighted share of pool liquidity rises by $4.20$ percentage points. The thinner depth translates into higher trading costs. We measure the effective spread of every swap in the pool, arbitrage and non-toxic flow alike, as its fee-exclusive execution price relative to the pool's pre-swap marginal price, the AMM analog of the prevailing mid-quote,\footnote{The pool price moves only when a swap executes, so the post-swap price of the immediately preceding swap is the exact prevailing quote at execution, and the measure is invariant to the block-time change itself. Because trading costs on an AMM scale with trade size, and the upgrade shifts the trade-size distribution, we estimate the difference-in-differences within a fixed size band rather than pooling across sizes. Appendix~\ref{app:eff-spread} reports the full size profile.} and estimate the DiD for trades of $\sim$1 ETH (0.8--1.25 ETH) at the chain-day level. The effective spread widens by $0.07$~bps on Base relative to Arbitrum: Base rises from $0.17$ to $0.25$~bps (Figure~\ref{fig:flashblocks_puzzle_spread}) while Arbitrum stays flat, and the widening generally grows with trade size (Appendix~\ref{app:eff-spread}). Worsened liquidity reduces volume and thus gains from trade too: Base ETH volume falls $\sim$18\% relative to Arbitrum (log volume $= -0.20$).\footnote{Because the regression is in logs with date fixed effects, the ETH price, common to both chains on any given day, drops out: the estimate is identical whether volume is measured in ETH or in dollars.}

Taken together, the evidence lines up with all three predictions. Faster blocks move arbitrage rents from the sequencer to arbitrageurs. The active LP rebalances faster, cream-skims by sharing fees while avoiding toxic flow, and earns more. And the pool itself becomes thinner: depth falls, spreads widen, and volume falls, even as pool prices track the reference price more closely.

\begin{table}[t!]
    \centering\small
    \caption{\textbf{Market quality: difference-in-differences estimates.} This table reports chain-day difference-in-differences estimates of price efficiency and market liquidity. TWA dev.\ = daily time-weighted average absolute CEX--DEX log price deviation (bps); Log depth = log daily time-weighted average pool liquidity at the active tick; Eff.\ spread $\sim$1 ETH = daily volume-weighted effective spread on trades of 0.8--1.25 ETH, measured against the pre-swap marginal pool price (bps); Top-1 TW liq.\ shr.\ = the largest single LP's time-weighted share of pool liquidity (\%); Log vol.\ = log daily ETH volume. All outcomes at the chain-day level. Standard errors clustered by date. Sample: WETH--USDC 5~bp pool on Base and Arbitrum, May~16--September~15, 2025 (July~7--16 excluded).}
    \label{tab:did-consolidated-mq}
    \begin{tabular}{lD{.}{.}{2.2}D{.}{.}{2.2}D{.}{.}{2.2}D{.}{.}{2.2}D{.}{.}{2.2}}
\toprule
 & \multicolumn{1}{c}{(1)} & \multicolumn{1}{c}{(2)} & \multicolumn{1}{c}{(3)} & \multicolumn{1}{c}{(4)} & \multicolumn{1}{c}{(5)} \\
 & \multicolumn{1}{c}{\makecell{TWA dev. \\ (bps)}} & \multicolumn{1}{c}{Log depth} & \multicolumn{1}{c}{\makecell{Eff. spread \\ $\sim$1 ETH (bps)}} & \multicolumn{1}{c}{\makecell{Top-1 TW \\ liq. shr. (\%)}} & \multicolumn{1}{c}{Log vol. (ETH)} \\
\midrule
Post $\times$ Base & -0.11^{***} & -0.21^{***} & 0.07^{***} & 4.20^{**} & -0.20^{***} \\
 & (0.03) & (0.04) & (0.01) & (2.11) & (0.04) \\[4pt]
\midrule
Chain FE & \multicolumn{1}{c}{Yes} & \multicolumn{1}{c}{Yes} & \multicolumn{1}{c}{Yes} & \multicolumn{1}{c}{Yes} & \multicolumn{1}{c}{Yes} \\
Date FE   & \multicolumn{1}{c}{Yes} & \multicolumn{1}{c}{Yes} & \multicolumn{1}{c}{Yes} & \multicolumn{1}{c}{Yes} & \multicolumn{1}{c}{Yes} \\
$N$   & \multicolumn{1}{c}{226} & \multicolumn{1}{c}{226} & \multicolumn{1}{c}{226} & \multicolumn{1}{c}{226} & \multicolumn{1}{c}{226} \\
$R^2$ & 0.157 & 0.301 & 0.592 & 0.065 & 0.339 \\
\bottomrule
\end{tabular}

\end{table}

\section{Conclusion}\label{sec:conclusion}

In this paper, we model how block speed shapes competition for execution priority and liquidity provision on blockchain-based exchanges.
Slow blocks make arbitrageurs clash in the same block and bid the arbitrage prize away to the sequencer; fast blocks make them race across blocks, and more of the prize stays with whoever arrives first.
Fast blocks also let an active liquidity provider withdraw ahead of stale-price arbitrage while keeping the liquidity-trader fees.
Liquidity is deepest at an intermediate speed.
Past that point faster blocks thin the pool, and beyond a threshold they let the active provider cream-skim: she earns more while passive capital retreats and pool depth declines further.

Empirical evidence from the Base Flashblocks rollout is consistent with these predictions. When blocks become much faster, gas bidding falls sharply on the largest arbitrage trades, and the net arbitrage markout rises.
The active liquidity provider rebalances faster, shifts her exposure away from toxic flow, and earns a positive net payoff.
Pool depth falls, liquidity concentrates, effective spreads widen, and volume declines, even as pool prices track the external market more closely.

The evidence does not locate the liquidity-maximizing block rate, and we draw no welfare conclusion.
It does point to a tension at the center of blockchain market design: faster blocks improve price discovery but can leave a thinner pool.
Under the SEC's exemption, tokenized U.S. stocks can now trade in AMM pools on public blockchains, and those pools inherit their chain's block speed.
Regulators and market participants should be mindful of this trade-off.

\printbibliography

\clearpage
\appendix
\begin{center}{\LARGE\bfseries Appendix}\end{center}
\setcounter{table}{0}\renewcommand{\thetable}{A\arabic{table}}\renewcommand{\theHtable}{A\arabic{table}}
\setcounter{figure}{0}\renewcommand{\thefigure}{A\arabic{figure}}\renewcommand{\theHfigure}{A\arabic{figure}}
\setcounter{equation}{0}\renewcommand{\theequation}{A\arabic{equation}}\renewcommand{\theHequation}{A\arabic{equation}}
\section{Parallel Trends: Daily Difference Plots}\label{app:parallel-trends}

This appendix provides daily diagnostics relevant to the parallel-trends assumption underlying the difference-in-differences design. We plot selected outcomes from the main tables over the sample window. Gray points are pre-upgrade days and blue points post-upgrade days; horizontal lines and shaded bands show the pre- and post-period means of the daily difference and their 95\% confidence intervals from the chain-day regression with date-clustered standard errors. The vertical dashed line marks the start of the post-upgrade period (July~17, 2025), and the gray vertical band the excluded transition window (July~7--16). These plots describe the observed time patterns; they do not establish the counterfactual parallel-trends assumption.

\begin{figure}[t]
    \centering
    \includegraphics[width=0.86\textwidth]{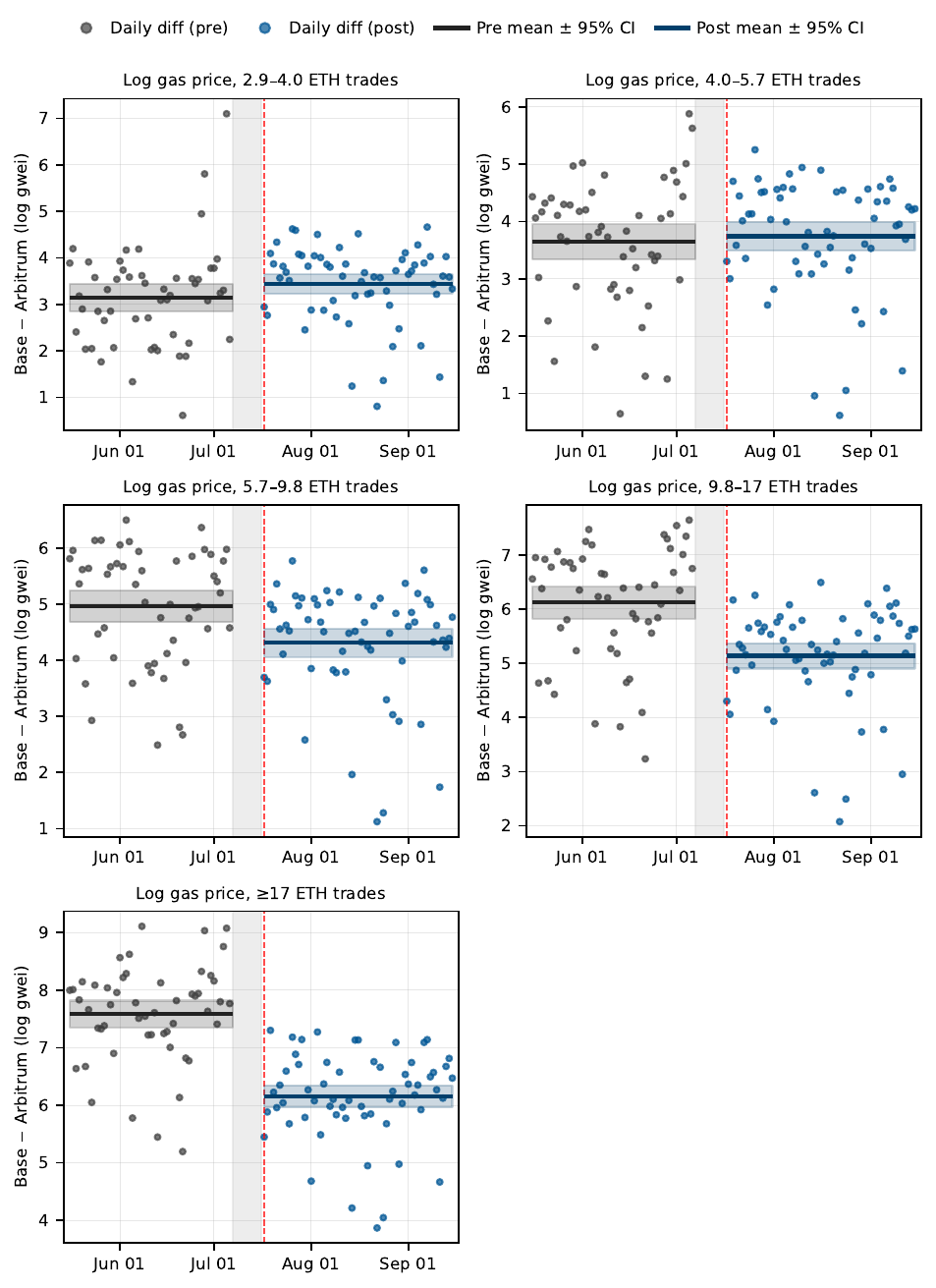}
    \caption{\textbf{Parallel trends: arbitrageur gas price} (Table~\ref{tab:did-consolidated-arb-behavior}). Daily Base $-$ Arbitrum difference in the mean log gas price on arbitrage swaps (signed CEX--DEX gap $> 10$~bp), within the five trade-size bins of Table~\ref{tab:did-consolidated-arb-behavior}. Sample: WETH--USDC 5~bp pool on Base and Arbitrum, May~16--September~15, 2025 (July~7--16 excluded).}
    \label{fig:app-pt-arb-gas}
\end{figure}

\begin{figure}[t]
    \centering
    \includegraphics[width=0.86\textwidth]{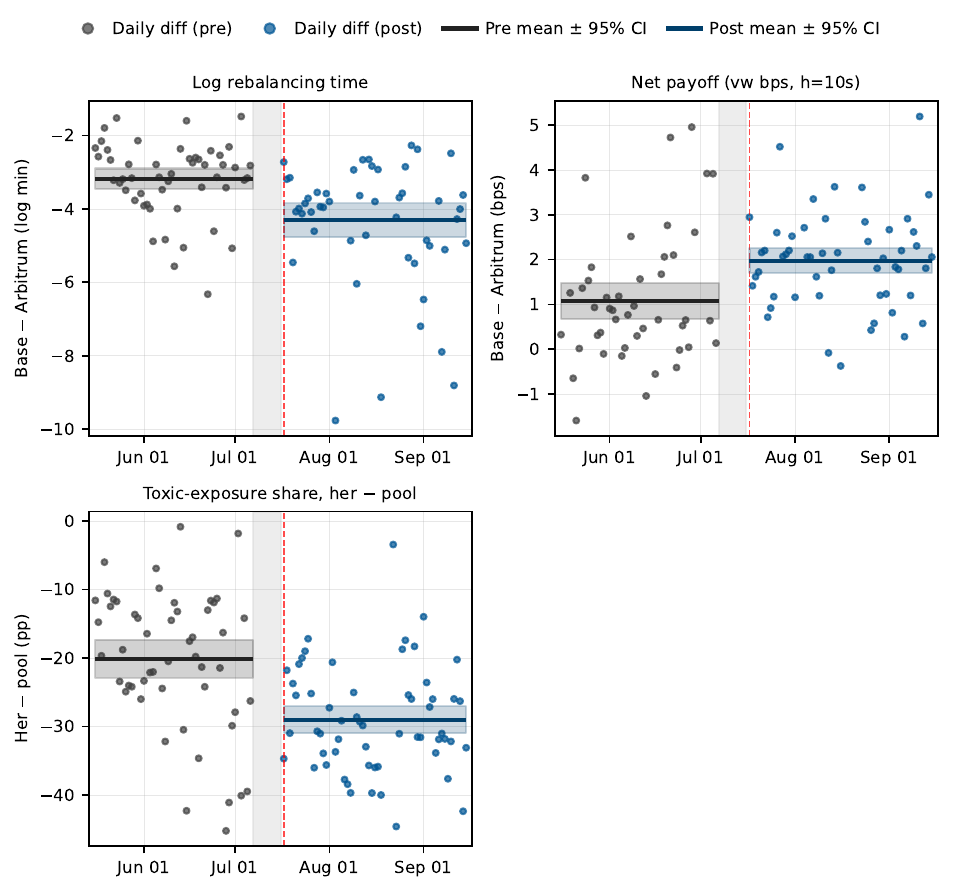}
    \caption{\textbf{Parallel trends: liquidity provision} (Tables~\ref{tab:did-consolidated-lp} and~\ref{tab:did-lp-payoff}; Figure~\ref{fig:lp-cream-skimming}). Top row: daily Base $-$ Arbitrum difference in the log median rebalancing time and in the active LPs' per-dollar net payoff (volume-weighted bps, 10-second horizon), each averaged across active LPs within a chain-day. Bottom panel: daily difference between the active Base LP's toxic-exposure share and the pool's toxic-flow share (percentage points; signed execution-time CEX--DEX gap $> 5$~bp); this outcome is a within-Base contrast, so the plotted series is the outcome itself rather than a Base $-$ Arbitrum difference. Sample: WETH--USDC 5~bp pool on Base and Arbitrum, May~16--September~15, 2025 (July~7--16 excluded).}
    \label{fig:app-pt-lp}
\end{figure}

Two reading notes. First, the gas-price and LP tables use swap-level and LP-day observations, respectively, whereas the figures aggregate to chain-day values with equal day weights. The plotted shifts therefore need not equal the table coefficients. Figure~\ref{fig:app-pt-arb-gas} shows all five large-trade bins of Table~\ref{tab:did-consolidated-arb-behavior}. Second, the bottom panel of Figure~\ref{fig:app-pt-lp} is a within-Base contrast: the active LP's toxic-exposure share minus the pool's toxic-flow share. This series has sharply negative observations late in the pre-period; Appendix~\ref{app:window-robustness} reports its sensitivity to the event window and a placebo date.

\begin{figure}[t]
    \centering
    \includegraphics[width=0.86\textwidth]{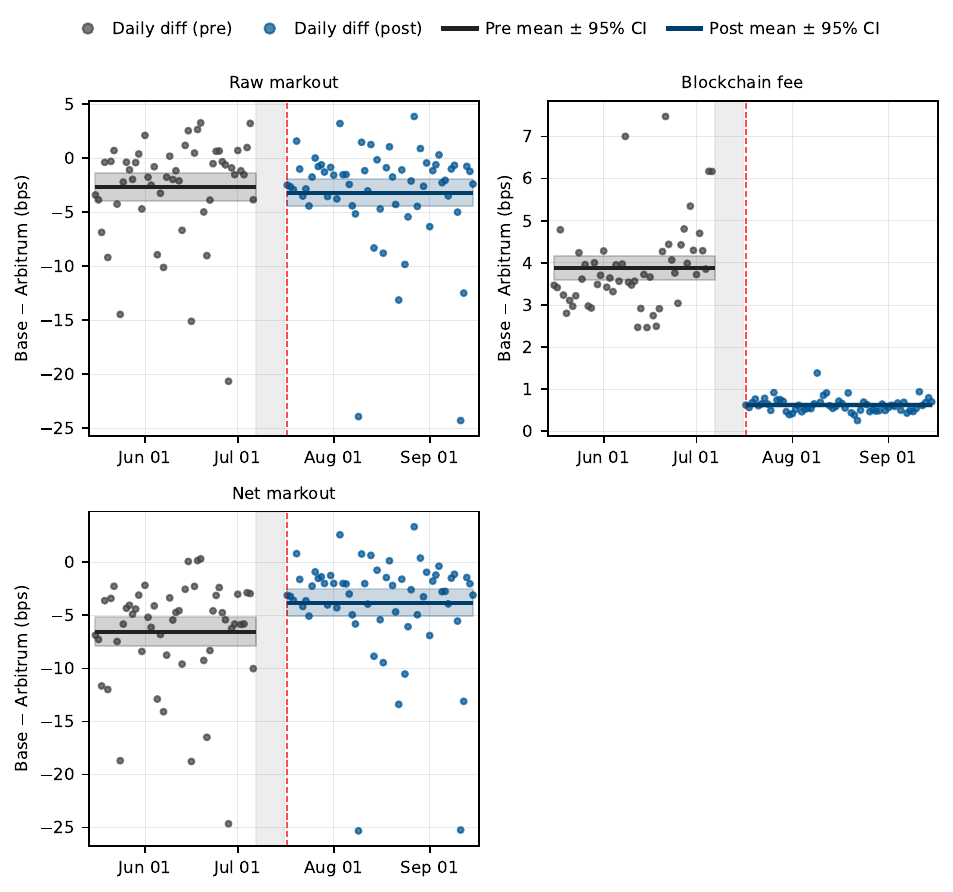}
    \caption{\textbf{Parallel trends: arbitrageur profit} (Table~\ref{tab:did-consolidated-arb-profit}). Daily Base $-$ Arbitrum difference in the volume-weighted raw markout, blockchain fee, and net markout on arbitrage swaps (bps, 10-second horizon). Sample: WETH--USDC 5~bp pool on Base and Arbitrum, May~16--September~15, 2025 (July~7--16 excluded).}
    \label{fig:app-pt-arb-profit}
\end{figure}

\begin{figure}[t]
    \centering
    \includegraphics[width=0.86\textwidth]{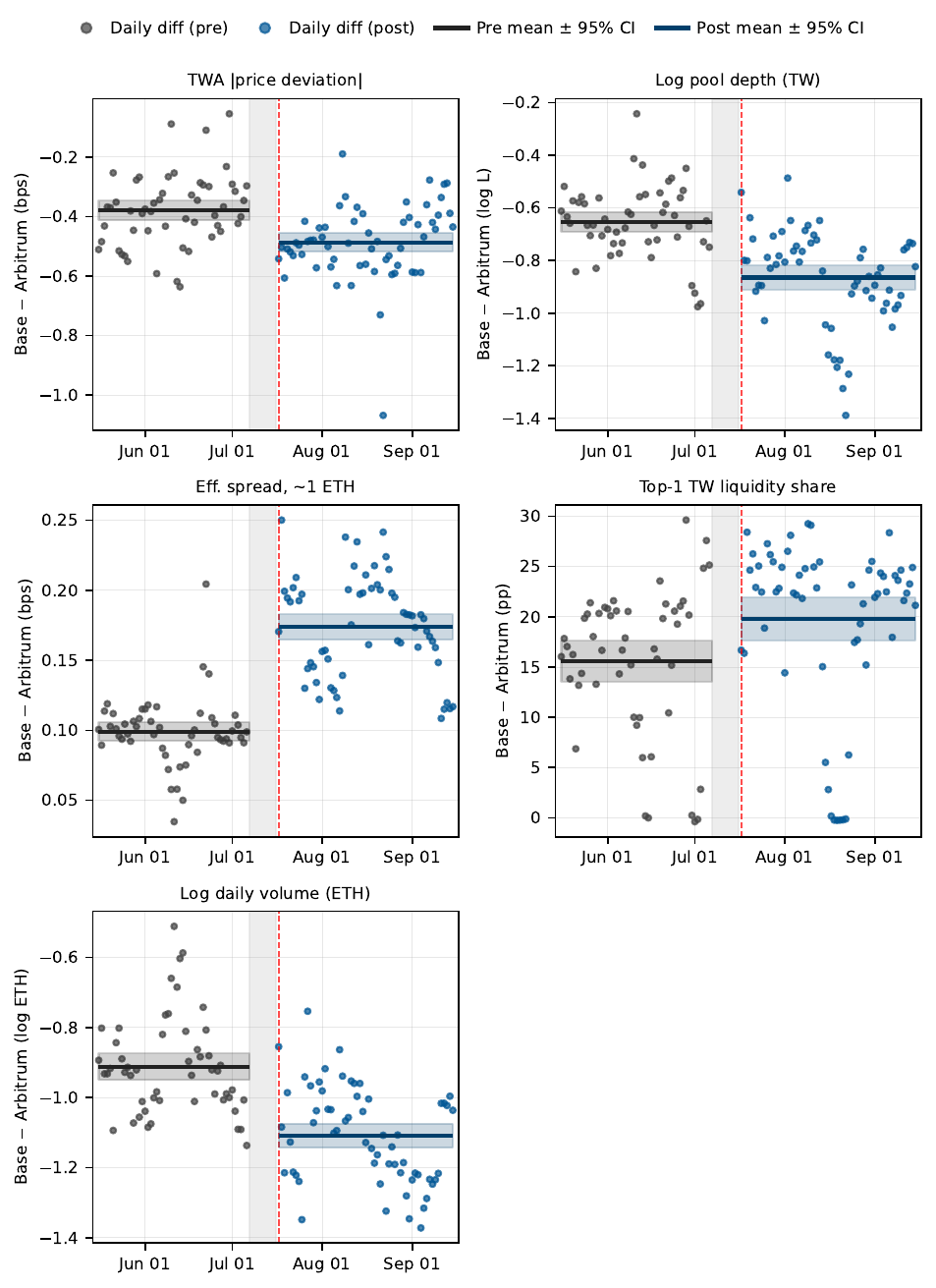}
    \caption{\textbf{Parallel trends: market quality} (Table~\ref{tab:did-consolidated-mq}). Daily Base $-$ Arbitrum difference in the time-weighted average absolute CEX--DEX deviation, log pool depth, the $\sim$1 ETH effective spread, the top-1 time-weighted liquidity share, and log ETH volume. Sample: WETH--USDC 5~bp pool on Base and Arbitrum, May~16--September~15, 2025 (July~7--16 excluded).}
    \label{fig:app-pt-mq}
\end{figure}

\clearpage
\section{Robustness}\label{app:robustness}

This appendix collects the robustness checks on the main estimates: the threshold that classifies arbitrage swaps together with the full trade-size profile of the gas-price estimates; alternative event windows, a placebo launch date, and the exclusion of the days on which the active Base LP was inactive; and the effective-spread estimates across the full trade-size distribution.

\subsection{Arbitrage classification threshold and gas-price size profile}\label{app:arb-classification}

This subsection documents the operational arbitrage classification of Section~\ref{sec:emp-arb}, based on the signed CEX--DEX gap in Equation~\ref{eq:arblike}. It reports the gap distribution that motivates the baseline $\tau=10$~bp, the sensitivity of the arbitrage-side estimates to $\tau=6$ and $8$~bp, and the gas-price estimates across the full trade-size distribution.

\paragraph{Gap distribution and the choice of $\tau$.}
Table~\ref{tab:app-gap-distribution} reports the signed-gap distribution by chain and period. In the pre-upgrade sample, $2.2\%$ of swaps on each chain exceed $10$~bp. This cutoff therefore selects similar pre-period fractions on the two chains, although equal fractions need not identify economically identical trades.

At the $5$~bp cutoff, the pre-period fractions differ substantially: $52.8\%$ on Arbitrum and $27.1\%$ on Base, whose median gaps are $5.08$ and $1.77$~bp. Gas-cost differences may help explain this contrast, but the table does not separate economic and measurement explanations. The previous-block DEX reference can differ from the immediately prevailing pool price, so the classification is a proxy for arbitrage rather than a guarantee that a swap moves the current pool price toward the CEX price. Arbitrum's median is unchanged across periods, while the fraction above $10$~bp falls on both chains: from $2.2\%$ to $1.3\%$ on Arbitrum and to $1.9\%$ on Base.

\begin{table}[htbp]
    \centering\small
    \caption{\textbf{Distribution of the signed CEX--DEX reference gap, by chain and period.} The gap is the log difference between the Binance mid-quote and the reference DEX price (the last swap price in the previous block), times trade direction, in bps. Positive values align trade direction with the CEX deviation from that previous-block reference, which need not equal the immediately prevailing pool price. $P(\text{gap}>\tau)$ is the share of swaps exceeding $\tau$~bp. Sample: WETH--USDC 5~bp pool on Base and Arbitrum, May~16--September~15, 2025 (July~7--16 excluded).}
    \label{tab:app-gap-distribution}
    \begin{tabular}{lcccc}
\hline\hline
 & \multicolumn{2}{c}{Base} & \multicolumn{2}{c}{Arbitrum} \\
\cmidrule(lr){2-3}\cmidrule(lr){4-5}
 & Pre & Post & Pre & Post \\
\hline
Median signed gap (bp) & 1.77 & 1.21 & 5.08 & 5.08 \\
Mean $|$gap$|$ (bp) & 4.26 & 3.87 & 5.23 & 5.12 \\
90th pctile (bp) & 6.96 & 6.59 & 6.92 & 6.73 \\
99th pctile (bp) & 12.52 & 11.63 & 12.97 & 10.58 \\
\hline
$P(\text{gap} > 5$~bp$)$ (\%) & 27.1 & 22.0 & 52.8 & 53.0 \\
$P(\text{gap} > 10$~bp$)$ (\%) & 2.2 & 1.9 & 2.2 & 1.3 \\
$P(\text{gap} > 20$~bp$)$ (\%) & 0.3 & 0.1 & 0.3 & 0.1 \\
\hline
$N$ swaps & 1,583,918 & 5,597,991 & 1,733,370 & 2,639,332 \\
\hline\hline
\end{tabular}

\end{table}

\paragraph{Threshold robustness.}
Tables~\ref{tab:app-arb-gas-tau} and~\ref{tab:app-arb-profit-tau} re-estimate the arbitrage-side difference-in-differences at $\tau\in\{6,8,10\}$, holding size-bin boundaries fixed. Gas-price estimates are negative and statistically significant in P70--P80, P80--P90, and P90--P100 at all three thresholds; estimates in P50--P70 depend on the threshold. At each threshold the blockchain-fee estimate is negative, the raw-markout estimate is statistically indistinguishable from zero, and the net-markout estimate is positive and significant. The estimated fee reduction and net-markout increase are larger at the higher thresholds.

\begin{table}[htbp]
    \centering\small
    \caption{\textbf{Arbitrageur gas price: threshold robustness.} Difference-in-differences estimates (Post $\times$ Base) of the swap-level log gas price on arbitrage swaps classified at alternative gap thresholds $\tau$, estimated within the five trade-size bins of Table~\ref{tab:did-consolidated-arb-behavior} (ETH boundaries held fixed at their baseline values), with chain and date fixed effects and date-clustered standard errors (in parentheses). The $\tau = 10$ column corresponds to Table~\ref{tab:did-consolidated-arb-behavior}. Sample: WETH--USDC 5~bp pool on Base and Arbitrum, May~16--September~15, 2025 (July~7--16 excluded).}
    \label{tab:app-arb-gas-tau}
    \begin{tabular}{lcD{.}{.}{2.2}@{\hskip 1.4em}lD{.}{.}{2.2}@{\hskip 1.4em}lD{.}{.}{2.2}@{\hskip 1.4em}l}
\toprule
Bin & Size (ETH) & \multicolumn{2}{c}{$\tau = 6$} & \multicolumn{2}{c}{$\tau = 8$} & \multicolumn{2}{c}{$\tau = 10$} \\
\midrule
P50--P60 & 2.9--4.0 & -0.34^{***} & (0.12) & 0.10 & (0.18) & 0.37 & (0.26) \\
P60--P70 & 4.0--5.7 & -0.48^{***} & (0.18) & -0.32 & (0.27) & 0.04 & (0.37) \\
P70--P80 & 5.7--9.8 & -0.84^{***} & (0.18) & -0.85^{***} & (0.26) & -0.75^{**} & (0.35) \\
P80--P90 & 9.8--17 & -0.96^{***} & (0.20) & -0.85^{***} & (0.24) & -0.86^{***} & (0.32) \\
P90--P100 & $\geq$17 & -1.41^{***} & (0.19) & -1.29^{***} & (0.19) & -1.31^{***} & (0.22) \\
\bottomrule
\end{tabular}

\end{table}

\begin{table}[htbp]
    \centering\small
    \caption{\textbf{Arbitrageur markout: threshold robustness.} Difference-in-differences estimates (Post $\times$ Base) of the chain-day volume-weighted raw markout, blockchain fee, and net markout (bps, 10-second horizon) on arbitrage swaps classified at alternative gap thresholds $\tau$, with chain and date fixed effects and date-clustered standard errors (in parentheses). The $\tau = 10$ row corresponds to Table~\ref{subtab:arb-markout}. Sample: WETH--USDC 5~bp pool on Base and Arbitrum, May~16--September~15, 2025 (July~7--16 excluded).}
    \label{tab:app-arb-profit-tau}
    \begin{tabular}{cD{.}{.}{2.2}@{\hskip 1.4em}lD{.}{.}{2.2}@{\hskip 1.4em}lD{.}{.}{2.2}@{\hskip 1.4em}lc}
\toprule
$\tau$ (bp) & \multicolumn{2}{c}{Raw markout} & \multicolumn{2}{c}{BC fee} & \multicolumn{2}{c}{Net markout} & $N$ swaps \\
\midrule
6 & -0.41 & (0.37) & -1.35^{***} & (0.12) & 0.93^{**} & (0.38) & 1,971,141 \\
8 & -0.14 & (0.89) & -2.21^{***} & (0.15) & 2.07^{**} & (0.93) & 537,943 \\
10 & -0.53 & (1.30) & -3.26^{***} & (0.21) & 2.73^{**} & (1.35) & 213,909 \\
\bottomrule
\end{tabular}

\end{table}

\paragraph{Gas-price size profile.}
Table~\ref{tab:app-arb-gas-profile} extends the gas-price comparison to the full trade-size distribution and splits the top decile. Pre-upgrade Base mean log gas prices rise from $-3.30$ in the bottom half to $4.29$ in the top five percent, while Arbitrum's profile is much flatter. Base's size gradient narrows after the upgrade but remains positive. The bottom-half difference-in-differences is positive ($1.19$ log points). In the upper tail, both P90--P95 and P95--P100 have negative estimates ($-1.03$ and $-1.47$). Splitting the top decile shows that the decline is present in both slices; it does not eliminate possible changes in composition within each slice.

\begin{table}[htbp]
    \centering\small
    \caption{\textbf{Arbitrageur gas price: full trade-size profile.} Difference-in-differences estimates (Post $\times$ Base) of the swap-level log gas price on arbitrage swaps (signed CEX--DEX gap $>10$~bp) within trade-size bins spanning the full pre-upgrade size distribution, with chain and date fixed effects and date-clustered standard errors (in parentheses). Base and Arbitrum columns report cell mean log gas prices (log gwei) by period. The P50--P90 bins match the first four columns of Table~\ref{tab:did-consolidated-arb-behavior}; the final two bins split its P90--P100 column. Sample: WETH--USDC 5~bp pool on Base and Arbitrum, May~16--September~15, 2025 (July~7--16 excluded).}
    \label{tab:app-arb-gas-profile}
    \begin{tabular}{lcD{.}{.}{2.2}D{.}{.}{2.2}D{.}{.}{2.2}D{.}{.}{2.2}D{.}{.}{2.2}@{\hskip 1.4em}lc}
\toprule
 & & \multicolumn{2}{c}{Base} & \multicolumn{2}{c}{Arbitrum} & \multicolumn{2}{c}{} & \\
\cmidrule(lr){3-4}\cmidrule(lr){5-6}
Bin & Size (ETH) & \multicolumn{1}{c}{Pre} & \multicolumn{1}{c}{Post} & \multicolumn{1}{c}{Pre} & \multicolumn{1}{c}{Post} & \multicolumn{2}{c}{DiD} & $N$ \\
\midrule
P0--P50 & $<$2.9 & -3.30 & -2.12 & -3.15 & -3.09 & 1.19^{***} & (0.21) & 148,436 \\
P50--P60 & 2.9--4.0 & -0.68 & -0.17 & -3.47 & -3.10 & 0.37 & (0.26) & 16,631 \\
P60--P70 & 4.0--5.7 & 0.15 & 0.22 & -3.02 & -2.62 & 0.04 & (0.37) & 15,073 \\
P70--P80 & 5.7--9.8 & 1.45 & 0.78 & -3.05 & -2.70 & -0.75^{**} & (0.35) & 13,685 \\
P80--P90 & 9.8--17 & 2.55 & 1.48 & -2.90 & -2.95 & -0.86^{***} & (0.32) & 10,834 \\
P90--P95 & 17--24 & 3.37 & 2.13 & -3.22 & -3.37 & -1.03^{***} & (0.24) & 4,670 \\
P95--P100 & $\geq$24 & 4.29 & 2.85 & -3.46 & -3.41 & -1.47^{***} & (0.23) & 4,580 \\
\bottomrule
\end{tabular}

\end{table}

\subsection{Event window, placebo date, and outage exclusion}\label{app:window-robustness}

This subsection reports three checks on the headline estimates. The first narrows the event window; the second assigns a placebo launch date inside the pre-period; the third excludes the days on which the active Base LP was inactive. Table~\ref{tab:window-robustness} reports all three, re-estimating the headline coefficient of every main exhibit with the specification held fixed (same panels, fixed effects, clustering, trade-size bin edges, and arbitrage threshold as the corresponding main table). Column~(2) narrows the window to four weeks on each side of the excluded transition (June~9--July~6 versus July~17--August~13). Column~(3) uses pre-upgrade data only and assigns a placebo launch date at the midpoint of the pre-period, June~11. Column~(4) drops the eleven days on which the active Base LP held less than one percent of Base pool liquidity, June~14--15, June~30--July~2, and August~17--22, from both chains.

\begin{table}[htbp]
    \centering\small
    \caption{\textbf{Event-window robustness and placebo tests.} Each row re-estimates the headline difference-in-differences coefficient of one main exhibit under alternative event windows. Column~(1): the baseline full-sample estimate (May~16--September~15, July~7--16 excluded). Column~(2): pre-period June~9--July~6, post-period July~17--August~13. Column~(3): pre-upgrade data only, with a placebo launch at the pre-period midpoint (June~11). Column~(4): full sample excluding the eleven days on which the active Base LP held less than one percent of Base pool liquidity (June~14--15, June~30--July~2, August~17--22), dropped on both chains. Specifications, fixed effects, and clustering follow the corresponding main tables; the trade-size bin edges and the $\tau = 10$~bp arbitrage threshold are held at their baseline values. Standard errors in parentheses, clustered by date. Sample: WETH--USDC 5~bp pool on Base and Arbitrum, May~16--September~15, 2025 (July~7--16 excluded).}
    \label{tab:window-robustness}
    \begin{tabular}{lD{.}{.}{3.2}D{.}{.}{3.2}D{.}{.}{3.2}D{.}{.}{3.2}}
\toprule
 & \multicolumn{1}{c}{(1)} & \multicolumn{1}{c}{(2)} & \multicolumn{1}{c}{(3)} & \multicolumn{1}{c}{(4)} \\
 & \multicolumn{1}{c}{Full sample} & \multicolumn{1}{c}{$\pm$4 weeks} & \multicolumn{1}{c}{Placebo} & \multicolumn{1}{c}{Drop outage days} \\
\midrule
Arb gas (log), top size decile & -1.31^{***} & -0.94^{***} & -0.28 & -1.17^{***} \\
 & (0.22) & (0.26) & (0.32) & (0.21) \\[3pt]
Arb raw markout (bps) & -0.53 & -0.36 & 0.90 & -0.05 \\
 & (1.30) & (1.89) & (1.87) & (1.39) \\[3pt]
Arb blockchain fee (bps) & -3.26^{***} & -3.36^{***} & 0.37 & -3.26^{***} \\
 & (0.21) & (0.32) & (0.41) & (0.22) \\[3pt]
Arb net markout (bps) & 2.73^{**} & 3.00 & 0.53 & 3.21^{**} \\
 & (1.35) & (1.98) & (1.98) & (1.43) \\[3pt]
LP gas price (gwei) & 0.03 & 0.04 & -0.02 & 0.03 \\
 & (0.02) & (0.03) & (0.02) & (0.03) \\[3pt]
LP log rebalancing time & -1.07^{***} & -0.64 & -0.33 & -1.11^{***} \\
 & (0.33) & (0.41) & (0.29) & (0.31) \\[3pt]
LP net payoff (bps) & 0.73^{**} & 0.39 & 0.62 & 0.73^{**} \\
 & (0.36) & (0.49) & (0.49) & (0.36) \\[3pt]
Cream-skim: toxic-exp.\ share (pp) & -8.93^{***} & -6.81^{***} & -4.81^{*} & -9.75^{***} \\
 & (1.74) & (2.57) & (2.78) & (1.68) \\[3pt]
TWA price deviation (bps) & -0.11^{***} & -0.13^{***} & 0.03 & -0.08^{***} \\
 & (0.03) & (0.04) & (0.05) & (0.03) \\[3pt]
Log pool depth & -0.21^{***} & -0.11^{**} & 0.01 & -0.19^{***} \\
 & (0.04) & (0.06) & (0.06) & (0.04) \\[3pt]
Eff.\ spread $\sim$1 ETH (bps) & 0.07^{***} & 0.08^{***} & -0.00 & 0.07^{***} \\
 & (0.01) & (0.01) & (0.01) & (0.01) \\[3pt]
Top-1 TW liq.\ share (\%) & 4.20^{**} & 9.82^{***} & -2.97 & 4.71^{***} \\
 & (2.11) & (2.70) & (2.89) & (1.56) \\[3pt]
Log volume (ETH) & -0.20^{***} & -0.15^{***} & 0.03 & -0.19^{***} \\
 & (0.04) & (0.06) & (0.05) & (0.04) \\[3pt]
\bottomrule
\end{tabular}

\end{table}

In the narrower window, the arbitrage, market-quality, and concentration coefficients retain their signs, although their magnitudes change. For example, the log-depth estimate changes from $-0.21$ to $-0.11$ and the concentration estimate from $4.20$ to $9.82$ percentage points, while the net markout keeps its sign and magnitude but is no longer statistically significant because its standard error grows. The LP log-rebalancing-time and net-payoff estimates retain their signs but are not statistically significant; reduced precision and differences in the estimates across windows both contribute. The cream-skimming contrast remains negative and significant at the $1\%$ level ($-6.81$ percentage points).

Under the placebo launch date, the coefficients for the arbitrage, market-quality, concentration, and LP outcomes are not statistically distinguishable from zero; one placebo date alone does not establish event specificity. The cream-skimming contrast deserves particular caution: its placebo estimate is $-4.81$ percentage points, significant at the $10\%$ level. The final weeks of the pre-period contain sharply negative days, visible in Figure~\ref{fig:app-pt-lp}. The narrow window still contains those days, so the estimate survives their inclusion.

The third check, in column~(4), excludes the eleven days on which the active Base LP was inactive. Every estimate keeps its sign and significance. The concentration estimate rises from $4.20$ to $4.71$ percentage points, because on the excluded days the measured top provider is a smaller runner-up, and the cream-skimming contrast widens from $-8.93$ to $-9.75$ percentage points, since those days contribute no exposure for her to avoid.

\subsection{Effective spread by trade size}\label{app:eff-spread}

Table~\ref{tab:app-eff-profile} reports the effective-spread difference-in-differences by decile of the pooled trade-size distribution, with the specification of Section~\ref{sec:emp-mq}. The widening is positive and significant at the $1\%$ level in each of the first nine deciles (the first-decile estimate rounds to zero at the table's precision) and grows with trade size, from $0.002$~bps in the second decile to $0.155$~bps in the ninth. In the top decile, trades above about $3$~ETH, the estimate is small, negative, and not statistically significant.

\begin{table}[htbp]
    \centering\small
    \caption{\textbf{Effective spread by trade size: difference-in-differences estimates.} For each decile of the pooled trade-size distribution, the table reports the chain-day panel difference-in-differences of the daily volume-weighted effective spread (bps, pre-swap anchor) with chain and date fixed effects and date-clustered standard errors, together with period means of the daily values by chain. Sample: WETH--USDC 5~bp pool on Base and Arbitrum, May~16--September~15, 2025 (July~7--16 excluded).}
    \label{tab:app-eff-profile}
    \begin{tabular}{lcD{.}{.}{2.3}D{.}{.}{2.3}D{.}{.}{2.3}D{.}{.}{2.3}D{.}{.}{2.3}@{\hskip 1.5em}l}
\toprule
 & & \multicolumn{2}{c}{Base} & \multicolumn{2}{c}{Arbitrum} & \multicolumn{2}{c}{} \\
\cmidrule(lr){3-4}\cmidrule(lr){5-6}
Decile & Size range (ETH) & \multicolumn{1}{c}{Pre} & \multicolumn{1}{c}{Post} & \multicolumn{1}{c}{Pre} & \multicolumn{1}{c}{Post} & \multicolumn{2}{c}{DiD} \\
\midrule
D1 & 0.001--0.005 & 0.001 & 0.001 & 0.001 & 0.001 & 0.000^{***} & (0.000) \\
D2 & 0.005--0.055 & 0.005 & 0.007 & 0.002 & 0.003 & 0.002^{***} & (0.000) \\
D3 & 0.055--0.24 & 0.022 & 0.032 & 0.011 & 0.012 & 0.009^{***} & (0.001) \\
D4 & 0.24--0.45 & 0.057 & 0.091 & 0.026 & 0.027 & 0.033^{***} & (0.003) \\
D5 & 0.45--0.79 & 0.105 & 0.154 & 0.044 & 0.047 & 0.046^{***} & (0.005) \\
D6 & 0.79--1.23 & 0.169 & 0.249 & 0.071 & 0.078 & 0.074^{***} & (0.008) \\
D7 & 1.23--1.60 & 0.234 & 0.345 & 0.101 & 0.107 & 0.105^{***} & (0.011) \\
D8 & 1.60--2.21 & 0.325 & 0.452 & 0.136 & 0.142 & 0.121^{***} & (0.015) \\
D9 & 2.21--3.07 & 0.446 & 0.622 & 0.174 & 0.195 & 0.155^{***} & (0.020) \\
D10 & 3.07--1,319 & 2.301 & 2.261 & 0.986 & 0.974 & -0.028 & (0.254) \\
\bottomrule
\end{tabular}

\end{table}
\clearpage
\begin{center}{\Huge Internet Appendix}\end{center}
\setcounter{section}{0}
\renewcommand{\thesection}{IA.\arabic{section}}
\renewcommand{\theHsection}{IA.\arabic{section}}
\setcounter{table}{0}\renewcommand{\thetable}{IA.\arabic{table}}\renewcommand{\theHtable}{IA.\arabic{table}}
\setcounter{figure}{0}\renewcommand{\thefigure}{IA.\arabic{figure}}\renewcommand{\theHfigure}{IA.\arabic{figure}}
\setcounter{equation}{0}\renewcommand{\theequation}{IA.\arabic{equation}}\renewcommand{\theHequation}{IA.\arabic{equation}}
\section{Foundations: Game, Controls and Passive Entry}\label{ia:foundations}

This section collects the formal objects that the later proofs use and proves the three foundations on which those proofs rest: the equilibrium of the percentage auction behind Proposition~\ref{prop:rent-share}, the active LP's optimal strategy together with its verification against all strategies, and the passive-entry accounting of Proposition~\ref{prop:passive}.
Every result in this section holds at all positive values of the seven primitives $J$, $\lambda$, $\delta$, $k$, $\ell$, $\rbar$ and $q$, so no result in this section uses Assumption~\ref{ass:main}.

\subsection{Objects used in the proofs}\label{ia:objects}

The proofs apply the transaction rules of Section~\ref{sec:timing} and the equilibrium concept of Definition~\ref{def:equilibrium} without change, and this subsection adds only the formal objects they need.
Recall that $J$ is the public information arrival rate, $\lambda$ the message speed of each arbitrageur and of the active LP, and $\delta$ the deadline rate of liquidity traders.
Recall also that $k$ is the passive holding cost per unit of capital and time, $\ell$ the loss per unit of deposited capital at each executed arbitrage trade, $q$ the active LP's capital, $P$ passive capital and $r=\rbar B/(B+\delta)$ the fee flow served, where the next sentences recall the block rate and the offered fee flow.
Only the block rate $B>0$ varies, and every other primitive is a fixed positive number.
Each private message, whether an arbitrage transaction or the active LP's withdrawal or redeposit, arrives after an independent exponential latency with the common rate $\lambda$, so the first of the two arbitrage transactions for an opportunity arrives at the combined rate $S=2\lambda$.
Liquidity traders arrive at rate $\nu$, and each pays the per-trade fee $\phi$ if its transaction executes, so the offered fee flow is $\rbar=\nu\phi$.
Only this product enters the game among liquidity providers.

\paragraph{Strategies.}
The active LP's four canonical strategies are always deposited, early withdrawal, late withdrawal and permanent exit.
Under always deposited, she never withdraws.
Under early withdrawal, the active LP sends a withdrawal when a small gap opens, and that withdrawal stays valid if the gap becomes profitable.
Under late withdrawal, she sends a fresh withdrawal when the gap becomes profitable.
Under permanent exit, she never redeposits after a withdrawal.
After a withdrawal, she sends a redeposit as soon as the gap returns to zero, provided that her optimal long-run average trading income is positive.
At the cutoff between late withdrawal and always deposited, a deterministic waiting timer implements every intermediate occupation: the deposited-time fraction and the corresponding rates of exposed trades and successful withdrawals. Proposition~\ref{prop:ia-entry} gives the construction.
A pure strategy is deterministic but may depend on her whole history, so it is not limited to the four named strategies.
The alternatives against which her strategy is judged also include randomized strategies.

\paragraph{Occupations and flows.}
For a strategy $j$, write $\pi_j$ for the fraction of calendar time her capital is deposited, and $e_j$ and $\omega_j$ for the rates of her exposed trades and of her successful withdrawals, where an exposed trade is an executed arbitrage trade that occurs while her capital is deposited and a successful withdrawal is one that executes.
Thus $\pi_N$, $\pi_E$, $\pi_L$ and $\pi_O$ are her deposited fractions under always deposited, early withdrawal, late withdrawal and permanent exit, and the same subscripts mark every other strategy-specific quantity, so a strategy letter never stands alone as a symbol.
Write $e$ for the rate of all executed arbitrage trades, together with the zero-size price resets that close a stale gap when the pool is empty, $T$ for the mean time between two such events at instant blocks, and $e_\infty$ for the limit of $e$ as blocks speed up.
Write $C$ for the mean time that news takes to open a profitable gap from a zero gap, $T_H$ for the mean life of a profitable gap until the block that closes it, and $d_c$ for a passive unit's required fee yield, its holding cost plus its arbitrage losses per unit of time:
\begin{equation}\label{eq:ia-D3}
    \begin{gathered}
        C=\frac{4}{J},\qquad
        T=C+\frac{1}{S},\qquad
        T_H=\frac{1}{S}+\frac{1}{B},\qquad
        e=\frac{B}{1+TB},\\
        e_\infty=T^{-1},\qquad
        \kappa=\frac{k}{\ell},\qquad
        d_c=k+\ell e,\qquad
        p_0=\frac{r}{d_c}.
    \end{gathered}
\end{equation}
Here $p_0$ is the competitive benchmark, the passive capital that breaks even when it is continuously available and alone in the pool.
An actual occupation $(\pi,e_A,\omega)$ records the same three long-run quantities for whatever strategy the active LP follows, named or not.
For such an occupation, write $D$ for time-average depth, $U_A$ for her trading income as in~\eqref{eq:lp-income}, $\sigma$ for her share of the pool's capital while she is deposited, $\Phi$ for her time-average liquidity share, $G_A$ for the arbitrage prize extracted per unit of time, and $\mathcal{R}_A$ and $\mathcal{S}_A$ for the dollars per unit of time that the arbitrageurs retain and the sequencer receives:
\begin{equation}\label{eq:ia-D4}
    D=P+q\pi,\qquad
    \sigma=\frac{q}{P+q},\qquad
    \Phi=\pi\sigma,\qquad
    U_A=r\Phi-\ell qe_A,
\end{equation}
\begin{equation}\label{eq:ia-D5}
    G_A=\ell\,(eP+qe_A),\qquad
    \mathcal{R}_A=(1-s)\,G_A,\qquad
    \mathcal{S}_A=s\,G_A.
\end{equation}
Here $s$ is the sequencer's share of the arbitrage prize.
Dividing any of these dollar flows by the rate $e$ gives the amount per executed arbitrage trade or price reset.

\paragraph{Normalized formulas.}
The algebra measures every rate in units of the public information arrival rate $J$ and every capital stock in units of $q$ or of a capital scale, using the clock ratio $a$ of message speed to public information arrivals, the normalized block rate $\widehat{B}$, the normalized holding-cost ratio $c$, the normalized deadline rate $\delta_n$, the normalized offered fee flow $f_{\mathrm{cap}}$, the capital scale $L_0$, and the normalized rate $E$ at which arbitrage trades execute at instant blocks, with its reciprocal $T_n$,
\begin{equation}\label{eq:ia-N0}
    \begin{gathered}
        a=\frac{\lambda}{J},\qquad
        \widehat{B}=\frac{B}{J},\qquad
        c=\frac{\kappa}{J},\qquad
        \delta_n=\frac{\delta}{J},\\
        f_{\mathrm{cap}}=\frac{\rbar}{\ell qJ},\qquad
        L_0=\frac{\rbar}{\ell J},\qquad
        E=\frac{e_\infty}{J}=\frac{2a}{8a+1},\qquad
        T_n=\frac{1}{E}.
    \end{gathered}
\end{equation}
In these units the fee flow is $R_{\widehat{B}}=f_{\mathrm{cap}}\,\widehat{B}/(\widehat{B}+\delta_n)$, a passive unit's required fee yield is $d_{\widehat{B}}=c+2a\widehat{B}/G$ and the competitive benchmark in units of $q$ is $n_{\widehat{B}}=R_{\widehat{B}}/d_{\widehat{B}}=p_0/q$.
Here $G$ is the first of the following polynomials, all of which have positive coefficients:
\begin{equation}\label{eq:ia-N1}
    \begin{aligned}
        G   & =(8a+1)\widehat{B}+2a,\qquad
        H_L=(6a+1)\widehat{B}+12a^2+a+1,                                                  \\
        V   & =(6a^2+12a+1)\widehat{B}^2+(12a^2+15a+1)\widehat{B}+3a(a+1),                \\
        K   & =(6a^2+12a+1)\widehat{B}^2+(12a^3+36a^2+16a+2)\widehat{B}+24a^3+27a^2+4a+1, \\
        D_N & =(10a+1)\widehat{B}^2+(12a^2+11a+1)\widehat{B}+2a(a+1),                     \\
        D_E & =(12a^2+24a+2)\widehat{B}^3+(24a^3+80a^2+44a+3)\widehat{B}^2                \\
            & \qquad+(60a^3+82a^2+21a+1)\widehat{B}+12a^3+14a^2+2a,                       \\
        A_E & =2(a+1)\widehat{B}^2+(6a^2+10a+3)\widehat{B}+6a^2+7a+1.
    \end{aligned}
\end{equation}
In this appendix the symbols $G$, $H_L$, $V$, $K$, $D_N$, $D_E$ and $A_E$, written exactly as above, name these polynomials.
In this appendix the prize extracted per unit of time $G_A$, the prize values $V_{\mathrm{current}}$, $V_{\mathrm{IN}}$ and $V_{\mathrm{OUT}}$, the bias $V_h$, the renewal lengths $K_L$ and $K_E$, the factor $K_\pi$, the depth envelope $\overline{D}_q$, the advantage $A(g)$ and the biases $h$ with state subscripts are different objects from these polynomials, each named where it appears.
In particular $D_N$ and $D_E$ are not depths, and $H_L$ is not a gap level.
A bare $E$ is always the normalized rate at which arbitrage trades execute at instant blocks.
Every strategy letter, including that of early withdrawal, appears only as a subscript.
The occupations of late and early withdrawal are
\begin{equation}\label{eq:ia-DT1}
    \Bigl(\pi_L,\;\frac{e_L}{J},\;\frac{\omega_L}{J}\Bigr)
    =\frac{1}{GH_L}\Bigl(4a(\widehat{B}+2a)\bigl((12a+1)\widehat{B}+3a\bigr),\;8a^2\widehat{B}(\widehat{B}+3a),\;4a^2\widehat{B}^2\Bigr),
\end{equation}
\begin{equation}\label{eq:ia-N2}
    \Bigl(\pi_E,\;\frac{e_E}{J},\;\frac{\omega_E}{J}\Bigr)
    =\frac{1}{GK}\Bigl(4a(\widehat{B}+2a)V,\;8\widehat{B}a^2\bigl[\widehat{B}^2+(3a+1)\widehat{B}+3a(a+1)\bigr],\;4\widehat{B}^2a^2\bigl[(6a+1)\widehat{B}+12a^2+3a+1\bigr]\Bigr),
\end{equation}
while always deposited has $(\pi_N,e_N,\omega_N)=(1,e,0)$ and permanent exit has $(\pi_O,e_O,\omega_O)=(0,0,0)$.
Her fee-to-loss ratio $b_f=r/[\ell(P+q)]$ is the fee income of one deposited unit per unit of time while she is deposited, measured in units of the loss $\ell$ at one executed arbitrage trade.
Proposition~\ref{prop:ia-controls} shows that her optimal strategy changes when $b_f$ crosses the cutoffs $\theta_0$, $\theta_{EL}$ and $\theta_{LN}$, whose normalized values are
\begin{equation}\label{eq:ia-N3}
    z_0=\frac{\theta_0}{J}=\frac{2a\widehat{B}\bigl[\widehat{B}^2+(3a+1)\widehat{B}+3a(a+1)\bigr]}{(\widehat{B}+2a)V},\quad
    z_{EL}=\frac{2a\widehat{B}A_E}{D_E},\quad
    z_{LN}=\frac{2a\widehat{B}\bigl[(2a+1)\widehat{B}+a+1\bigr]}{D_N},
\end{equation}
so that $\theta_{EL}=Jz_{EL}$ and $\theta_{LN}=Jz_{LN}$.
These two normalized cutoffs differ from the normalized rate $e/J$ at which arbitrage trades execute by
\begin{equation}\label{eq:ia-N4}
    z_{LN}-\frac{e}{J}=\frac{32a^3\widehat{B}^3}{GD_N}>0,\qquad
    z_{EL}-\frac{e}{J}=\frac{4a^2\widehat{B}^3\bigl[(2a-3)\widehat{B}+12a^2+5a-3\bigr]}{GD_E}.
\end{equation}
Primes on normalized formulas denote derivatives with respect to $\widehat{B}$, and the derivative with respect to $B$ multiplies the derivative with respect to $\widehat{B}$ by the positive factor $1/J$.

\subsection{The percentage auction and the sequencer's share}\label{ia:auction}

Proposition~\ref{prop:rent-share} rests on the following description of the percentage auction, in which $b_i\in[0,1]$ is the percentage of the arbitrage prize that arbitrageur $i$ pays the sequencer if its transaction executes.

\begin{proposition}[Percentage auction]\label{prop:ia-auction}
    \begin{enumerate}[label=(\roman*),ref=\theproposition(\roman*),leftmargin=*,nosep]
        \item In a symmetric equilibrium of the auction, each arbitrageur draws its percentage from the distribution function $F_B$, and the sequencer's share of the arbitrage prize is $s(B)$, where
              \begin{equation}\label{eq:ia-P1}
                  F_B(b_i)=\frac{B}{2\lambda}\,\frac{b_i}{1-b_i},\quad
                  0\leq b_i\leq\frac{2\lambda}{B+2\lambda};\qquad
                  s(B)=\frac{\lambda}{B+\lambda},\quad
                  s'(B)<0.
              \end{equation}
        \item Against the randomized percentage of part~(i), each arbitrageur sends its transaction as soon as a profitable gap opens, which we call prompt dispatch, and offers it for execution at the first block after it arrives.
              No admissible unilateral deviation in timing does better, given that each percentage stays fixed for the whole opportunity.
              The prize extracted per unit of time divides between the arbitrageurs and the sequencer as in~\eqref{eq:ia-D5}, even when the active LP withdraws her capital.
        \item The odds of the sequencer's share are $s/(1-s)=\lambda/B$.
              Hence an $m$-fold increase in the block rate moves the sequencer's share from its value $s_-$ at the slower rate to $s_+=s_-/[m-(m-1)s_-]$ at the faster rate.
    \end{enumerate}
\end{proposition}

The proposition concerns shares, so it does not imply that the sequencer's dollar receipts fall or that the arbitrageurs' retained dollars rise, because both flows scale with the prize extracted per unit of time, $G_A$, which changes with the rate at which arbitrage trades execute and with capital.
Nor does it assign any part of the prize to the active LP, who can only avoid losses on her own capital.

\begin{proof}[Proof of Proposition~\ref{prop:rent-share} and Proposition~\ref{prop:ia-auction}]
    At the onset of a profitable gap and after every block that leaves the gap open, both arbitrageurs' remaining latencies are again exponential with rate $\lambda$, by the memorylessness of the exponential latency.
    A block that leaves the gap open reveals only that the rival's promptly sent transaction had not yet arrived, not the rival's percentage.
    Until the next block the current prize $V_{\mathrm{current}}=\ell Q$ is constant.
    Here $\ell$ is the loss per unit of deposited capital, $Q=P+qI$ is the capital in the pool, and $I$ equals one when her capital is deposited and zero otherwise.
    Write $u_1$ for the probability that, by the next block, an arbitrageur's own transaction has arrived and its rival's has not, and $u_2$ for the probability that both have arrived:
    \begin{equation}\label{eq:ia-u}
        u_1=\frac{B\lambda}{(B+\lambda)(B+2\lambda)},\qquad
        u_2=\frac{2\lambda^2}{(B+\lambda)(B+2\lambda)}.
    \end{equation}
    Against the rival's mixed bid, every percentage in the support of~\eqref{eq:ia-P1} earns the same expected amount,
    \begin{equation}\label{eq:ia-indifference}
        \bigl(u_1+u_2F_B(b_i)\bigr)(1-b_i)\,V_{\mathrm{current}}=u_1V_{\mathrm{current}},
    \end{equation}
    and a higher percentage outside the support pays more for the same advantage in winning the arbitrage trade.
    The probability that at least one arbitrage transaction has arrived by the next block is $2\lambda/(B+2\lambda)$.
    Conditional on the capital at the start of the interval, the arbitrageurs' combined expected retained dollars and the expected gross extraction stand in the ratio $B/(B+\lambda)$.
    Summing this conditional equality over all intervals proves~\eqref{eq:ia-P1} and the factorization~\eqref{eq:ia-D5}, even though the pool's capital changes when her capital moves.
    The share odds and the $m$-fold formula follow directly from $s(B)=\lambda/(B+\lambda)$.
    Geometric renewal at each return of the gap to zero gives $C=4/J$, $T_H=1/S+1/B$ and $e=1/(C+T_H)$.

    It remains to verify bidding and prompt dispatch for the canonical LP
    strategies and the deterministic timers used below.
    Write $\gamma\geq0$ for one arbitrageur's average income and
    $Z_{\mathrm{IN}}$ for its continuation bias at a zero gap with the active
    capital deposited.
    For an aligned outside state with timer or return phase $\zeta_O$, write
    $Z_{\mathrm{OUT}}(\zeta_O)$ for the corresponding bias.
    These are relative values in the expected-average-income problem.
    For each canonical strategy, coupling initially deposited and outside systems
    until their capital statuses agree gives
    $Z_{\mathrm{IN}}\geq Z_{\mathrm{OUT}}$.
    For a finite timer, the same comparison follows by decomposing the
    arbitrageur's reward into its passive-capital and active-capital components.
    The passive component follows the same environmental process in either
    capital state.
    If $\gamma_q\geq0$ is the average active-capital component and $T_R$ is the
    first redeposit time from an aligned outside state, its expected waiting
    time is finite, it ends at alignment, and no active-capital reward is earned
    before it. Hence
    \[
        Z_{\mathrm{IN}}-Z_{\mathrm{OUT}}(\zeta_O)
        =\gamma_q\,\mathbb E_{\zeta_O}T_R\geq0.
    \]
    The permanent-exit endpoint uses its canonical comparison.
    During an unresolved profitable gap capital cannot increase, and an
    outside return timer does not advance before alignment.
    Thus a withdrawal cannot improve the aligned successor bias.

    For a fixed supported percentage $b_i$, let
    $M_{b_i,\iota}=(1-b_i)V_\iota$, where
    $\iota\in\{\mathrm{IN},\mathrm{OUT}\}$ denotes the current capital status
    and $V_\iota$ is its current prize.
    Let $h_{b_i}^{R}$ and $h_{b_i}^{U}$ be the continuation biases when the
    transaction is ready and unready, respectively; their state argument
    includes the relevant LP timer or instruction phase.
    Before the next correction at most one current margin can be earned.
    The nonincreasing prize and the aligned-bias comparison therefore imply
    \[
        h_{b_i}^{R},\ h_{b_i}^{U}\ \leq\ Z_\iota+M_{b_i,\iota}.
    \]
    Here an outside $Z_\iota$ includes its fixed outside phase.

    Let $u$ be the elapsed time since the profitable gap opened or the last
    noncorrecting block, and put $p_u=\exp(-\lambda u)$.
    Even along a unilateral timing deviation, a noncorrecting block proves
    that the prompt rival was unready, so $p_u$ is its conditional probability
    of still being unready; its bid retains distribution $F_B$.
    At a ready block, participation rather than abstention has advantage
    \[
        (1-p_u)F_B(b_i)M_{b_i,\iota}
        +p_u\bigl[M_{b_i,\iota}+Z_\iota
        -h_{b_i}^{R}(0,\zeta^-)\bigr]\geq0,
    \]
    where $\zeta^-$ is the continuation state if no arbitrage executes and any
    valid LP withdrawal is processed.
    The same bound verifies receipt timing. With
    $T_B\sim\operatorname{Exp}(B)$ the next block wait,
    \[
        \begin{aligned}
            h_{b_i}^{R}(u,\zeta)-h_{b_i}^{U}(u,\zeta)
            =\mathbb E\Bigl[ e^{-\lambda T_B}\Bigl\{
             & [p_{u+T_B}+(1-p_{u+T_B})F_B(b_i)]M_{b_i,\iota} \\
             & +p_{u+T_B}[Z_\iota-h_{b_i}^{U}(0,\zeta^-)]
            \Bigr\}\Bigr]\geq0.
        \end{aligned}
    \]
    The two continuations agree if the initially unready transaction arrives
    by the block; the displayed expectation covers its nonarrival.
    Suppressing receipt or refusing execution therefore cannot improve income.

    At an interval opening, equation~\eqref{eq:ia-indifference} and the fact that the
    noncorrection probability and successor state do not depend on the
    supported bid show that all supported bids have the same continuation
    value.
    For a bidder that has delayed its initial send until age $u$, put
    $\bar b=2\lambda/(B+2\lambda)$.
    First-block conditioning then gives
    \begin{equation}\label{eq:ia-delay}
        h_{\bar b}^{U}(u,\zeta)-h_{b_i}^{U}(u,\zeta)
        =\frac{B}{2(B+\lambda)}(1-e^{-\lambda u})
        (\bar b-b_i)V_\iota\geq0.
    \end{equation}
    Bids above $\bar b$ pay more without improving eligibility or winning
    against the rival's bids.
    Use $h_{\bar b}^{U}$ as the value of an unsent transaction.
    Waiting instead of sending has generator advantage
    $-\lambda(h_{\bar b}^{R}-h_{\bar b}^{U})\leq0$.
    Immediate sending is therefore optimal, and at the opening any supported
    bid, including a draw from $F_B$, attains this value.
    The prompt rival bounds the remaining correction wait by one exponential
    delivery and one block wait, and the canonical and finite-timer return
    times have finite means.
    Their bounded biases make the integrated generator, submission and block
    inequalities valid for limit-superior expected average income, including
    adapted timing and randomized deviations under the locked-percentage rule.
    Proposition~\ref{prop:ia-entry} supplies an equilibrium with one of these
    canonical LP continuations at every block rate; the same argument verifies
    the timer implementations at ties.
\end{proof}

\subsection{The active LP's optimal controls}\label{ia:controls}

The active LP solves a different problem from the arbitrageurs', because a withdrawal avoids the arbitrage loss on her own capital while giving up fees for the rest of the profitable gap and during the redeposit delay, and a withdrawal sent at a small gap also risks leaving before news eliminates it.
Recall that $b_f=r/[\ell(P+q)]$ is her fee-to-loss ratio and that $\theta_0$, $\theta_{EL}$ and $\theta_{LN}$ are the cutoffs whose normalized values appear in~\eqref{eq:ia-N3}.

\begin{proposition}[Optimal controls of the active LP]\label{prop:ia-controls}
    Hold passive capital $P\geq0$ fixed, using at $P=0$ the prompt zero-size price reset of Section~\ref{sec:timing}.
    \begin{enumerate}[label=(\roman*),ref=\theproposition(\roman*),leftmargin=*,nosep]
        \item The cutoffs, the rate at which arbitrage trades execute and her exposed trades per unit of deposited time under late withdrawal satisfy the following strict inequalities:
              \begin{equation}\label{eq:ia-P2}
                  0<\theta_0<\frac{e_L}{\pi_L}<\theta_{EL}<\theta_{LN},\qquad
                  \theta_0<e<\theta_{LN}.
              \end{equation}
        \item Her optimal strategy is permanent exit when $b_f<\theta_0$, early withdrawal when $\theta_0<b_f<\theta_{EL}$, late withdrawal when $\theta_{EL}<b_f<\theta_{LN}$ and always deposited when $b_f>\theta_{LN}$.
              Only adjacent strategies tie at a cutoff.
              Her optimal trading income is
              \begin{equation}\label{eq:ia-P2-income}
                  U_A=\ell q\max\bigl\{0,\;b_f\pi_E-e_E,\;b_f\pi_L-e_L,\;b_f-e\bigr\}.
              \end{equation}
        \item These strategies are optimal against every admissible deviation, including history-dependent and randomized choices of waiting, submission, redeposit and participation.
              The four named strategies therefore do not restrict her choice set.
    \end{enumerate}
\end{proposition}

Unlike this proposition, the break-even and depth formulas of Section~\ref{ia:entry} require positive passive capital.

\begin{proof}[Proof of Proposition~\ref{prop:ia-controls}]
    \emph{Occupations.}
    Write $t_0$ for her mean wait outside the pool from a zero gap until a redeposit executes, counting redeposits that new small gaps cancel, and $m_0$ for the mean time from the opening of a small gap until the gap next returns to zero.
    Write $p_x$ for the probability that a withdrawal sent when a profitable gap opens executes before the arbitrage trade, $m_x$ for the probability that a withdrawal sent when a small gap opens executes while the small gap is still open, and $n_x$ for the probability that it executes after the gap has become profitable but before the arbitrage trade:
    \begin{equation}\label{eq:ia-P3}
        \begin{aligned}
            t_0 & =\frac{J(1+TB)(B+J+\lambda)}{2\lambda B^2},\qquad
            m_0=\frac{1}{J}+\frac{T_H}{2},\qquad
            p_x=\frac{B}{3(B+2\lambda)},                            \\
            m_x & =\frac{\lambda B}{(\lambda+J)(B+J)},\qquad
            n_x=\frac{J\lambda B(B+J+3\lambda)}{2(\lambda+J)(B+2\lambda)(3\lambda)(B+J)}.
        \end{aligned}
    \end{equation}
    A renewal, a cycle that starts and ends with her deposited at a zero gap, has mean length $K_L$ under late withdrawal and $K_E$ under early withdrawal, with expected rewards listed as deposited time, exposed trades and successful withdrawals:
    \begin{equation}\label{eq:ia-renewals}
        \begin{aligned}
            K_L & =C+T_H+p_xt_0,                   &  & \bigl(C+(1-p_x)T_H,\;1-p_x,\;p_x\bigr),                                    \\
            K_E & =\frac{C+T_H}{2}+(m_x+n_x)\,t_0, &  & \Bigl(\frac{C+T_H}{2}-m_xm_0-n_xT_H,\;\frac{1-m_x}{2}-n_x,\;m_x+n_x\Bigr).
        \end{aligned}
    \end{equation}
    Dividing each expected reward by the mean length of its renewal gives the occupations~\eqref{eq:ia-DT1} and~\eqref{eq:ia-N2} by exact cancellation.
    These occupations already count failed redeposits and races lost within a block.

    \emph{Gains and advantages.}
    Write $g_j=b_f\pi_j-e_j$ for her normalized gain $U_A/(\ell q)$ under strategy $j$, and drop the subscript for a generic value of this gain.
    The auxiliary ratio $h_x=Sn_x/m_x$ multiplies the advantage of a withdrawal executed during a profitable gap within the advantage of sending one at a small gap, both defined next.
    A withdrawal that executes during a profitable gap has the advantage $A(g)$ over staying deposited: the avoided arbitrage loss net of the fees forgone for the rest of the gap and the gain forgone during the redeposit wait.
    Sending a withdrawal at a small gap has the advantage $R_M(g)$ over waiting:
    \begin{equation}\label{eq:ia-advantages}
        A(g)=1-b_fT_H-gt_0,\qquad
        R_M(g)=h_xA(g)-b_f-\frac{J}{2}\,g\,t_0.
    \end{equation}
    Exact renewal identities give
    \begin{equation}\label{eq:ia-P4}
        g_L-g_N=\frac{ep_x}{1+ep_xt_0}\,A(g_N),\qquad
        g_E-g_L=\frac{m_x}{JK_E}\,R_M(g_L).
    \end{equation}
    As functions of $b_f$ the gains are affine with slopes $\pi_j$, where $\pi_E<\pi_L<1$ and $e_E<e_L<e$.
    At $b_f=e_L/\pi_L$ the gain of early withdrawal is strictly positive because $h_xC-(1-p_x)>0$.
    At $b_f=\theta_{LN}$ we have $A=0$, $g_L=g_N>0$ and $R_M<0$.
    These facts prove the strict cutoff order in~\eqref{eq:ia-P2} and the affine envelope~\eqref{eq:ia-P2-income}.

    \emph{Explicit certificate.}
    Put $K_\pi=(12a+1)\widehat{B}+3a$, and write $r_L=(e_L/J)/\pi_L$ for the normalized value of $e_L/\pi_L$.
    Direct subtraction gives
    \begin{equation}\label{eq:ia-certificate}
        \begin{aligned}
            r_L-z_0            & =\frac{12a^3\widehat{B}^3(\widehat{B}+3a+2)}{(\widehat{B}+2a)K_\pi V},                &
            z_{EL}-r_L         & =\frac{4a^2\widehat{B}^3(\widehat{B}+3a+2)H_L}{(\widehat{B}+2a)K_\pi D_E},              \\
            z_{LN}-z_{EL}      & =\frac{4a^2\widehat{B}^3\bigl[(2a+3)\widehat{B}+3a+3\bigr]H_L}{D_ND_E},               &
            \frac{e}{J}-z_0    & =\frac{4a^2\widehat{B}^3\bigl[(3a+2)\widehat{B}+6a^2+6a+2\bigr]}{(\widehat{B}+2a)GV},   \\
            \widehat{B}-z_{LN} & =\frac{\widehat{B}^2\bigl[(10a+1)\widehat{B}+8a^2+9a+1\bigr]}{D_N}>0.
        \end{aligned}
    \end{equation}
    Every numerator and denominator is positive, and~\eqref{eq:ia-N4} supplies $z_{LN}-e/J>0$.
    These identities establish the exact cutoff order, including the inequality $z_{LN}<\widehat{B}$ at all positive $a$ and $\widehat{B}$, which later results use.

    \emph{Optimality against all admissible strategies.}
    Section~\ref{ia:verification} states the full verification.
    The verification uses finite biases, which are relative values of her states, normalized to zero in the deposited state at a zero gap.
    In the states in which a withdrawal or redeposit is pending, these biases solve the equations of the verification.
    A free state is a state at a small gap or a profitable gap in which she is deposited with no withdrawal pending.
    In a free state her strategy chooses between waiting and submitting.
    The difference between submitting and waiting has the sign of the withdrawal advantage $A$ at a profitable gap and of the small-gap advantage $R_M$ at a small gap.
    A delayed redeposit is strictly worse when $g>0$.
    Below $\theta_0$ the biases of early withdrawal at zero gain are supersolutions, so zero bounds her average income from above.
    Eventual exit attains that bound.
    Granting her knowledge of whether an arbitrageur is ready only enlarges her information.
    The optimizing controls, however, use no such knowledge, so they attain this upper bound in the actual game, in which she does not observe arrivals.
    The liquidity-trader corrector of Section~\ref{ia:verification} proves that $r$ is the exact fee accounting even under deviations that condition on the number of pending requests.
\end{proof}

\subsection{Full control verification and liquidity-trader accounting}\label{ia:verification}

This subsection completes the proof of Proposition~\ref{prop:ia-controls} at a fixed level of passive capital.
To form an upper bound on the active LP's income, we reveal to her whether an arbitrage transaction has arrived, a fact she does not actually observe.
The strategies that attain the bound do not use this information, so her actual information is unchanged.

\paragraph{Biases.}
Fix a positive normalized gain $g=U_A/(\ell q)$ and describe each state by its bias, its relative value in her long-run average income problem in units of $\ell q$, set to zero in the deposited state at a zero gap.
Subscripts on a bias $h$ are local labels for states: $O$ for outside the pool, $I$ for inside, $K$ for at least one ready arbitrageur, one whose transaction has arrived, $u$ for a withdrawal in transit, $r$ for a delivered withdrawal and $D$ for a delivered redeposit, combined with the gap level $0$, $M$ or $H$.
The label $f$ marks a free state, whose bias her strategy selects.
The label $w$ marks waiting at a small gap without a pending withdrawal.
The ratio $h_x$ of the previous proof is not a bias.
With the quantities of~\eqref{eq:ia-D3} and~\eqref{eq:ia-P3}, the finite biases are
\begin{equation}\label{eq:ia-A1}
    \begin{gathered}
        h_{O0}=-gt_0,\qquad
        h_{OM}=-g(t_0+m_0),\qquad
        h_{OH}=-g(t_0+T_H),\qquad
        h_{OK}=-g\Bigl(t_0+\frac{1}{B}\Bigr),\\
        h_D=-g\,\frac{1+J(t_0+m_0)}{B+J},\qquad
        h_{IK}=\frac{b_f-g}{B}-1.
    \end{gathered}
\end{equation}
With the advantage $A(g)$ of~\eqref{eq:ia-advantages}, write $V_h$ for the bias of staying deposited through a profitable gap and put
\begin{equation}\label{eq:ia-A2}
    V_h=(b_f-g)\,T_H-1,\qquad
    h_{Hr}=V_h+\frac{B}{B+S}\,A(g),\qquad
    h_{Hu}=V_h+p_xA(g).
\end{equation}
The free bias at a profitable gap is $h_{Hf}=h_{Hu}$ under early or late withdrawal and $h_{Hf}=V_h$ under always deposited.
At a small gap,
\begin{equation}\label{eq:ia-A3}
    h_{Mr}=\frac{b_f-g+\frac{J}{2}h_{Hr}+Bh_{OM}}{J+B},\qquad
    h_{Mu}=\frac{b_f-g+\frac{J}{2}h_{Hu}+\lambda h_{Mr}}{J+\lambda},\qquad
    h_{Mw}=\frac{b_f-g+\frac{J}{2}h_{Hf}}{J},
\end{equation}
and the free bias at a small gap is $h_{Mf}=h_{Mu}$ under early withdrawal and $h_{Mf}=h_{Mw}$ under late withdrawal or always deposited.

\paragraph{Generator equations.}
A generator equation says that her normalized gain equals the flow reward in a state plus the expected rate of change of its bias.
Substituting $g=g_E$, $g_L$ or $g_N$ verifies every generator equation for a state with a pending withdrawal or redeposit and the deposited zero-gap equation $g=b_f+Jh_{Mf}$.
For example, a withdrawal that has arrived at a profitable gap executes at the next block unless an arbitrage trade executes first, and it satisfies
\begin{equation}\label{eq:ia-Hr}
    g=b_f+S\,(h_{IK}-h_{Hr})+B\,(h_{OH}-h_{Hr}),
\end{equation}
and a withdrawal in transit at a profitable gap replaces the exit term $B\,(h_{OH}-h_{Hr})$ by $\lambda\,(h_{Hr}-h_{Hu})$.
Both equations include the loss from an arbitrage trade that executes before her withdrawal and the loss from every race that she loses.
Outside the pool at a zero gap, the generator gives
\begin{equation}\label{eq:ia-out}
    g=J\,(h_{OM}-h_{O0})+\lambda\,(h_D-h_{O0}),\qquad
    g=J\,(h_{OM}-h_D)-B\,h_D,
\end{equation}
where the second equation applies once her redeposit has arrived.
Together the two equations price both the risk that a new small gap cancels a redeposit and the fresh latency of the redeposit at the next zero gap.

\paragraph{Comparisons.}
Under early and late withdrawal,
\begin{equation}\label{eq:ia-A4}
    h_{Mu}-h_{Mw}=\frac{m_x}{J}\,R_M(g),\qquad
    R_M(g_E)=R_M(g_L)\,\frac{K_L}{2K_E}.
\end{equation}
Under always deposited, submission at a small gap has the additional term $p_xA(g)/2$, which has the sign of $A(g)$.
The strict cutoff order of Proposition~\ref{prop:ia-controls} gives $R_M(g_E)\geq0$ and $A(g_E)>0$ under early withdrawal, $A(g_L)\geq0$ and $R_M(g_L)\leq0$ under late withdrawal, and $A(g_N)\leq0$ under always deposited, so every comparison with immediate submission has the optimal sign.
Waiting has the generator advantage $S\,(V_h-h_{Hf})$ at a free profitable gap and $J\,(h_{Mw}-h_{Mf})$ at a free small gap, and both are nonpositive.
Delaying a redeposit outside the pool at a zero gap has the advantage $-g\,(1+Jm_0)<0$.
Once an arbitrageur is ready, a fresh withdrawal cannot execute before the arbitrage trade, so submitting it is harmless rather than profitable.
The comparisons above cover delays that depend on elapsed time as well as randomized strategies.
They impose no restriction to finitely many actions.

\paragraph{Low fee-to-loss ratios.}
At $b_f=\theta_0$ the biases of early withdrawal remain finite with $g=0$.
At every smaller $b_f$ these biases are supersolutions, because a lower ratio reduces only the fee rewards earned while deposited.
At every such ratio she withdraws within a finite mean time, as under early withdrawal, and staying out thereafter gives her zero average income.
These two conclusions also hold when a withdrawal or redeposit is pending or her capital is deposited at the start.

\paragraph{No passive capital.}
The verification also applies at $P=0$, where $b_f=r/(\ell q)$, the fee constant $c_A$ of the next paragraph equals $\phi$, and every displayed bias remains finite.
While she is outside, a stale gap is closed by the prompt zero-size price reset of Section~\ref{sec:timing}.
This justifies the optimality of always deposited that the corner condition~\eqref{eq:ia-P7} requires, and it does not substitute $P=0$ into the passive-entry or depth formulas.

\paragraph{Liquidity-trader accounting.}
We derive her fee reward from the arrivals of liquidity-trader requests, not from an assumed fee stream.
Let $N_{\mathrm{req}}$ be the number of pending requests, $c_A=\phi q/(P+q)$ her fee per served request while she is deposited, $I$ her deposit indicator and $\mathcal{L}$ the generator of the state process.
The corrector $\mathcal{F}$ is a function of the pending request count chosen so that her request-by-request fees plus its expected rate of change equal her share of the fee flow served, and it satisfies
\begin{equation}\label{eq:ia-A5}
    \mathcal{F}=\frac{B}{B+\delta}\,c_AIN_{\mathrm{req}},\qquad
    \mathcal{L}\mathcal{F}+Bc_AIN_{\mathrm{req}}=\frac{B\nu}{B+\delta}\,c_AI.
\end{equation}
The identity follows from request arrivals, individual deadlines and the reset at the first block, and capital changes occur after liquidity traders execute and pending requests reset, so they add no omitted jump to the corrector.
Since the expected number of live requests at every date is at most $N_{\mathrm{req}}(0)+\nu/\delta$, the expected contributions of the corrector at the endpoints vanish after division by time.
Integrating the generator and impulse inequalities, adding~\eqref{eq:ia-A5} and dividing by the horizon then gives an upper bound on her limit-superior expected average income, because the biases are bounded and so is the expected request count.
The canonical controls and the timers whose actions are equalities attain this bound, so conditioning on request counts or on unobserved arrivals leaves the cutoffs unchanged.

\paragraph{Ties.}
At the late-withdrawal/always-deposited cutoff, waiting and sending at a profitable gap are equality actions.
The deterministic timer in Proposition~\ref{prop:ia-entry} therefore attains every intermediate occupation on this segment and the same optimal value.
Passive capital is determined before trading and remains fixed under a unilateral deviation.

\subsection{Passive entry, tied controls, existence and the depth envelope}\label{ia:entry}

\begin{proposition}[Passive entry and depth]\label{prop:passive}
    Fix a block rate $B$ and an equilibrium in which the active LP is deposited a fraction $\pi$ of the time.
    \begin{enumerate}[label=(\roman*),ref=\theproposition(\roman*),leftmargin=*,nosep]
        \item If passive capital is positive, each passive unit breaks even:
              \[
                  r(B)\left[\frac{\pi}{P+q}+\frac{1-\pi}{P}\right]=k+\ell e.
              \]
        \item If she is always deposited and $P>0$, then $P=p_0-q$ and $D=p_0$, which requires $p_0>q$.
              If she exits permanently, then $P=D=p_0$, so her permanent exit never leaves the pool empty.
        \item If no passive capital enters, she has the always-deposited occupation $\pi=1$ and depth equals her capital, $D=q$, which requires $p_0\leq q$.
        \item If passive capital is positive, depth is at least the benchmark:
              \[
                  D-p_0=\frac{\pi(1-\pi)\,q^2}{P+(1-\pi)q}\geq0.
              \]
    \end{enumerate}
\end{proposition}

Proposition~\ref{prop:passive} states how passive capital and depth follow from the break-even condition, and the next proposition completes that statement with the positive root, the conditions for an empty passive pool, the tied equilibria, existence and the envelope of depth.
In it, $P_\theta$ and $\pi_\theta$ are the passive capital and her deposited fraction of time at a cutoff $\theta$, $t_w$ is the waiting time of a waiting-timer strategy, $w_T$ is the time weight that such a strategy places on late withdrawal, and $\overline{D}_q$ is the upper envelope of depth.

\begin{proposition}[Passive entry, tied controls, existence and the depth envelope]\label{prop:ia-entry}
    Fix a block rate $B>0$.
    \begin{enumerate}[label=(\roman*),ref=\theproposition(\roman*),leftmargin=*,nosep]
        \item If $0\leq\pi<1$, the break-even condition in part~(i) of Proposition~\ref{prop:passive} has the sole positive root
              \begin{equation}\label{eq:ia-P6}
                  P_\pi=\frac{p_0-q+\sqrt{(p_0-q)^2+4p_0q(1-\pi)}}{2}.
              \end{equation}
              Under always deposited with $P>0$, her fee-to-loss ratio is $b_f=\kappa+e$.
              Under permanent exit $P=D=p_0>0$ at every finite block rate.
        \item An equilibrium with $P=0$ has the always-deposited occupation $\pi=1$ and satisfies
              \begin{equation}\label{eq:ia-P7}
                  \theta_{LN}\leq\frac{r}{\ell q}\leq\kappa+e.
              \end{equation}
        \item At the cutoff $\theta=\theta_{LN}$ between late withdrawal and always deposited, put
              \begin{equation}\label{eq:ia-P8}
                  P_\theta=\frac{r}{\ell\theta}-q>0,\qquad
                  \pi_\theta=\frac{P_\theta+q}{q}\left(1-\frac{P_\theta}{p_0}\right).
              \end{equation}
              These form a tied equilibrium exactly when $\pi_\theta\in[\pi_L,1]$.
              Put $w_T=(1-\pi_\theta)/(1-\pi_L)$.
              The waiting-timer strategy that sends a withdrawal once a profitable gap has lasted the waiting time
              \begin{equation}\label{eq:ia-P9}
                  t_w=\frac{1}{S}\log\frac{1+ep_xt_0(1-w_T)}{w_T},\qquad 0<w_T\leq1,
              \end{equation}
              implements this tie, and $w_T=0$ means always deposited.
        \item At every block rate an equilibrium exists in which she follows one of the four strategies (always deposited, early withdrawal, late withdrawal or permanent exit) without randomizing between them.
              Tied equilibria supplement these equilibria rather than replace them.
        \item If $P>0$, depth obeys the sharp, nondecreasing envelope
              \begin{equation}\label{eq:ia-P11}
                  D\leq\overline{D}_q(p_0)=
                  \begin{cases}
                      q,                     & p_0\leq q/2, \\[2pt]
                      p_0+\dfrac{q^2}{4p_0}, & p_0\geq q/2,
                  \end{cases}
              \end{equation}
              and for $p_0>q/2$ equality requires $\pi=1/2+q/(4p_0)$ and $P=p_0-q/2>0$.
              With no passive capital, depth $D=q$ obeys the same bound.
    \end{enumerate}
\end{proposition}

\begin{proof}[Proof of Proposition~\ref{prop:passive} and Proposition~\ref{prop:ia-entry}]
    \emph{Break-even and the positive root.}
    A passive unit collects fees at the rate $r/(P+q)$ while the active LP's capital is deposited, a fraction $\pi$ of the time by the accounting of Section~\ref{ia:verification}, and at the rate $r/P$ while it is out.
    It also pays the holding cost $k$ and loses $\ell$ at each of the $e$ executed arbitrage trades per unit of time.
    For positive $P$ its zero-return condition is therefore
    \begin{equation}\label{eq:ia-P5}
        r\left[\frac{\pi}{P+q}+\frac{1-\pi}{P}\right]=k+\ell e,
        \qquad\text{equivalently}\qquad
        P(P+q)=p_0\bigl[P+(1-\pi)q\bigr],
    \end{equation}
    which is part~(i) of Proposition~\ref{prop:passive}.
    For $0\leq\pi<1$ its sole positive root is~\eqref{eq:ia-P6}.
    Under always deposited, $P=p_0-q$, which is positive if and only if $p_0>q$, and $D=P+q=p_0$ by~\eqref{eq:ia-D4}.
    Her fee-to-loss ratio is therefore $b_f=\kappa+e$.
    Under permanent exit, $P=D=p_0>0$ at every finite positive block rate, so permanent exit never leaves the pool empty.
    These two cases prove part~(ii) of Proposition~\ref{prop:passive}.

    \emph{Empty passive pool.}
    An equilibrium with $P=0$ must have the always-deposited occupation $\pi=1$ and obey~\eqref{eq:ia-P7}.
    If its limiting fraction of time outside were positive, an entrant of size $\varepsilon>0$ would have average fee income
    \[
        r(B)\left[\pi\frac{\varepsilon}{q+\varepsilon}+(1-\pi)\right],
    \]
    against holding costs and arbitrage losses $\varepsilon(k+\ell e)$.
    As $\varepsilon\downarrow0$, the former tends to $r(B)(1-\pi)>0$ and the latter to zero.
    Thus positive long-run absence cannot sustain an empty passive pool.
    This concerns the equilibrium descriptions of Definition~\ref{def:equilibrium}, not every transient or zero-frequency action of a representing strategy.
    The left inequality in~\eqref{eq:ia-P7} is the optimality of always deposited at $b_f=r/(\ell q)$, which Section~\ref{ia:verification} verifies at $P=0$, and the right inequality is equivalent to $p_0\leq q$ because $\kappa+e=(k+\ell e)/\ell$.
    Depth then equals her capital, $D=q$, which proves part~(iii) of Proposition~\ref{prop:passive}.

    \emph{Tied strategies.}
    At $\theta=\theta_{LN}$, the passive capital $P_\theta$ in~\eqref{eq:ia-P8} sets her fee-to-loss ratio equal to $\theta$, and $\pi_\theta$ is the deposited fraction of time at which $P_\theta$ breaks even in~\eqref{eq:ia-P5}.
    These two quantities therefore form a tied equilibrium exactly when $\pi_\theta\in[\pi_L,1]$.
    For the tie between late withdrawal and always deposited, she waits for her fixed waiting time after the gap becomes profitable before sending her withdrawal, so the probability that she sends it before any arbitrage transaction arrives is $\alpha_T=\exp(-S\,t_w)$.
    Her probability of a successful withdrawal per profitable gap is then $\alpha_Tp_x$, and a renewal lasts $1/e+\alpha_Tp_xt_0$ on average.
    The resulting time weight on late withdrawal is $\alpha_T(1+ep_xt_0)/(1+e\alpha_Tp_xt_0)=w_T$, which~\eqref{eq:ia-P9} solves for $t_w$.
    Waiting and sending are equality actions at this cutoff, so the timer attains the upper bound of the full control problem.
    Under Assumption~\ref{ass:main}, only this tie can occur in equilibrium, by Lemma~\ref{ia2:lem-implications}.

    \emph{Existence.}
    Fix the block rate and view the passive excess return, a passive unit's fee yield less its required yield $d_c$ of~\eqref{eq:ia-D3}, as a function of $P$.
    On each interval of passive capital over which one of the four strategies is optimal, the excess return strictly decreases in $P$.
    At a strategy switch induced by the fall in $b_f$ as $P$ rises, her deposited fraction of time drops and the passive excess return jumps upward.
    The excess return starts positive unless the corner with always deposited and no passive capital is feasible, and it is negative at sufficiently large $P$.
    It therefore crosses zero inside such an interval or at its endpoint, which establishes existence without a lottery, and tied occupations supplement these canonical roots rather than replace them.

    \emph{Depth and its envelope.}
    Let $\Xi$ denote the square root in~\eqref{eq:ia-P6}:
    \begin{equation}\label{eq:ia-Xi}
        \Xi=\sqrt{(p_0-q)^2+4p_0q(1-\pi)}=2P+q-p_0.
    \end{equation}
    Then, for a positive passive pool, \eqref{eq:ia-P5} gives
    \begin{equation}\label{eq:ia-P10}
        D-p_0=\frac{\pi(1-\pi)\,q^2}{P+(1-\pi)q}\geq0,\qquad
        D=\frac{p_0+\Xi+(2\pi-1)q}{2},
    \end{equation}
    which proves part~(iv) of Proposition~\ref{prop:passive}.
    At fixed $p_0$, $\partial D/\partial\pi=q(1-p_0/\Xi)$ and $\partial^2D/\partial\pi^2=-2p_0^2q^2/\Xi^3<0$, so depth is concave in her deposited fraction.
    Maximizing depth over $\pi$ yields the nondecreasing sharp envelope~\eqref{eq:ia-P11}.
    For $p_0>q/2$ equality requires $\pi=1/2+q/(4p_0)$ and $P=p_0-q/2>0$, and the corner $D=q$ also obeys the bound.
    The identity~\eqref{eq:ia-P10} holds for a positive passive pool.
\end{proof}

\section{The Fee Assumption and the Equilibrium Map}\label{ia:map}

This section proves the full equilibrium map and the income selection behind Theorem~\ref{thm:map}.

\subsection{Inputs to the equilibrium map}\label{ia2:units}

Use the normalized units in~\eqref{eq:ia-N0}, the polynomials in~\eqref{eq:ia-N1}, and the occupations in~\eqref{eq:ia-DT1}.
In particular, $a=\lambda/J$, $\widehat{B}=B/J$, $c=\kappa/J$, $\delta_n=\delta/J$ and $E=e_\infty/J$.
The normalized served fee flow and passive required yield are
\begin{equation}\label{eq:ia2-Rd}
    R_{\widehat{B}}=\frac{f_{\mathrm{cap}}\widehat{B}}{\widehat{B}+\delta_n}
    =\frac{r(B)}{\ell qJ},\qquad
    d_{\widehat{B}}=c+\frac{2a\widehat{B}}G=\frac{k+\ell e}{\ell J}.
\end{equation}
Thus $p_0/q=R_{\widehat{B}}/d_{\widehat{B}}$.
Write $d_\infty=k+\ell e_\infty$.
The clock and patience assumptions read $a>1/2$ and $\delta_n<c/20$.

Proposition~\ref{prop:ia-controls} supplies the full fixed-$P$ optimal-control envelope and its cutoffs.
At the late-withdrawal/always-deposited cutoff, equality of the two incomes gives
\begin{equation}\label{eq:ia2-tie-identity}
    e_L=e-(1-\pi_L)\theta_{LN}.
\end{equation}
The positive-entry root is~\eqref{eq:ia-P6}; the separate no-entry test is~\eqref{eq:ia-P7}.
Equations~\eqref{eq:ia-P8}--\eqref{eq:ia-P9} give the tied capital, occupation and deterministic timer at $\theta_{LN}$.
The timer's endpoint occupations are late withdrawal and always deposited, and installed passive capital remains fixed in every unilateral deviation.

\subsection{What the four assumptions imply}\label{ia2:lemma}

Write $F=e_\infty\lambda/(J+\lambda)$ for the comparison rate in the fee assumption.
The fee assumption restricts primitives only, and the next lemma derives the two consequences that the map uses: her fee-to-loss ratio never falls into the range in which early withdrawal or permanent exit pays, and her capital is small relative to the capital that fees support.
Its proof relies on the coefficient bounds of Section~\ref{ia2:certificate}, which also define the cost floor $\kappa_{\min}$ and the coefficient maximum $\Psi$ in part~(i).

\begin{lemma}[Implications of the four assumptions]\label{ia2:lem-implications}
    Suppose the primitives satisfy Assumption~\ref{ass:main}.
    Then:
    \begin{enumerate}[label=(\roman*),ref=(\roman*),leftmargin=*,nosep]
        \item $\kappa>F>\kappa_{\min}$ and $\rbar>q\Psi$, where $\kappa_{\min}$ and $\Psi$ are defined in \eqref{eq:ia2-floor};
        \item $\delta<e_\infty^2/(\kappa+3e_\infty)$;
        \item her capital, relative to the capital $\rbar/d_\infty$ that fees support at instant blocks, satisfies
              \begin{equation}\label{eq:ia2-load}
                  \eta_q:=\frac{q\,d_\infty}{\rbar}<\frac{a}{(a+1)(4a+1)}<\frac{1}{9}.
              \end{equation}
    \end{enumerate}
    Consequently, in every equilibrium with positive passive capital she uses neither permanent exit nor early withdrawal, nor a tie at $\theta_0$ or $\theta_{EL}$, and $\rbar>q\,d_\infty$.
\end{lemma}

\begin{proof}
    The patience assumption gives $\delta_n<c/20<c$, so the coefficient bounds of Section~\ref{ia2:certificate} apply and give
    \begin{equation}\label{eq:ia2-L2}
        \kappa>F>\kappa_{\min},\qquad
        \Psi<\frac{d_\infty^2}{k-\ell F}.
    \end{equation}
    The fee assumption, $\rbar(k-\ell F)>q\,d_\infty^2$, therefore gives $\rbar>q\Psi$, which completes part~(i).
    It also makes every coefficient of the exclusion polynomial in \eqref{eq:ia2-AF5} positive, which yields
    \begin{equation}\label{eq:ia2-L3}
        \frac{\underline{b}_f}{J}=\frac{R_{\widehat{B}}d_{\widehat{B}}}{R_{\widehat{B}}+d_{\widehat{B}}}>z_{EL},\qquad
        \underline{b}_f:=\frac{r(B)}{\ell(p_0+q)},
    \end{equation}
    where $\underline{b}_f$ is her fee-to-loss ratio when passive capital equals the benchmark $p_0$.
    Every equilibrium has $P\le p_0$ and hence $b_f\ge\underline{b}_f>\theta_{EL}$.
    The cutoff order of Proposition~\ref{prop:ia-controls} therefore excludes permanent exit, early withdrawal and the ties at $\theta_0$ and $\theta_{EL}$.
    An equilibrium without passive capital has her always deposited, by \eqref{eq:ia-P7} and Proposition~\ref{prop:passive}(iii).

    For the load bound, note that $\eta_q=(c+E)/f_{\mathrm{cap}}$ in normalized units.
    The argument uses the ratio $c/E$ of the normalized holding cost $c$ to the normalized rate $E$ at which arbitrage trades execute at instant blocks.
    The fee assumption implies $\eta_q<[c/E-a/(a+1)]/(c/E+1)$, a bound that strictly increases in $c/E$.
    The carry assumption bounds $c/E<2a/(1+2a)$, so the increasing bound yields the first inequality in \eqref{eq:ia2-load}, and the identity
    \[
        \frac19-\frac{a}{(a+1)(4a+1)}=\frac{(2a-1)^2}{9(a+1)(4a+1)}>0,
    \]
    which is strict because $a>1/2$, gives the second.
    Finally, the carry assumption gives $c<E$, and with $\delta_n<c/20$ this implies $\delta_n<E^2/(c+3E)$, which is part~(ii) in normalized units, because $(c/E)(c/E+3)<4<20$ when $0<c/E<1$.
\end{proof}

Two remarks delimit the lemma.
First, the bound \eqref{eq:ia2-load} on the size of her capital is partly a by-product of stating the fee assumption as one simple product.
The bound is not evidence that her strategy has a negligible effect on depth in general.
Second, early withdrawal and permanent exit remain admissible deviations at every block rate, and the lemma shows only that neither is her best response in any equilibrium under the four assumptions.

The economics of the exclusion is a floor on her fee-to-loss ratio.
Passive capital never exceeds the benchmark $p_0$, so her fee-to-loss ratio never falls below its value $\underline{b}_f$ at the benchmark, whereas early withdrawal or permanent exit can be optimal only if that ratio is at or below $\theta_{EL}$.
The fee assumption keeps this floor strictly above $\theta_{EL}$ at every block rate by requiring the fee pool to support her capital after a conservative allowance for cheap passive funding.
In this comparison $k+\ell e_\infty$ is the benchmark cost of passive capital at instant blocks, its holding cost plus its arbitrage losses, and she does not pay its holding cost $k$ on her own capital.
The rate $F$ is the rate $e_\infty$ at which arbitrage trades execute at instant blocks, multiplied by $\lambda/(J+\lambda)$, the probability that one message arrives before a fresh small gap opens; it is a comparison rate inside a sufficient bound, not the value of her withdrawal option.
A large fee pool alone does not suffice, because when passive capital is very cheap to hold, early withdrawal can remain her unique equilibrium strategy however large the fees.
The fee assumption also implies, with Assumptions~\ref{ass:clock} and~\ref{ass:carry}, that her capital is less than one ninth of $\rbar/(k+\ell e_\infty)$, the passive capital at which fees just cover holding cost and arbitrage losses at instant blocks.

\paragraph{Feasibility of the assumptions.}
Consider the economy with $J=1$, $\lambda=2$, $\delta=1/2000$, $k=9/50$, $\ell=1$, $\rbar=1$ and $q=1/10$.
Direct substitution gives the following positive slacks for the four assumptions:
\[
    \begin{gathered}
        2\lambda-J=3,\qquad e_\infty\frac{2\lambda}{J+2\lambda}-\kappa=\frac7{850},\\
        \frac\kappa{20}-\delta=\frac{17}{2000},\qquad \rbar(k-\ell F)-q(k+\ell e_\infty)^2=\frac{127673}{21675000}.
    \end{gathered}
\]

\subsection{The coefficient certificate for the fee assumption}\label{ia2:certificate}

This subsection proves the coefficient bounds used in Lemma~\ref{ia2:lem-implications}.
We use the clock ratio $a$, normalized holding cost $c$, normalized deadline rate $\delta_n$ and fee capacity $f_{\mathrm{cap}}$ of \eqref{eq:ia-N0}, and the normalized rate $E=2a/(1+8a)$ at which arbitrage trades execute at instant blocks.
Two further quantities enter.
The first, $F_n$, is the comparison rate $F=2J\lambda^2/[(J+8\lambda)(J+\lambda)]$, which appears in the fee assumption, measured in units of $J$.
The second, $d_{\infty n}$, is the normalized instant-block cost of a unit of passive capital, its holding cost plus its arbitrage losses, which differs from the finite-speed cost $d_{\widehat{B}}=c+e/J$ of \eqref{eq:ia2-Rd}:
\begin{equation}\label{eq:ia2-Fn}
    F_n=\frac{F}{J}=E\,\frac{a}{a+1},\qquad d_{\infty n}=\frac{d_\infty}{\ell J}=c+E.
\end{equation}
In these units the fee assumption reads exactly $f_{\mathrm{cap}}(c-F_n)>d_{\infty n}^2$.

\paragraph{Coefficient arrays.}
Write $G_0=2a$ and $G_1=8a+1$ for the coefficients of $G$, $A_{E,i}$ and $D_{E,i}$ for the coefficients of $\widehat{B}^i$ in $A_E$ and $D_E$, and $\mathcal{Z}_i$ for the coefficient of $\widehat{B}^i$ in the product $GD_E$, so that
\begin{equation}\label{eq:ia2-arrays}
    \begin{aligned}
         & A_{E,0}=6a^2+7a+1,\qquad A_{E,1}=6a^2+10a+3,\qquad A_{E,2}=2a+2,                                                \\
         & D_{E,0}=12a^3+14a^2+2a,\qquad D_{E,1}=60a^3+82a^2+21a+1,                                                        \\
         & D_{E,2}=24a^3+80a^2+44a+3,\qquad D_{E,3}=12a^2+24a+2,                                                           \\
         & \mathcal{Z}_0=G_0D_{E,0},\qquad \mathcal{Z}_1=G_1D_{E,0}+G_0D_{E,1},\qquad \mathcal{Z}_2=G_1D_{E,1}+G_0D_{E,2}, \\
         & \mathcal{Z}_3=G_1D_{E,2}+G_0D_{E,3},\qquad \mathcal{Z}_4=G_1D_{E,3}.
    \end{aligned}
\end{equation}
With $u_f=a(3-2a)$, $v_f=-a(3a-1)(4a+3)$, $\mathcal{X}_0=cG_0$ and $\mathcal{X}_1=cG_1+2a$, define the two arrays
\begin{equation}\label{eq:ia2-varpi-psi}
    \begin{aligned}
        (\varpi_0,\ldots,\varpi_4) & =\bigl(c\mathcal{Z}_0,\ c\mathcal{Z}_1,\ c\mathcal{Z}_2,\ c\mathcal{Z}_3+4av_f,\ c\mathcal{Z}_4+4au_f\bigr),   \\
        \psi_0                     & =2a\delta_n\mathcal{X}_0A_{E,0},                                                                               \\
        \psi_1                     & =2a\bigl[(\mathcal{X}_0+\delta_n\mathcal{X}_1)A_{E,0}+\delta_n\mathcal{X}_0A_{E,1}\bigr],                      \\
        \psi_2                     & =2a\bigl[\mathcal{X}_1A_{E,0}+(\mathcal{X}_0+\delta_n\mathcal{X}_1)A_{E,1}+\delta_n\mathcal{X}_0A_{E,2}\bigr], \\
        \psi_3                     & =2a\bigl[\mathcal{X}_1A_{E,1}+(\mathcal{X}_0+\delta_n\mathcal{X}_1)A_{E,2}\bigr],                              \\
        \psi_4                     & =2a\mathcal{X}_1A_{E,2}.
    \end{aligned}
\end{equation}
The cost floor and the coefficient maximum of Lemma~\ref{ia2:lem-implications} are
\begin{equation}\label{eq:ia2-floor}
    \frac{\kappa_{\min}}{J}=\max\Bigl\{0,\ \frac{4a^2(3a-1)(4a+3)}{192a^4+688a^3+480a^2+72a+3}\Bigr\},\qquad
    \frac{\Psi}{\ell J}=\max_{0\le i\le4}\frac{\psi_i}{\varpi_i}.
\end{equation}

\paragraph{A lower bound on the first of the two arrays.}
For every $a>0$ the exact identities
\begin{equation}\label{eq:ia2-AF-identities}
    F_n\mathcal{Z}_3+4av_f=\frac{2a^2(392a^3+414a^2+116a+9)}{(a+1)(8a+1)}>0,\qquad
    F_n\mathcal{Z}_4+4au_f=\frac{4a^2(4a^2+13a+4)}{a+1}>0
\end{equation}
hold, and together with the first three entries they give
\begin{equation}\label{eq:ia2-AF3}
    \varpi_i>(c-F_n)\mathcal{Z}_i\qquad(i=0,\ldots,4).
\end{equation}
Moreover, $F_n>\kappa_{\min}/J$.
Indeed, $\mathcal{Z}_3=192a^4+688a^3+480a^2+72a+3$, and the first identity in \eqref{eq:ia2-AF-identities} divided by $\mathcal{Z}_3$ is exactly
\[
    F_n-\frac{4a^2(3a-1)(4a+3)}{192a^4+688a^3+480a^2+72a+3}>0,
\]
while the zero inside the maximum in \eqref{eq:ia2-floor} also lies below $F_n$ because $F_n>0$.

\paragraph{An upper bound on the second array.}
It remains to show $\psi_i<d_{\infty n}^2\mathcal{Z}_i$ when $\delta_n<c$ and $c>F_n$.
Each $\psi_i$ is nondecreasing in $\delta_n$, so it suffices to show that
\[
    \mathcal{K}_i(c,a)=(c+E)^2\mathcal{Z}_i-\psi_i\big|_{\delta_n=c}
\]
is positive, and the first three entries expand as
\begin{equation}\label{eq:ia2-H012}
    \begin{aligned}
        \mathcal{K}_0 & =\frac{16a^3(a+1)(6a+1)(8ca+c+a)}{(8a+1)^2},                                  \\
        \mathcal{K}_1 & =\frac{2a}{(8a+1)^2}\Bigl[c^2(3072a^5+4736a^4+2000a^3+366a^2+31a+1)           \\
                      & \qquad+c(1920a^5+2864a^4+1000a^3+116a^2+4a)+432a^5+576a^4+144a^3+8a^2\Bigr],  \\
        \mathcal{K}_2 & =\frac{1}{(8a+1)^2}\Bigl[c^2(27648a^6+51456a^5+28336a^4+6744a^3+790a^2+45a+1) \\
                      & \qquad+c(7680a^6+14784a^5+7456a^4+1276a^3+86a^2+2a)                           \\
                      & \qquad+576a^6+1328a^5+624a^4+48a^3\Bigr].
    \end{aligned}
\end{equation}
Every coefficient in these three expressions is positive, so $\mathcal{K}_0$, $\mathcal{K}_1$ and $\mathcal{K}_2$ are positive.
For the other two entries it suffices to show positivity at the lower bound $c=F_n$ and a positive derivative in $c$:
\begin{equation}\label{eq:ia2-H34}
    \begin{aligned}
        \mathcal{K}_3(F_n,a)    & =\frac{4a^3(800a^4+1384a^3+886a^2+234a+17)}{(a+1)^2(8a+1)^2}>0, \\
        \mathcal{K}_4(F_n,a)    & =\frac{8a^3(2a+1)(4a^2+13a+4)}{(a+1)^2(8a+1)}>0,                \\
        \partial_c\mathcal{K}_3 & =\frac{2}{8a+1}\bigl[c(1536a^5+5440a^4+4208a^3+988a^2+92a+3)    \\
                                & \qquad+576a^4+530a^3+78a^2+3a\bigr]>0,                          \\
        \partial_c\mathcal{K}_4 & =4\bigl[c(48a^3+102a^2+20a+1)+4a^3+15a^2+a\bigr]>0.
    \end{aligned}
\end{equation}
Consequently all five entries satisfy
\begin{equation}\label{eq:ia2-AF4}
    \psi_i<(c+E)^2\mathcal{Z}_i,\qquad
    \frac{\psi_i}{\varpi_i}<\frac{(c+E)^2}{c-F_n}<f_{\mathrm{cap}},
\end{equation}
where the last inequality is the fee assumption in normalized units.

\paragraph{The exclusion polynomial.}
The bounds \eqref{eq:ia2-AF4} make every coefficient $f_{\mathrm{cap}}\varpi_i-\psi_i$ positive, and these are exactly the coefficients of the all-speed exclusion polynomial.
To connect them with the exclusion statement, write $\mathcal{X}=\mathcal{X}_0+\mathcal{X}_1\widehat{B}$, recall that $G=G_0+G_1\widehat{B}$, $D_E=\sum_iD_{E,i}\widehat{B}^i$ and $A_E=\sum_iA_{E,i}\widehat{B}^i$, and let $\underline{b}_f$ be the fee-to-loss ratio at the benchmark from \eqref{eq:ia2-L3}.
Exact subtraction gives
\begin{equation}\label{eq:ia2-AF5}
    \frac{\underline{b}_f}{J}-z_{EL}
    =\frac{\widehat{B}\,\bigl[f_{\mathrm{cap}}\sum_{i=0}^4\varpi_i\widehat{B}^i-\sum_{i=0}^4\psi_i\widehat{B}^i\bigr]}
    {D_E\,\bigl[f_{\mathrm{cap}}\,\widehat{B}G+(\widehat{B}+\delta_n)\mathcal{X}\bigr]}.
\end{equation}
The bracketed numerator has degree four and strictly positive coefficients by \eqref{eq:ia2-AF4}, and the denominator is positive at every finite $\widehat{B}>0$, so $\underline{b}_f>\theta_{EL}$ at every block rate.
Because every installed-capital candidate has $b_f\ge\underline{b}_f$, early withdrawal and permanent exit cannot be optimal.
The fee assumption is therefore a sufficient condition on primitives for this exclusion, not an assumption about the equilibrium outcome.

\subsection{The equilibrium map}\label{ia2:map}

The theorem below states the full equilibrium map behind Theorem~\ref{thm:map}, including the equilibria the main text does not select, with the equations that locate its three boundaries.
Recall that $R_{\widehat{B}}$ and $d_{\widehat{B}}$ in \eqref{eq:ia2-Rd} are the normalized fee flow served and the normalized cost of a unit of passive capital, $\pi_L$ in \eqref{eq:ia-DT1} is her deposited fraction of time under late withdrawal, and $z_{LN}$ in \eqref{eq:ia-N3} is her normalized cutoff between late withdrawal and always deposited.
As in Section~\ref{sec:model}, an equilibrium is described by its passive capital, her deposited fraction of time and exposed-trade rate.

\begin{theorem}[Equilibrium map]\label{ia2:thm-map}
    Suppose the primitives satisfy Assumption~\ref{ass:main}.
    There are exactly three boundary block rates, $0<B_0<B_1<B_2$, and in normalized units $\widehat{B}_i=B_i/J$:
    \begin{enumerate}[label=(\roman*),ref=(\roman*),leftmargin=*,nosep]
        \item $\widehat{B}_0$ is the unique positive root of $R_{\widehat{B}}=d_{\widehat{B}}$.
              Clearing the positive denominators of this equation gives the entry polynomial, a quadratic in $\widehat{B}$ with constant term $-2ac\delta_n$, leading coefficient $\varsigma_2=(8a+1)(f_{\mathrm{cap}}-c)-2a$ and linear coefficient $\varsigma_1=2a(f_{\mathrm{cap}}-c)-\delta_n\bigl[(8a+1)c+2a\bigr]$, and its positive root is
              \begin{equation}\label{eq:ia2-b0}
                  \widehat{B}_0=\frac{-\varsigma_1+\sqrt{\varsigma_1^2+8ac\delta_n\,\varsigma_2}}{2\varsigma_2};
              \end{equation}
        \item $\widehat{B}_2$ is the unique positive root of the cubic equation
              \begin{equation}\label{eq:ia2-b2}
                  cGD_N-32a^3\widehat{B}^3=0;
              \end{equation}
        \item $\widehat{B}_1$ is the unique root in $(\widehat{B}_0,\widehat{B}_2)$ of
              \begin{equation}\label{eq:ia2-b1}
                  \mathcal{H}_L(\widehat{B}):=R_{\widehat{B}}\,(d_{\widehat{B}}-z_{LN})-z_{LN}\,(d_{\widehat{B}}-\pi_Lz_{LN})=0.
              \end{equation}
    \end{enumerate}
    Clearing the denominators of $\mathcal{H}_L$ and removing its factors of $\widehat{B}$ leaves a numerator of degree six for almost all primitives, whose unique root in the interval of part~(iii) is $\widehat{B}_1$.
    The theorem claims neither that this numerator has a single positive root nor that the middle boundary of part~(iii) can be written with arithmetic operations and roots.
    Table~\ref{ia2:tab-map} lists, for each block rate, every equilibrium that exists there.
\end{theorem}

\begin{table}[htbp]
    \centering
    \caption{\textbf{Equilibria at each block rate under Assumption~\ref{ass:main}.}}\label{ia2:tab-map}
    \begin{tabularx}{\textwidth}{@{}lX@{}}
        \toprule
        Block rate   & Every distinct equilibrium                                                                                         \\
        \midrule
        $0<B\le B_0$ & Always deposited without passive capital, the only equilibrium: $P=0$ and $D=q$.                                   \\
        $B_0<B<B_1$  & Always deposited with passive capital, the only equilibrium: $P=p_0-q$ and $D=p_0$.                                \\
        $B=B_1$      & Always deposited and late withdrawal; the waiting-timer strategy at this endpoint coincides with late withdrawal.  \\
        $B_1<B<B_2$  & Always deposited, late withdrawal, and one waiting-timer strategy strictly between them.                           \\
        $B=B_2$      & Always deposited and late withdrawal; the waiting-timer strategy at this endpoint coincides with always deposited. \\
        $B>B_2$      & Late withdrawal, the only equilibrium, with passive capital $P_\pi$ from \eqref{eq:ia-P6} at $\pi=\pi_L$.          \\
        \bottomrule
    \end{tabularx}
\end{table}

\begin{proof}
    \emph{Exclusions.}
    By Lemma~\ref{ia2:lem-implications}, no equilibrium has her exit permanently, withdraw early or tie at $\theta_0$ or $\theta_{EL}$.

    \emph{Completeness of the occupation descriptions.}
    Consider an optimal active-LP strategy with the limiting occupations
    retained by Definition~\ref{def:equilibrium}, holding $P$ fixed.
    The integrated verification inequalities imply that every strictly
    positive residual has zero average contribution: otherwise the strategy's
    limiting income would be strictly below the attained bound.
    Every subsequential state and action occupation measure therefore satisfies
    the balance equations using only equality actions.
    Away from $b_f=\theta_{LN}$ these equations give the appropriate canonical
    occupation, although zero-frequency or payoff-irrelevant actions need not
    coincide path by path with its canonical representative.

    At $b_f=\theta_{LN}$, the optimal normalized gain is
    $g=\theta_{LN}-e>0$.
    Small-gap submission is strictly inferior, and delaying a permitted
    redeposit is strictly inferior; only withdrawal timing in a profitable
    gap is indifferent.
    The cycle calculation below may be applied to an equality-action realization of each limiting physical state/action flow; its cycles need not be the individual cycles of the original optimal history.
    To identify all resulting occupations, set $H_0=1/e$ and start a cycle
    aligned and deposited.
    End it at the next arbitrage correction if no withdrawal succeeds, and
    otherwise at the first subsequent redeposit.
    For any equality-action timing rule, including one depending on its past,
    let $v_n$ be the probability of a successful withdrawal in cycle $n$,
    conditional on the history at its start.
    Then $0\leq v_n\leq p_x$: coupling a delayed fresh withdrawal with one
    sent at the profitable gap's onset shows that its success event is a
    subset of the latter's success event.
    At a successful withdrawal block no arbitrage transaction is ready, so
    the residual profitable-gap lifetime has mean $T_H$ and the subsequent
    return wait has mean $t_0$.
    Consequently the conditional expected cycle length, deposited time,
    exposed-trade count and withdrawal count are
    \[
        \bigl(H_0+v_n t_0,\ H_0-v_nT_H,\ 1-v_n,\ v_n\bigr).
    \]
    The environmental and prompt-return times have uniformly finite moments
    at each fixed positive block rate; hence endpoint contributions vanish
    when these conditional balances are averaged over a long horizon.
    Take any convergent subsequence of the bounded expected cycle and withdrawal
    counts per unit time, and denote its limits by $c_R$ and $\omega$.
    The conditional balances give
    \[
        1=c_RH_0+\omega t_0,\qquad
        \pi=c_RH_0-\omega T_H,\qquad
        e_A=c_R-\omega,\qquad
        0\leq\omega\leq p_xc_R.
    \]
    The first two equations give
    $\omega=(1-\pi)/(t_0+T_H)$ and
    $c_R=(1-\omega t_0)/H_0$.
    Thus every subsequence has the same limits, proving convergence of these
    auxiliary rates without an additional assumption; in particular $c_R>0$.
    Put $v=\omega/c_R\in[0,p_x]$. Then every such occupation is
    \[
        (\pi,e_A,\omega)
        =\frac{\bigl(H_0-vT_H,\ 1-v,\ v\bigr)}{H_0+vt_0},
    \]
    which is exactly the line segment between always deposited and late
    withdrawal, with late-withdrawal weight
    $w=v(1+ep_xt_0)/[p_x(1+evt_0)]$.
    Every point is attained by the deterministic timer
    \eqref{eq:ia-P9}.
    The remaining entry tests therefore exhaust the retained capital and occupation descriptions; no uniqueness of their implementing histories is asserted.

    \emph{Entry.}
    Lemma~\ref{ia2:lem-implications} also gives $\rbar>q\,d_\infty$, which in normalized units reads $f_{\mathrm{cap}}>c+E$, so $\varsigma_2>0$.
    The entry polynomial, obtained by clearing the positive denominators of $R_{\widehat{B}}=d_{\widehat{B}}$, has the negative constant term $-2ac\delta_n$ and the positive leading coefficient $\varsigma_2$, so by Descartes' rule of signs it has exactly one positive root, and the quadratic formula gives \eqref{eq:ia2-b0}.

    \emph{Extinction of always deposited.}
    The wedge $h_n=z_{LN}-e/J=32a^3\widehat{B}^3/(GD_N)$, the normalized distance between her cutoff $\theta_{LN}$ and the rate $e$ at which arbitrage trades execute, increases strictly with the block rate from zero toward its upper bound $h_{\infty n}=32a^3/[(8a+1)(10a+1)]$.
    To verify the monotonicity, write $GD_N=\sum_{i=0}^{3}[GD_N]_i\,\widehat{B}^i$, where every coefficient $[GD_N]_i$ is positive; differentiating $\widehat{B}^3/(GD_N)$ leaves the numerator $3[GD_N]_0\widehat{B}^2+2[GD_N]_1\widehat{B}^3+[GD_N]_2\widehat{B}^4>0$.
    The carry assumption implies $c<h_{\infty n}$, because it bounds the normalized holding cost by the denominator of the ratio below, and that ratio exceeds one:
    \begin{equation}\label{eq:ia2-M21}
        \frac{h_{\infty n}}{E\cdot2a/(1+2a)}=\frac{8a(1+2a)}{1+10a}>1\qquad(a>1/2),
    \end{equation}
    by the clock assumption $a>1/2$.
    Thus the cubic in \eqref{eq:ia2-b2}, which is the equation $h_n=c$ with its denominators cleared, has a negative cubic coefficient and positive lower coefficients, so it has exactly one positive root $\widehat{B}_2$.

    \emph{Appearance of late withdrawal.}
    Let $b_{f,L}$ denote her fee-to-loss ratio when she uses late withdrawal and passive capital equals the break-even amount $P_\pi$ at $\pi=\pi_L$.
    On $0<\widehat{B}<\widehat{B}_2$, put $A_{\widehat{B}}=d_{\widehat{B}}-z_{LN}=c-h_n$, the excess of the normalized cost of a passive unit over her normalized cutoff.
    Also put $V_{\widehat{B}}=(1-\pi_L)z_{LN}$, her normalized cutoff scaled by her fraction of time outside the pool under late withdrawal.
    The function that locates where late withdrawal appears is
    \begin{equation}\label{eq:ia2-M22}
        \Gamma_{\widehat{B}}=\frac{R_{\widehat{B}}}{z_{LN}}\,\frac{A_{\widehat{B}}}{A_{\widehat{B}}+V_{\widehat{B}}}.
    \end{equation}
    As the endpoint tests below show, $\Gamma_{\widehat{B}}>1$ exactly when $b_{f,L}>\theta_{LN}$, that is, when her break-even fee-to-loss ratio under late withdrawal exceeds her cutoff.
    The function $\Gamma_{\widehat{B}}$ strictly decreases on this interval.
    Indeed, $A_{\widehat{B}}'<0$ because the wedge increases strictly, as shown above.
    The ratio $R_{\widehat{B}}/z_{LN}$ of the normalized fee flow served to her normalized cutoff strictly decreases because $\delta_n<E$, which holds since $\delta_n<c/20$ and $c<E$.
    The negative of the derivative of this ratio has a quadratic numerator in $\widehat{B}$ whose coefficients $\Omega_2$, $\Omega_1$ and $\Omega_0$ are all strictly positive, because $E-\delta_n>0$:
    \[
        \begin{aligned}
            \Omega_2 & =\frac{16a^2(12a^2+11a+1)}{8a+1}+(E-\delta_n)(2a+1)(10a+1), \\
            \Omega_1 & =\frac{64a^3(a+1)}{8a+1}+2(E-\delta_n)(a+1)(10a+1),         \\
            \Omega_0 & =(E-\delta_n)(a+1)^2(8a+1),
        \end{aligned}
    \]
    and that negative derivative is
    \begin{equation}\label{eq:ia2-Uprime}
        -\Bigl(\frac{R_{\widehat{B}}}{z_{LN}}\Bigr)'=\frac{f_{\mathrm{cap}}\,(\Omega_2\widehat{B}^2+\Omega_1\widehat{B}+\Omega_0)}{2a(\widehat{B}+\delta_n)^2\bigl[(2a+1)\widehat{B}+a+1\bigr]^2}.
    \end{equation}
    Also $V_{\widehat{B}}'>0$, because its reduced derivative numerator is
    \[
        4a^2\bigl\{(96a^3+56a^2-4a-1)\widehat{B}^2+(48a^3+28a^2+6a+2)\widehat{B}+12a^3+13a^2+2a+1\bigr\},
    \]
    and writing $a=1/2+\xi$, with $\xi>0$ by the clock assumption, turns its only coefficient of uncertain sign into $96\xi^3+200\xi^2+124\xi+23>0$.
    This step uses no sign of the derivative of $\pi_L$.

    Next, $\widehat{B}_0<\delta_n$, because at $\widehat{B}=\delta_n$, $R_{\widehat{B}}=f_{\mathrm{cap}}/2>(9/2)(c+E)>d_{\widehat{B}}$ by \eqref{eq:ia2-load}.
    At $\widehat{B}_0$, where $R_{\widehat{B}}=d_{\widehat{B}}$, the identity $z_{LN}<\widehat{B}$ of \eqref{eq:ia-certificate} and $c>20\delta_n$ give
    \[
        \frac{b_{f,L}}{J}=\frac{d_{\widehat{B}}}{1+\sqrt{1-\pi_L}}>\frac{d_{\widehat{B}}}{2}>\frac{c}{2}>\widehat{B}_0>z_{LN},
    \]
    hence $\Gamma_{\widehat{B}}>1$ at $\widehat{B}=\widehat{B}_0$.
    As $\widehat{B}$ approaches $\widehat{B}_2$ from below, $A_{\widehat{B}}$ vanishes and $\Gamma_{\widehat{B}}$ tends to zero.
    There is therefore exactly one crossing $\Gamma_{\widehat{B}}=1$ in $(\widehat{B}_0,\widehat{B}_2)$, and this crossing is equivalent to \eqref{eq:ia2-b1}.
    Clearing the denominators gives degree six; all denominators are positive, and the crossing is simple because $\Gamma_{\widehat{B}}'<0$.

    \emph{Endpoint tests.}
    For $0\le\pi<1$ and any cutoff $z<d_{\widehat{B}}$, the normalized fee-to-loss ratio $b_f/J$ that passive entry produces is the smaller root of
    \[
        \pi\Bigl(\frac{b_f}{J}\Bigr)^2-(R_{\widehat{B}}+d_{\widehat{B}})\,\frac{b_f}{J}+R_{\widehat{B}}d_{\widehat{B}}=0,
    \]
    which is its unique root in $(0,d_{\widehat{B}})$, and $b_f/J-z$ has the sign of $R_{\widehat{B}}(d_{\widehat{B}}-z)-z(d_{\widehat{B}}-\pi z)$.
    If $z\ge d_{\widehat{B}}$, then $b_f/J<z$ automatically.
    Always deposited with positive passive capital does not use this sign rule: there passive entry sets $b_f/J=d_{\widehat{B}}$, and passive capital is positive exactly when $R_{\widehat{B}}>d_{\widehat{B}}$.
    For $0<\widehat{B}\le \widehat{B}_0<\delta_n$, $R_{\widehat{B}}/\widehat{B}=f_{\mathrm{cap}}/(\widehat{B}+\delta_n)>f_{\mathrm{cap}}/(2\delta_n)>1$, hence $R_{\widehat{B}}>\widehat{B}>z_{LN}$; together with $R_{\widehat{B}}\le d_{\widehat{B}}$, this verifies the corner test \eqref{eq:ia-P7} throughout that interval.
    Applying the sign rule with $z=z_{LN}$, the corner test \eqref{eq:ia-P7} and the tie formula \eqref{eq:ia-P8} yields every remaining cell and endpoint of Table~\ref{ia2:tab-map}.
    A late-withdrawal equilibrium with positive passive capital is impossible below $\widehat{B}_0$ because $\Gamma_{\widehat{B}}>1$ there, and above $\widehat{B}_2$ every candidate has $b_f/J\le d_{\widehat{B}}<z_{LN}$.
    Because these tests hold on each interval, a root of the degree-six numerator outside $(\widehat{B}_0,\widehat{B}_2)$ changes no equilibrium.
\end{proof}

\subsection{The equilibrium selected by active-LP income}
\label{ia2:income-selection}

\begin{proof}[Proof of Theorem~\ref{thm:map}]
    Under Assumption~\ref{ass:main}, use the arbitrageur strategies of
    Proposition~\ref{prop:rent-share}, as required by its scope restriction.
    The full correspondence is Theorem~\ref{ia2:thm-map}; it remains to apply
    the income selector in Section~\ref{sec:model}.
    Fix a block rate in $[B_1,B_2]$.
    The map gives $p_0>q$, positive passive capital in every equilibrium,
    and deposited fractions ordered as
    \[
        1=\pi_N\geq\pi_T\geq\pi_L>0,
    \]
    where the subscript $T$ denotes the tied timer occupation.
    Both inequalities are strict in $(B_1,B_2)$.
    At $B_1$ the timer duplicates late withdrawal, and at $B_2$ it duplicates
    always deposited.
    Differentiating the entry root~\eqref{eq:ia-P6} gives
    \[
        \frac{\partial P_\pi}{\partial\pi}
        =-\frac{p_0q}{\sqrt{(p_0-q)^2+4p_0q(1-\pi)}}<0.
    \]
    This also holds at $\pi=1$ here because $p_0>q$.
    Thus $P_N<P_T<P_L$ in the open coexistence interval,
    $P_N<P_T=P_L$ at $B_1$, and $P_N=P_T<P_L$ at $B_2$.

    For a fixed installed stock $P$, let $\mathcal V(P)$ be the optimal
    long-run income in~\eqref{eq:ia-P2-income}.
    The verification in Section~\ref{ia:verification} bounds the prescribed
    expected-limit-superior objective, and canonical strategies attain ordinary
    renewal limits.
    The canonical strategies have the same physical occupations and exposed-trade
    rates at every positive $P$: neither their submission rules nor the
    prescribed arbitrageur timing depends on the size of passive capital.
    Consequently a canonical strategy attaining $\mathcal V(P_b)$ can also be
    used at $P_a<P_b$, while each comparison holds its own installed stock
    fixed. If that strategy has deposited fraction $\pi_j>0$, then
    \[
        \mathcal V(P_a)\geq\mathcal V(P_b)
        +r(B)q\pi_j\left(\frac1{P_a+q}-\frac1{P_b+q}\right)
        >\mathcal V(P_b).
    \]
    The cutoff order and Lemma~\ref{ia2:lem-implications} exclude a zero-value
    optimum at every coexistence stock, so a canonical maximizer has positive
    occupation; at the timer stock either
    adjacent canonical strategy attains the same value.
    It follows that their incomes satisfy
    \[
        \begin{array}{ll}
            U_N>U_T>U_L, & B_1<B<B_2, \\
            U_N>U_T=U_L, & B=B_1,     \\
            U_N=U_T>U_L, & B=B_2.
        \end{array}
    \]
    For an equivalent direct check, always deposited with positive passive capital gives
    $U_N=kq$, the timer gives
    $U_T=\ell q(\theta_{LN}-e)$, and the tie identity
    \eqref{eq:ia2-tie-identity} gives
    \[
        U_N-U_T=\ell q(\kappa+e-\theta_{LN}),\qquad
        U_T-U_L=\ell q\pi_L(\theta_{LN}-b_{f,L}).
    \]
    The map and the capital ordering give exactly the strict signs and
    endpoint equalities above.

    Outside coexistence the map has only one capital/occupation outcome.
    For $0<B\leq B_0$, that outcome is always deposited with $P=0$:
    the separate corner test~\eqref{eq:ia-P7} verifies both active
    optimality under the prescribed prompt zero-size correction continuation
    and passive nonentry. Its income is $r(B)-\ell qe(B)$; at $B_0$ it equals
    $kq$ because $p_0=q$.
    No expression containing $1/P$ is used at this corner.
    For $B_0<B\leq B_2$, the selected outcome is always deposited with
    $P=p_0-q$.
    For $B>B_2$, only late withdrawal remains, and its positive passive stock
    is the unique entry root at $\pi=\pi_L$.
    These are exactly the strategy and entry profile in Theorem~\ref{thm:map}.
    Uniqueness refers to the capital/occupation descriptions of
    Definition~\ref{def:equilibrium}; different histories with the same
    description are not distinct selected outcomes.
    The comparison selects among operating equilibria and does not allow a
    unilateral deviation to change installed passive capital.
\end{proof}

\section{The Liquidity Peak, the Late Region and Further Results}\label{ia:results}

This section proves the depth, cream-skimming, duration and price-tracking results of Section~\ref{sec:equilibrium}.
Assumption~\ref{ass:main} holds throughout, except where a statement says that it uses only the model of Section~\ref{sec:model}.

\subsection{Normalized units and facts from the earlier parts}\label{ia3:notation}

Recall that $B$ is the block rate, $\lambda$ the message speed of each arbitrageur and of the active LP, $J$ the public information arrival rate and $\delta$ the deadline rate of liquidity traders.
Passive capital pays the holding cost $k$, each executed arbitrage trade costs every unit of deposited capital the loss $\ell$, liquidity traders offer the fee flow $\rbar$, and the active LP's capital is $q$.
Recall also the fee flow served $r(B)=\rbar B/(B+\delta)$ and the rate $e=B/(1+TB)$ at which arbitrage trades execute, where $T=4/J+1/(2\lambda)$ is the mean time between executed arbitrage trades at instant blocks and $e_\infty=1/T$ is that rate there.
Recall finally the active LP's deposited fraction of time $\pi$, passive capital $P$, time-average depth $D=P+q\pi$, and the competitive benchmark $p_0=r(B)/(k+\ell e)$ of equation~\eqref{eq:ia-D3}.

The proofs measure every rate relative to the public information arrival rate $J$ and write the clock ratio $a$, the normalized block rate $\widehat{B}$, the normalized cost ratio $c$, the normalized deadline rate $\delta_n$, the fee capacity $f_{\mathrm{cap}}$, the capital scale $L_0$, and the normalized rate $E$ at which arbitrage trades execute at instant blocks, with its reciprocal $T_n$, as
\begin{equation}\label{ia3:eq-units}
    \begin{gathered}
        a=\frac{\lambda}{J},\qquad \widehat{B}=\frac BJ,\qquad c=\frac{\kappa}{J},\qquad \delta_n=\frac{\delta}{J},\\
        f_{\mathrm{cap}}=\frac{\rbar}{\ell qJ},\qquad L_0=\frac{\rbar}{\ell J},\qquad E=\frac{e_\infty}{J}=\frac{2a}{8a+1},\qquad T_n=\frac1E.
    \end{gathered}
\end{equation}
Here $a$ is the clock ratio of message speed to the public information arrival rate, and $\widehat{B}$, $c$ and $\delta_n$ are the block rate, the cost ratio $\kappa=k/\ell$ and the deadline rate in the same units.
The capital scale $L_0$ is the fee flow divided by the loss rate $\ell J$, the fee capacity $f_{\mathrm{cap}}=L_0/q$ measures that scale in units of $q$, and $E$ and $T_n$ are the instant-block rate at which arbitrage trades execute and the mean time between them, in normalized units.
This change of units is only a normalization and loses no generality.
In these units the clock assumption reads $a>1/2$ and the patience assumption reads $\delta_n<c/20$, while the carry assumption implies $c<E$, so $cT_n<1$.

In the same units, the fee flow served per unit of the loss rate on her capital, a passive unit's holding cost plus arbitrage losses, and the benchmark measured in units of $q$ are
\begin{equation}\label{ia3:eq-rdn}
    R_{\widehat{B}}=\frac{f_{\mathrm{cap}}\,\widehat{B}}{\widehat{B}+\delta_n},\qquad d_{\widehat{B}}=c+\frac{2a\widehat{B}}{G},\qquad n_{\widehat{B}}=\frac{R_{\widehat{B}}}{d_{\widehat{B}}}=\frac{p_0}{q},
\end{equation}
where $G$ is the following polynomial, and $H_L$ and $D_N$ are two further polynomials that the displays below use, all three with positive coefficients:
\[
    \begin{gathered}
        G=(8a+1)\widehat{B}+2a,\qquad H_L=(6a+1)\widehat{B}+12a^2+a+1,\\
        D_N=(10a+1)\widehat{B}^2+(12a^2+11a+1)\widehat{B}+2a(a+1).
    \end{gathered}
\]
Under late withdrawal, her deposited fraction of time $\pi_L$, her exposed-trade rate $e_L$ (the rate of executed arbitrage trades while her capital is deposited) and the rate $\omega_L$ of her successful withdrawals are
\begin{equation}\label{ia3:eq-occ}
    \Big(\pi_L,\ \frac{e_L}{J},\ \frac{\omega_L}{J}\Big)=\frac{\big(4a(\widehat{B}+2a)[(12a+1)\widehat{B}+3a],\ 8a^2\widehat{B}(\widehat{B}+3a),\ 4a^2\widehat{B}^2\big)}{GH_L}.
\end{equation}
In the proofs, a prime on a normalized quantity denotes the derivative in $\widehat{B}$, and since $\widehat{B}=B/J$, its sign is that of the derivative in $B$.

The proofs use the following facts, which the earlier sections of this Internet Appendix establish.
\begin{enumerate}[label=(F\arabic*),ref=(F\arabic*),leftmargin=*]
    \item\label{ia3:F1} \emph{Her optimal strategy.}
          Let $b_f=r(B)/[\ell(P+q)]$ be her fee-to-loss ratio, her fee income per unit of time while deposited divided by the loss $\ell q$ that one executed arbitrage trade imposes on her capital.
          Her optimal strategy is permanent exit, early withdrawal, late withdrawal or always deposited as $b_f$ rises across three cutoffs $\theta_0<\theta_{EL}<\theta_{LN}$, with ties only between adjacent strategies.
          The rate at which arbitrage trades execute satisfies $\theta_0<e<\theta_{LN}$.
          We call $\theta_{LN}$, at which she is indifferent between late withdrawal and staying deposited, the late cutoff.
          Under late withdrawal, her trading income is $U_A=\ell q(b_f\pi_L-e_L)$.
          The late cutoff, which is $z_{LN}=\theta_{LN}/J$ in normalized units, satisfies
          \begin{equation}\label{ia3:eq-zln}
              \begin{gathered}
                  z_{LN}=\frac{2a\widehat{B}[(2a+1)\widehat{B}+a+1]}{D_N},\qquad z_{LN}-\frac eJ=\frac{32a^3\widehat{B}^3}{GD_N}>0,\\
                  \widehat{B}-z_{LN}=\frac{\widehat{B}^2[(10a+1)\widehat{B}+8a^2+9a+1]}{D_N}>0.
              \end{gathered}
          \end{equation}
    \item\label{ia3:F2} \emph{The bound on active capital and the excluded strategies.}
          With $d_\infty=k+\ell e_\infty$, the cost that a passive unit bears at instant blocks, the four assumptions bound her capital relative to the capital that fees alone support at instant blocks:
          \begin{equation}\label{ia3:eq-lot}
              \eta_q=\frac{qd_\infty}{\rbar}<\frac{a}{(a+1)(4a+1)}<\frac19,\qquad \frac{q}{L_0}=\frac{\eta_q}{c+E}.
          \end{equation}
          Every equilibrium has $P\le p_0$, so her fee-to-loss ratio is at least $\underline{b}_f=r(B)/[\ell(p_0+q)]$, which in normalized units is $\underline{b}_f/J=R_{\widehat{B}}d_{\widehat{B}}/(R_{\widehat{B}}+d_{\widehat{B}})$ and which exceeds $\theta_{EL}$ at every block rate.
          Early withdrawal, permanent exit and ties involving either are therefore never equilibria.
          An equilibrium without passive capital also has her always deposited.
    \item\label{ia3:F3} \emph{Entry accounting.}
          With positive passive capital, the entry condition of Proposition~\ref{prop:passive} is equivalent to $P(P+q)=p_0[P+(1-\pi)q]$, whose positive root for $0\le\pi<1$ is, writing $\Xi$ for its square-root term as in~\eqref{eq:ia-Xi},
          \begin{equation}\label{ia3:eq-root}
              P=\frac{p_0-q+\Xi}{2},\qquad \Xi=\sqrt{(p_0-q)^2+4p_0q(1-\pi)}=2P+q-p_0.
          \end{equation}
          Depth then satisfies
          \begin{equation}\label{ia3:eq-env}
              \begin{gathered}
                  D-p_0=\frac{\pi(1-\pi)q^2}{P+(1-\pi)q}\ge0,\qquad D=\frac{p_0+\Xi+(2\pi-1)q}{2},\\
                  D\le\overline{D}_q(p_0)=\begin{cases}q,&p_0\le q/2,\\ p_0+q^2/(4p_0),&p_0\ge q/2,\end{cases}
              \end{gathered}
          \end{equation}
          where the envelope $\overline{D}_q$ is nondecreasing in $p_0$ and also bounds the no-entry corner, the equilibrium without passive capital, at which $D=q$.
    \item\label{ia3:F4} \emph{The equilibrium map.}
          The equilibrium map of Theorem~\ref{ia2:thm-map} has $0<B_0<B_1<B_2$ with $B_0<\delta$.
          Up to $B_0$ she is alone in the pool, and on $(B_0,B_1)$ she is always deposited with positive passive capital.
          On $[B_1,B_2]$ both always deposited and late withdrawal are equilibria.
          Inside $(B_1,B_2)$ a third equilibrium has her follow a waiting-timer strategy, withdrawing once a profitable gap lasts a fixed time.
          Above $B_2$, late withdrawal is the unique equilibrium.
\end{enumerate}

\subsection{The common depth peak}\label{ia3:peak}

This subsection proves Proposition~\ref{prop:liquidity-maximum}.
A selection of equilibria chooses one equilibrium at each block rate where several exist.

\begin{proposition}[Common global maximum of depth]\label{ia3:prop-peak}
    Let Assumption~\ref{ass:main} hold, and let $A_d=\ell-\delta T(kT+\ell)$, the margin whose positivity gives the benchmark a turning point at a finite block rate.
    Then $A_d>0$.
    Under every selection of equilibria, including waiting-timer strategies and the equilibrium without passive capital, the time-average depth $D$ attains a unique global maximum, at the block rate
    \begin{equation}\label{ia3:eq-Bd}
        B_d=\frac{T\delta k+\sqrt{T^2\delta^2k^2+A_d\delta k}}{A_d}=\frac{\sqrt{k\delta}}{\sqrt{\ell(1-\delta T)}-T\sqrt{k\delta}}.
    \end{equation}
    This block rate and the peak depth satisfy
    \begin{equation}\label{ia3:eq-height}
        \delta<B_d<\frac\kappa2,\qquad D(B_d)=p_0(B_d)=\frac{\rbar}{\big[\sqrt{k(1-\delta T)}+\sqrt{\ell\delta}\big]^2}.
    \end{equation}
    At every block rate in $[\delta,\kappa]$, every equilibrium has the active LP always deposited with positive passive capital.
    The proposition does not say that depth under an arbitrary selection falls monotonically beyond $B_d$.
\end{proposition}

The proof rests on one exact identity, which writes the offered fee flow divided by the benchmark in terms of the rate at which arbitrage trades execute.

\begin{lemma}[Benchmark height]\label{ia3:lem-height}
    At every block rate, the rate $e=e(B)$ at which arbitrage trades execute lies in $(0,1/T)$, with $B=e/(1-Te)$, and
    \begin{equation}\label{ia3:eq-C1}
        \frac{\rbar}{p_0(B)}=k(1-\delta T)+\ell\delta+\frac{k\delta}{e}+\ell(1-\delta T)e.
    \end{equation}
    The last two terms sum to $2\sqrt{k\delta\ell(1-\delta T)}$ plus the square $\big(\sqrt{k\delta/e}-\sqrt{\ell(1-\delta T)e}\big)^2$.
    The square vanishes at $e^*=\sqrt{k\delta/[\ell(1-\delta T)]}$, which lies in $(0,1/T)$ exactly when $A_d>0$.
    The benchmark is then maximized at $B_d=e^*/(1-Te^*)$, with the value in \eqref{ia3:eq-height}.
\end{lemma}

\begin{proof}
    Substituting $B=e/(1-Te)$ into $p_0=r(B)/(k+\ell e)$ and dividing gives \eqref{ia3:eq-C1}.
    The square is nonnegative and vanishes only at $e^*$, so when $e^*<1/T$ the inverse benchmark attains its smallest value exactly there.
    Converting $e^*$ to a block rate gives \eqref{ia3:eq-Bd}, and substituting it gives the height in \eqref{ia3:eq-height}.
\end{proof}

\begin{proof}[Proof of Proposition~\ref{ia3:prop-peak}]
    In normalized units, write the benchmark in units of $L_0$ as $f_{\mathrm{bench}}$.
    Its formula has a positive linear factor $W$ in the denominator, and the sign of its slope is set by a quadratic $N_{\widehat{B}}$ whose leading coefficient is $\mathcal{A}=1-\delta_n T_n(1+cT_n)$:
    \begin{equation}\label{ia3:eq-fbench}
        W=c+(1+cT_n)\widehat{B},\qquad f_{\mathrm{bench}}(\widehat{B})=\frac{p_0}{L_0}=\frac{\widehat{B}(1+T_n\widehat{B})}{(\widehat{B}+\delta_n)W},\qquad N_{\widehat{B}}=\mathcal{A}\widehat{B}^2-2T_n\delta_n c\widehat{B}-\delta_n c.
    \end{equation}
    The derivative of $f_{\mathrm{bench}}$ is $-N_{\widehat{B}}/[(\widehat{B}+\delta_n)^2W^2]$.
    Recall that $cT_n<1$ and $\delta_n/c<1/20$.
    The leading coefficient $\mathcal{A}$, which equals $A_d/\ell$, is positive.
    Moreover $N_{\widehat{B}}(\delta_n)<\delta_n(\delta_n-c)<0$, and
    \[
        \frac{N_{\widehat{B}}}{\widehat{B}^2}\Big|_{\widehat{B}=c/2}=1-\frac{\delta_n}{c}\,(c^2T_n^2+5cT_n+4)>\frac12.
    \]
    So the quadratic $N_{\widehat{B}}$ has one positive root, and it lies in $(\delta_n,c/2)$.
    This root is the benchmark's unique turning point and gives \eqref{ia3:eq-Bd}.
    Lemma~\ref{ia3:lem-height} gives the height.

    Next take $\delta_n\le \widehat{B}\le c$, an interval we call the bracket.
    Then $R_{\widehat{B}}\ge f_{\mathrm{cap}}/2>\tfrac92(c+E)>9c$ by \eqref{ia3:eq-lot} and $c<E$, whereas $d_{\widehat{B}}<2c$, so $n_{\widehat{B}}=R_{\widehat{B}}/d_{\widehat{B}}>1$.
    Moreover $d_{\widehat{B}}-\widehat{B}>0$, and $\widehat{B}d_{\widehat{B}}/(d_{\widehat{B}}-\widehat{B})$ strictly increases in $\widehat{B}$, since
    \[
        \frac{d}{d\widehat{B}}\,\frac{\widehat{B}d_{\widehat{B}}}{d_{\widehat{B}}-\widehat{B}}=\frac{c(T_n\widehat{B}+1)(T_n\widehat{B}c+2\widehat{B}+c)}{(T_n\widehat{B}^2-T_n\widehat{B}c-c)^2}>0,\qquad \frac{\widehat{B}d_{\widehat{B}}}{d_{\widehat{B}}-\widehat{B}}\Big|_{\widehat{B}=c}=c(2+cT_n)<3c.
    \]
    So the fee flow exceeds this ratio's maximum on the bracket, which gives $\underline{b}_f/J=R_{\widehat{B}}d_{\widehat{B}}/(R_{\widehat{B}}+d_{\widehat{B}})>\widehat{B}>z_{LN}$, the last inequality by \eqref{ia3:eq-zln}.
    By \ref{ia3:F1} and \ref{ia3:F2}, every equilibrium with positive passive capital has her always deposited.
    The no-entry corner is ruled out because $n_{\widehat{B}}>1$, which means $p_0>q$.
    This proves the statement about $[\delta,\kappa]$.

    Outside $[\delta_n,c]$, the benchmark lies below its value at the nearer end of the bracket, because it has one turning point.
    At $\widehat{B}=\delta_n$, $f_{\mathrm{bench}}<1/(2c)$, and at $\widehat{B}=c$,
    \[
        f_{\mathrm{bench}}(c)=\frac{1+cT_n}{c(1+\delta_n/c)(2+cT_n)}<\frac{2}{3c}.
    \]
    Let $q_n=q/L_0=\eta_q/(c+E)<1/(18c)$ be her capital in units of $L_0$.
    The nondecreasing envelope of \ref{ia3:F3} therefore bounds depth in every equilibrium outside the bracket by
    \begin{equation}\label{ia3:eq-outside}
        \frac{D}{L_0}\le\overline{D}_{q_n}\Big(\frac2{3c}\Big)=\frac2{3c}+\frac{q_n^2}{4\cdot 2/(3c)}<\frac{577}{864c}.
    \end{equation}
    At the comparison block rate $\widehat{B}_*=\sqrt{c\delta_n}$, which lies inside the bracket, depth equals the benchmark and
    \begin{equation}\label{ia3:eq-inside}
        \frac{D}{L_0}=f_{\mathrm{bench}}(\widehat{B}_*)>\frac{1}{c\,(1+\sqrt{\delta_n/c})^2}>\frac{20}{(21+4\sqrt5)c}>\frac{577}{864c}.
    \end{equation}
    The last comparison holds because $5163>2308\sqrt5$, as $5163^2-5\cdot2308^2=22249>0$.
    With $v=\sqrt{\delta_n/c}$, the left side of the first comparison exceeds the right side by $cT_n v^2/[c(1+v)^2(1+v+cT_n v)]$.
    Hence every equilibrium outside the bracket has depth strictly below a value that is attained inside the bracket.
    Inside the bracket, every equilibrium has $D=p_0$, and the benchmark has its unique maximum at $B_d$.
    This proves that the maximum is global, unique and attained.
\end{proof}

Faster blocks first save liquidity traders whose deadlines would otherwise pass, and they also make arbitrage trades execute more often, which raises the arbitrage losses that every passive unit bears.
The benchmark weighs these saved trades against these losses.
Under the four assumptions it has one interior maximum, at the block rate $B_d$ in \eqref{ia3:eq-Bd}.
On a bracket around that block rate, fees are large relative to the active LP's small capital, so in every equilibrium she stays deposited and depth equals the benchmark.
Outside the bracket, depth stays below a fixed bound that lies below the benchmark's peak, however long she is deposited, so her strategy cannot lift depth to that peak.

\subsection{Depth in the late-withdrawal region}\label{ia3:late}

This subsection proves Proposition~\ref{prop:cream}(ii) and the depth decline discussed in Section~\ref{sec:liquidity-maximum}.

\begin{proposition}[Late-region decline]\label{ia3:prop-late}
    Let Assumption~\ref{ass:main} hold.
    \begin{enumerate}[label=(\roman*),leftmargin=*]
        \item Every block rate at which late withdrawal or a waiting-timer strategy is an equilibrium exceeds $\kappa$.
        \item At every block rate at which late withdrawal is an equilibrium, its passive capital and depth, given by the entry root with $\pi=\pi_L$, satisfy $P'(B)<0$ and $D'(B)<0$, whatever the sign of $\pi_L'(B)$.
    \end{enumerate}
    The proposition asserts no derivative for waiting-timer equilibria and no sign at a switch between equilibria.
\end{proposition}

\begin{proof}
    For part (i), the map in \ref{ia3:F4} and the result of Proposition~\ref{ia3:prop-peak} that every equilibrium on $[\delta,\kappa]$ is always deposited imply $B_1>\kappa$.
    So every late-withdrawal or waiting-timer equilibrium has $\widehat{B}>c$.

    For part (ii), take $\widehat{B}\ge c$.
    The exact identity
    \begin{equation}\label{ia3:eq-pinf}
        f_{\mathrm{bench}}(\widehat{B})-\frac1{c+E}=\frac{(\mathcal{A}/T_n)\widehat{B}-\delta_n c}{(c+E)(\widehat{B}+\delta_n)W}
    \end{equation}
    has a positive right-hand side, whose numerator is at least $(c/T_n)[1-\delta_n T_n(2+cT_n)]>17c/(20T_n)>0$.
    Undoing the normalization gives $p_0>\rbar/d_\infty$, and hence $q/p_0<\eta_q<1/9$, at every block rate with $\widehat{B}\ge c$, including every late-withdrawal and waiting-timer equilibrium.
    For $\widehat{B}\ge c$, the polynomial $N_{\widehat{B}}$ of \eqref{ia3:eq-fbench} also satisfies
    \begin{equation}\label{ia3:eq-slope}
        \frac{N_{\widehat{B}}}{\widehat{B}^2}>1-5\,\frac{\delta_n}{c},\qquad -f_{\mathrm{bench}}'(\widehat{B})>\frac{1-5\delta_n/c}{(1+\delta_n/c)^2W^2}>\frac{100}{147\,W^2},
    \end{equation}
    where $100/147=(1-5/20)/(1+1/20)^2$.

    To bound the response of her deposited fraction of time to the block rate, factor $\pi_L=A_0C_0$ with
    \[
        A_0=\frac{1+(4+1/(3a))\widehat{B}}{1+T_n\widehat{B}},\qquad C_0=\frac{\widehat{B}+2a}{(1+1/(6a))\widehat{B}+2a+(1+a)/(6a)}.
    \]
    Both factors lie in $(0,1)$.
    Their derivatives are, in order, $-1/[6a(1+T_n\widehat{B})^2]$ and $(1-a)/\{6a[(1+1/(6a))\widehat{B}+2a+(1+a)/(6a)]^2\}$.
    With $d_{\infty n}=d_\infty/(\ell J)=c+E$ denoting, in normalized units, the instant-block cost $d_\infty$ of a passive unit, direct ratio bounds give
    \begin{equation}\label{ia3:eq-pislope}
        \frac{W^2|\pi_L'|}{d_{\infty n}}<\frac{2}{3(8a+1)}+\frac{6|a-1|(8a+1)}{(6a+1)^2}<\frac43.
    \end{equation}
    For $a\ge1$, $4/3$ exceeds the middle expression by $2(15a+2)(44a+5)/[3(6a+1)^2(8a+1)]>0$.
    For $1/2<a\le1$ the excess has numerator $2(1152a^3-204a^2-107a-8)$, and substituting $a=1/2+\xi$ gives twice $1152\xi^3+1524\xi^2+553\xi+63/2>0$.
    This bound uses only $a>1/2$ and $cT_n<1$, and it covers both signs of $\pi_L'$.

    Differentiating the depth identity in \eqref{ia3:eq-env}, which contains the square root $\Xi$ of~\eqref{ia3:eq-root}, with respect to $p_0$ and $\pi$ gives
    \begin{equation}\label{ia3:eq-sens}
        \begin{gathered}
            D_{p_0}=\frac12\Big(1+\frac{p_0-(2\pi-1)q}{\Xi}\Big)\ge1-\frac12\Big(\frac q{p_0}\Big)^2,\qquad |D_\pi|\le\frac{q^2}{p_0-q},\\
            P'=D_{p_0}p_0'-\frac{p_0q}{\Xi}\,\pi'.
        \end{gathered}
    \end{equation}
    For the first bound, write $\tilde q=q/p_0<1$ for her capital relative to the benchmark and $\tilde\pi=2\pi-1$.
    The ratio $(1-\tilde q\tilde\pi)/\sqrt{1-2\tilde q\tilde\pi+\tilde q^2}$ in the first sensitivity is smallest at $\tilde\pi=\tilde q$, since its derivative in $\tilde\pi$ has the sign of $\tilde\pi-\tilde q$.
    Hence $D_{p_0}\ge[1+\sqrt{1-\tilde q^2}]/2\ge1-\tilde q^2/2$.
    The second bound uses $\Xi^2=(p_0-(2\pi-1)q)^2+4\pi(1-\pi)q^2$, $|\Xi-p_0|\le q$ and $\Xi\ge p_0-q$.
    Combining \eqref{ia3:eq-slope}, \eqref{ia3:eq-pislope} and \eqref{ia3:eq-sens} yields
    \begin{equation}\label{ia3:eq-dom}
        \begin{gathered}
            \frac{D'}{L_0}<\frac1{W^2}\Big[-\frac{100}{147}\Big(1-\frac{\eta_q^2}2\Big)+\frac{4\eta_q^2}{3(1-\eta_q)}\Big]<0,\\
            \frac{P'}{L_0}<\frac1{W^2}\Big[-\frac{100}{147}\Big(1-\frac{\eta_q^2}2\Big)+\frac{4\eta_q}{3(1-\eta_q)}\Big]<0.
        \end{gathered}
    \end{equation}
    Both brackets increase with $\eta_q$, and at $\eta_q=1/9$ they equal $-2237/3402$ and $-1733/3402$, so both are negative under \eqref{ia3:eq-lot}.
    Both primed quantities are derivatives in $\widehat{B}$, so the derivatives in $B$ have the same signs.
\end{proof}

In the late-withdrawal region, every passive unit still bears the loss at every executed arbitrage trade, while the service gain from faster blocks has largely run its course, so faster blocks lower the fee-supported benchmark.
If she spends more time deposited, she takes a larger part of the fees and can push passive capital out further, and if she spends less, passive units collect extra fees during her absences, which partly offsets their higher losses.
The fee assumption keeps her capital small enough that neither change in fee sharing can overturn the decline of the benchmark.
At a fixed block rate, equilibrium depth still weakly exceeds the benchmark, because her absences open fee income to passive entrants.
So two effects are distinct: the fall in depth as blocks speed up, and the small gap between depth and the benchmark at one block rate.

\subsection{Late-withdrawal timing and the cream-skimming core}\label{ia3:core}

This subsection proves Proposition~\ref{prop:cream}(i).
Part (i) of the next proposition describes the timing of late withdrawal and uses only the model, while part (ii) gives the cream-skimming result beyond an explicit onset under the four assumptions.
The proposition makes no claim about the direction in which her deposited-time fraction or her share of liquidity-trader fees moves.

\begin{proposition}[Late timing and cream-skimming]\label{ia3:prop-core}
    \begin{enumerate}[label=(\roman*),leftmargin=*]
        \item For all positive primitives, when she withdraws late, the slope of her exposed-trade rate in the normalized block rate has the sign opposite to that of the quadratic $\mathcal{C}_e\widehat{B}^2-\mathcal{M}_e\widehat{B}-\mathcal{N}_e$, whose coefficients are the following polynomials in the clock ratio $a$:
              \[
                  \mathcal{C}_e=48a^3+10a^2-8a-1,\qquad \mathcal{M}_e=48a^3+4a^2+4a,\qquad \mathcal{N}_e=72a^4+6a^3+6a^2.
              \]
              Her exposed-trade and successful-withdrawal rates then change with the normalized block rate as
              \begin{equation}\label{ia3:eq-dt3}
                  \begin{gathered}
                      \Big(\frac{e_L}J\Big)'=-\frac{8a^2(\mathcal{C}_e\widehat{B}^2-\mathcal{M}_e\widehat{B}-\mathcal{N}_e)}{(GH_L)^2},\\
                      \Big(\frac{\omega_L}J\Big)'=\frac{4\widehat{B}a^2[(96a^3+32a^2+11a+1)\widehat{B}+48a^3+4a^2+4a]}{(GH_L)^2}>0.
                  \end{gathered}
              \end{equation}
              The fraction of all executed arbitrage trades that hit her deposited capital satisfies
              \begin{equation}\label{ia3:eq-chi}
                  \chi_L=\frac{e_L}e=\frac{4a(\widehat{B}+3a)}{H_L},\qquad \chi_L'=-\frac{4a(6a^2+2a-1)}{H_L^2}<0\quad(a>1/2).
              \end{equation}
              When $a>1/2$, $\mathcal{C}_e>0$, and her exposed-trade rate rises with $\widehat{B}$ up to
              \begin{equation}\label{ia3:eq-be}
                  \widehat{B}_e=\frac{\mathcal{M}_e+\sqrt{\mathcal{M}_e^2+4\mathcal{C}_e\mathcal{N}_e}}{2\mathcal{C}_e}
              \end{equation}
              and falls beyond it.
              Let $\widehat{B}_U$ be the unique positive root of the cubic
              \begin{equation}\label{ia3:eq-QU}
                  \mathcal{Q}_U=(24a^2+12a+1)\widehat{B}^3-(34a^2+6a+1)\widehat{B}^2-(36a^3+35a^2+5a)\widehat{B}-6a^2(a+1).
              \end{equation}
        \item Let Assumption~\ref{ass:main} hold, and let $\bar B=\max\{B_2,J\widehat{B}_e,J\widehat{B}_U\}$, which is finite.
              At every $B>\bar B$ the unique equilibrium is late withdrawal.
              As $B$ rises above $\bar B$, passive capital $P$, depth $D$, her exposed-trade rate $e_L$ and her arbitrage losses $\ell qe_L$ fall, while her successful-withdrawal rate $\omega_L$, her share of the pool's capital while deposited, $q/(P+q)$, and her trading income $U_A$ rise.
              In particular, the passive LPs' share of the pool while she is deposited, $P/(P+q)$, falls.
    \end{enumerate}
\end{proposition}

\begin{proof}
    Differentiating \eqref{ia3:eq-occ} gives \eqref{ia3:eq-dt3}, \eqref{ia3:eq-chi} and the derivative of her deposited fraction of time,
    \begin{equation}\label{ia3:eq-pid}
        \pi_L'=-\frac{4a[(96a^3-70a^2-18a-1)\widehat{B}^2+(72a^3-40a^2-4a)\widehat{B}+24a^4+8a^3-4a^2]}{(GH_L)^2},
    \end{equation}
    which the proof uses only through the identity \eqref{ia3:eq-income} below, never through its sign.
    The cubic $\mathcal{C}_e$ has one sign change in descending powers, a negative constant and a positive leading coefficient, so it has exactly one positive root.
    This root lies in $(3/8,2/5)$, so $\mathcal{C}_e>0$ for $a>1/2$.
    The quadratic $\mathcal{C}_e\widehat{B}^2-\mathcal{M}_e\widehat{B}-\mathcal{N}_e$ then has leading and constant terms of opposite signs and exactly one positive root, $\widehat{B}_e$, which gives the shape of $e_L$ in part (i).
    The cubic $\mathcal{Q}_U$ has a positive leading coefficient and negative lower coefficients, so it has exactly one positive root, $\widehat{B}_U$.

    Now let Assumption~\ref{ass:main} hold and $B>\bar B$.
    By \ref{ia3:F4} the unique equilibrium is late withdrawal, and part (ii) of Proposition~\ref{ia3:prop-late} gives $P'<0$ and $D'<0$.
    Because $\widehat{B}>\widehat{B}_e$, her exposed-trade rate and her arbitrage losses $\ell qe_L$ fall.
    By \eqref{ia3:eq-dt3}, $\omega_L$ rises, and $q/(P+q)$ rises because $P$ falls.
    For her income, $U_A=\ell q(b_f\pi_L-e_L)$ from \ref{ia3:F1} gives $U_A'=\ell q[b_f'\pi_L+b_f\pi_L'-e_L']$.
    Her fee-to-loss ratio $b_f=r(B)/[\ell(P+q)]$ rises because $r'>0$ and $P'<0$.
    If $\pi_L'<0$, the inequality $b_f\le\theta_{LN}$, which holds under late withdrawal by \ref{ia3:F1}, implies $b_f\pi_L'-e_L'\ge\theta_{LN}\pi_L'-e_L'$, and in normalized units
    \begin{equation}\label{ia3:eq-income}
        z_{LN}\pi_L'-\Big(\frac{e_L}J\Big)'=\frac{16a^3\mathcal{Q}_U}{G^2H_LD_N}>0\qquad(\widehat{B}>\widehat{B}_U).
    \end{equation}
    If $\pi_L'\ge0$, then $b_f\pi_L'\ge0$ and $-e_L'>0$ because $\widehat{B}>\widehat{B}_e$.
    In both cases $U_A'>0$ above the onset.
    All thresholds are finite, and no limiting approximation enters any sign above the onset.
\end{proof}

A late withdrawal succeeds before the arbitrage trade with probability $B/[3(B+2\lambda)]<1/3$, so she loses more than two thirds of her races against the arbitrageurs.
Each success protects her capital from at least the next arbitrage loss, and she earns fees whenever she is deposited.
As blocks speed up, the success probability rises, so a smaller fraction of executed arbitrage trades hits her capital.
Her exposed trades per unit of time peak at $\widehat{B}_e$ and fall beyond it.
Falling passive capital and a rising fee flow served raise her fee income per unit of time while deposited.
If her deposited time falls, the fee income forgone per unit of lost deposited time is at most the late cutoff times her loss per executed arbitrage trade, since late withdrawal requires $b_f\le\theta_{LN}$.
Beyond $\widehat{B}_U$, \eqref{ia3:eq-income} shows that the arbitrage losses she saves dominate this forgone income, so her income still rises.
Her income gain comes from the arbitrage losses she avoids, not from a share of the prize.

\subsection{Duration of liquidity provision}\label{ia3:spells}

This subsection proves Proposition~\ref{prop:spells}, which uses only the model and late withdrawal.
The duration of liquidity provision starts when her redeposit executes at a block with a zero gap and ends at her next successful withdrawal.

\begin{proposition}[Duration of liquidity provision shortens]\label{ia3:prop-spells}
    Let $J,\lambda,B>0$, let her strategy be late withdrawal, and ignore a finite initial period.
    Write $X_B$ for the duration of liquidity provision at block rate $B$.
    For every $B^{+}>B$, $X_{B^{+}}$ is strictly shorter than $X_B$ in the sense of first-order stochastic dominance: at every duration $\upsilon>0$, the duration at the faster block rate is more likely to be at most $\upsilon$,
    \[
        \Pr(X_{B^{+}}\le\upsilon)>\Pr(X_B\le\upsilon).
    \]
    Both distribution functions are continuous and strictly increasing on $(0,\infty)$, so every population quantile with probability in $(0,1)$ strictly decreases as $B$ rises.
    The mean duration and its derivative are
    \begin{equation}\label{ia3:eq-EM2}
        \E X_B=\frac{\pi_L}{\omega_L}=\frac{12}J+\frac1\lambda+\frac{24\lambda+5J}{JB}+\frac{6\lambda}{B^2},\qquad \frac{d}{dB}\E X_B=-\frac{24\lambda+5J}{JB^2}-\frac{12\lambda}{B^3}<0.
    \end{equation}
    The result describes late withdrawal itself, so it holds in every recurrent late-withdrawal equilibrium without Assumption~\ref{ass:main}.
    It does not sign durations across a switch between equilibria, and it does not describe daily sample medians.
\end{proposition}

\begin{proof}
    Start deposited with a zero gap, and let $\Theta$ be the time until the gap first becomes profitable.
    The two first-step equations of the gap process give the Laplace transform $\mathcal{L}_\Theta$ of $\Theta$, written as a function of the transform argument $\zeta$:
    \begin{equation}\label{ia3:eq-EM4}
        \mathcal{L}_\Theta(\zeta)=\E\big[\exp(-\zeta\Theta)\big]=\frac{J^2}{J^2+4J\zeta+2\zeta^2}.
    \end{equation}
    Once the gap is profitable, she and the two arbitrageurs each send a transaction.
    The first of the three transactions arrives after a time $Y$ that is exponential with rate $3\lambda$.
    Its sender $\iota$ is independent of that time and equally likely to be any of the three.
    The next block arrives $Z/B$ after that arrival, where $Z$ is exponential with rate one.
    If $\iota$ is an arbitrageur, an arbitrage trade closes the gap at that block and her capital stays deposited.
    If $\iota$ is the active LP, let $\tilde Y$ be the further time from the arrival of her withdrawal to the arrival of the first arbitrage transaction.
    This time is exponential with rate $2\lambda$.
    Her capital leaves exactly when $Z/B<\tilde Y$.
    Otherwise, an arbitrage transaction arrives before that block and executes ahead of her withdrawal, so her withdrawal fails after her capital has borne the arbitrage loss.
    Her capital stays deposited when an arbitrageur arrives first or when her withdrawal fails.
    The episode, which runs from a zero gap to the executed arbitrage trade, then ends with her capital still deposited, and a new episode starts from a zero gap.

    Use independent and identically distributed trials $(\Theta_n,Y_n,Z_n,\iota_n,\tilde Y_n)$, one for each episode, with the same draws at every block rate.
    The duration $\Theta_n+Y_n+Z_n/B$ of trial $n$ falls as $B$ rises for every fixed set of draws, while its success indicator $\ind\{\iota_n=\text{LP},\ Z_n/B<\tilde Y_n\}$ rises for every fixed set of draws.
    So the index $n^*_B$ of the first successful trial weakly falls, and the duration $X_B=\sum_{n=1}^{n^*_B}(\Theta_n+Y_n+Z_n/B)$ weakly falls for every fixed set of draws.
    Strict dominance at each duration $\upsilon>0$ follows from a first-trial event of positive probability, in which $\iota_1$ is the active LP, $\Theta_1+Y_1<\upsilon$, $Z_1\in\big(B(\upsilon-\Theta_1-Y_1),B^{+}(\upsilon-\Theta_1-Y_1)\big]$ and $\tilde Y_1>Z_1/B$.
    On this event the duration ends in the first trial at both block rates, and $X_{B^{+}}\le\upsilon<X_B$.
    Positive densities of each trial's waiting times make each duration distribution function continuous and strictly increasing, so with strict dominance every quantile with probability in $(0,1)$ strictly falls.

    For an independent check by transforms, write $\mathcal{T}$ for the Laplace transform of a trial's duration and $\mathcal{T}_{\mathrm{succ}}$ for the transform of its duration on the event that the trial succeeds, both as functions of $\zeta$:
    \[
        \mathcal{T}(\zeta)=\mathcal{L}_\Theta(\zeta)\,\frac{3\lambda}{3\lambda+\zeta}\,\frac{B}{B+\zeta},\qquad \mathcal{T}_{\mathrm{succ}}(\zeta)=\mathcal{L}_\Theta(\zeta)\,\frac{\lambda}{3\lambda+\zeta}\,\frac{B}{B+2\lambda+\zeta}.
    \]
    The Laplace transform of the duration is
    \begin{equation}\label{ia3:eq-EM5}
        \E\big[\exp(-\zeta X_B)\big]=\frac{\mathcal{T}_{\mathrm{succ}}(\zeta)}{1-\mathcal{T}(\zeta)+\mathcal{T}_{\mathrm{succ}}(\zeta)},
    \end{equation}
    whose denominator, after cancellation, is a polynomial of degree five in $\zeta$.
    A trial succeeds with probability $\mathcal{T}_{\mathrm{succ}}(0)=B/[3(B+2\lambda)]$ and lasts $4/J+1/(3\lambda)+1/B$ on average.
    The stopping index is geometric with success probability $\mathcal{T}_{\mathrm{succ}}(0)>0$.
    The event $\{n^*_B\ge n\}$ depends only on the first $n-1$ trials, so summing over trials gives $\E X_B=\sum_n\Pr(n^*_B\ge n)\,[4/J+1/(3\lambda)+1/B]$, which equals the mean trial length divided by $\mathcal{T}_{\mathrm{succ}}(0)$.
    The renewal equation $\E X_B=[4/J+1/(3\lambda)+1/B]+(1-\mathcal{T}_{\mathrm{succ}}(0))\,\E X_B$ gives the same mean \eqref{ia3:eq-EM2}.
    Dividing $\pi_L$ by $\omega_L$ in \eqref{ia3:eq-occ} gives it too.
\end{proof}

Between a redeposit and her next successful withdrawal, faster blocks end every episode sooner and make it more likely that her withdrawal, once it arrives first, executes before an arbitrage transaction also arrives.
Both forces shorten the time until her next successful withdrawal.
Her deposited fraction of time also depends on how long she stays out after each withdrawal, so shorter durations do not by themselves determine that fraction.

\subsection{Price tracking and measurement identities}\label{ia3:tracking}

This subsection proves part (ii) of Lemma~\ref{lem:frequency}, that faster blocks shrink the calendar-time price gap.
It also shows that the gap sampled just after blocks rises with block speed.
Finally, it derives the mean inverse depth, the mean of one over depth, that executed liquidity-trader trades meet.
Parts (i) and (ii) below use only the model.
Part (iii) additionally requires an operating equilibrium with positive passive capital.

\begin{proposition}[Tracking and measurement]\label{ia3:prop-tracking}
    \begin{enumerate}[label=(\roman*),leftmargin=*]
        \item The stationary probabilities of a zero gap, a small gap, a profitable gap before any arbitrage transaction has arrived, and a profitable gap after one has arrived are
              \begin{equation}\label{ia3:eq-ME1}
                  \Big(\frac{2e}J,\ \frac{2e}J,\ \frac{e}{2\lambda},\ \frac eB\Big),\qquad e'(B)=\frac{1}{(1+TB)^2}>0.
              \end{equation}
        \item Let $\mathcal{E}_{\mathrm{cont}}$ be the mean absolute gap between the reference price and the pool price over calendar time, and let $\mathcal{E}_{\mathrm{end}}$ be the mean gap recorded just after each block executes, each block weighted by the length of the interval that follows it.
              Here $M$ and $H$ are the small and profitable gap levels of Section~\ref{sec:model}.
              Then
              \begin{equation}\label{ia3:eq-ME2}
                  \mathcal{E}_{\mathrm{cont}}=H-\frac{e(4H-2M)}J,\qquad \mathcal{E}_{\mathrm{end}}=e\Big(\frac{2M}J+\frac H{2\lambda}\Big),\qquad \mathcal{E}_{\mathrm{cont}}-\mathcal{E}_{\mathrm{end}}=\frac{He}B,
              \end{equation}
              and as $B$ rises, $\mathcal{E}_{\mathrm{cont}}$ falls while $\mathcal{E}_{\mathrm{end}}$ rises.
        \item In an operating equilibrium with positive passive capital, a liquidity trader whose transaction executes meets, on average, the time-average depth $D$.
              The capital in the pool that a trade meets is $Q=P+qI$, where $I$ equals one when the active LP's capital is deposited and zero otherwise, and the mean of one over this capital, which weights its two values by her deposited fraction of time and its complement, is
              \begin{equation}\label{ia3:eq-ME3}
                  \E[Q^{-1}]=\frac{\pi}{P+q}+\frac{1-\pi}{P}=\frac{k+\ell e}{r(B)}=\frac1{p_0},
              \end{equation}
              and, separately, liquidity-trader transactions execute at their arrival rate times $B/(B+\delta)$, so their execution rate rises with $B$.
    \end{enumerate}
\end{proposition}

\begin{proof}
    Every cycle of the gap, from a zero gap until a block closes it, spends a mean time $2/J$ at a zero gap and $2/J$ at a small gap, waits a mean $1/(2\lambda)$ until an arbitrage transaction arrives, and waits a mean $1/B$ more for the block that closes the gap.
    Dividing these mean times by the mean cycle length $1/e$ gives part (i), and capital does not change this law.
    A block that closes the gap records a zero gap, so the probabilities just after a block are $(2e/J+e/B,\ 2e/J,\ e/(2\lambda),\ 0)$, which give $\mathcal{E}_{\mathrm{end}}$, while part (i) gives $\mathcal{E}_{\mathrm{cont}}$.
    Since $e'>0$ and $4H>2M$, the two averages move in opposite directions, so a claim that this block-end statistic falls with block speed would be false.
    For example, at $J=\lambda=1$, $M=1$ and $H=3$, $\mathcal{E}_{\mathrm{end}}$ equals $7/11$ at $B=1$ and $7/10$ at $B=2$, while $\mathcal{E}_{\mathrm{cont}}$ equals $13/11$ and $1$, for any fixed profitability threshold between $1$ and $3$.
    For part (iii), capital is constant between successive blocks and liquidity traders execute before the next capital update.
    Conditional on the capital just after a block, the next Poisson block wait and future liquidity-trader arrivals and deadlines are independent of that capital.
    The expected number of requests served at the next block is therefore independent of its pre-update capital, just as the expected inter-block duration is.
    Consequently sampling at executed liquidity-trader transactions and calendar-time sampling give the same capital weights $\pi$ and $1-\pi$.
    Positive passive capital ensures that every surviving request is served.
    The entry condition in \ref{ia3:F3} equates the mean inverse depth with $(k+\ell e)/r(B)=1/p_0$.
    These identities do not specify AMM curvature, realized effective spreads or the composition of executed trade sizes.
    At the no-entry corner, where no passive capital enters, every successful liquidity trader meets $Q=q$.
    The inverse-depth formula, stated for positive passive capital, is not used there.
\end{proof}

Faster execution of arbitrage trades shortens the time that the pool price stays stale, so the calendar-time gap falls.
The block-end statistic, however, is recorded just after each block, and a block that executes an arbitrage trade leaves a zero gap.
At slow block rates this zero gap just after an arbitrage trade carries a disproportionate weight in the block-end statistic.
So the two statistics average different samples of the same gap process and move in opposite directions.
The empirical block-end tracking measure therefore needs explicit rules for blocks without swaps, for matching blocks to the reference price and for duration weights before it can be identified with either model statistic.
\clearpage
\end{document}